\documentclass[authoryear,preprint,11pt]{elsarticle}

\RequirePackage{tikz} % figures, diagrams, ..

\RequirePackage{amssymb}
\RequirePackage{amsthm}
\RequirePackage{enumerate}
\RequirePackage{color}
\RequirePackage{placeins}
\RequirePackage{pdfsync}
\RequirePackage{mathbbol}
\RequirePackage{natbib}
\RequirePackage{amsmath}
\RequirePackage{amsthm}
\RequirePackage[colorlinks]{hyperref}
\RequirePackage{xcolor}
\RequirePackage{array}
\RequirePackage{makecell}
\RequirePackage{booktabs}
\RequirePackage{fullpage}
\RequirePackage{graphicx} 
\RequirePackage{float} 
\RequirePackage{subcaption}
\RequirePackage{comment}
\RequirePackage{algorithm}
\usepackage{xr-hyper}
\usepackage{cleveref}
\RequirePackage{algpseudocode}
\definecolor{halfgray}{gray}{0.55} % chapter numbers will be semi transparent .5 .55 .6 .0
\definecolor{webgreen}{rgb}{0,.5,0}
\definecolor{webbrown}{rgb}{.6,0,0}
\definecolor{RoyalBlue}{rgb}{0,0.08,0.45}
\RequirePackage{hypernat}
\RequirePackage{multirow}
\RequirePackage{textcomp}
\usepackage{ulem}
\usepackage{rotating}

\usepackage{etoolbox}

\makeatletter
\newcommand*{\sortentry}[1]{%
  \if@filesw
    \immediate\write\@auxout{\string\uNAT@aux@sortentry{#1}}%
  \fi}
\newcommand*{\uNAT@aux@sortentry}{%
  \listgadd{\uNAT@bibsortlist}}
\newcommand*{\uNAT@bibsortlist}{}

\newcommand*{\uNAT@citekeys}{}

\newcommand*{\uNAT@writetocitelistsort}[1]{%
  \ifinlist{#1}{\uNAT@citekeys}
    {\ifdefvoid{\NAT@cite@list}
       {\def\NAT@cite@list{#1}}
       {\expandafter\def\expandafter\NAT@cite@list\expandafter{\NAT@cite@list,#1}}%
     \listgadd{\uNAT@foundkeys}{#1}}
    {}}

\newcommand*{\uNAT@writetocitelistforgotten}[1]{%
  \ifinlist{#1}{\uNAT@foundkeys}
    {}
    {\ifdefvoid{\NAT@cite@list}
       {\def\NAT@cite@list{#1}}
       {\expandafter\def\expandafter\NAT@cite@list\expandafter{\NAT@cite@list,#1}}}}

\newcommand*{\uNAT@sortcites}[1]{%
  \let\NAT@cite@list\@empty
  \let\uNAT@citekeys\@empty
  \let\uNAT@foundkeys\@empty
  \forcsvlist{\listadd{\uNAT@citekeys}}{#1}%
  \forlistloop{\uNAT@writetocitelistsort}{\uNAT@bibsortlist}%
  \forlistloop{\uNAT@writetocitelistforgotten}{\uNAT@citekeys}%
}

\def\NAT@citex%
  [#1][#2]#3{%
  \NAT@reset@parser
  \NAT@sort@cites{#3}%
  \uNAT@sortcites{#3}%<- this is new
  \NAT@reset@citea
  \@cite{\let\NAT@nm\@empty\let\NAT@year\@empty
    \@for\@citeb:=\NAT@cite@list\do
    {\@safe@activestrue
     \edef\@citeb{\expandafter\@firstofone\@citeb\@empty}%
     \@safe@activesfalse
     \@ifundefined{b@\@citeb\@extra@b@citeb}{\@citea%
       {\reset@font\bfseries ?}\NAT@citeundefined
                 \PackageWarning{natbib}%
       {Citation `\@citeb' on page \thepage \space undefined}\def\NAT@date{}}%
     {\let\NAT@last@nm=\NAT@nm\let\NAT@last@yr=\NAT@year
      \NAT@parse{\@citeb}%
      \ifNAT@longnames\@ifundefined{bv@\@citeb\@extra@b@citeb}{%
        \let\NAT@name=\NAT@all@names
        \global\@namedef{bv@\@citeb\@extra@b@citeb}{}}{}%
      \fi
     \ifNAT@full\let\NAT@nm\NAT@all@names\else
       \let\NAT@nm\NAT@name\fi
     \ifNAT@swa\ifcase\NAT@ctype
       \if\relax\NAT@date\relax
         \@citea\NAT@hyper@{\NAT@nmfmt{\NAT@nm}\NAT@date}%
       \else
         \ifx\NAT@last@nm\NAT@nm\NAT@yrsep
            \ifx\NAT@last@yr\NAT@year
              \def\NAT@temp{{?}}%
              \ifx\NAT@temp\NAT@exlab\PackageWarningNoLine{natbib}%
               {Multiple citation on page \thepage: same authors and
               year\MessageBreak without distinguishing extra
               letter,\MessageBreak appears as question mark}\fi
              \NAT@hyper@{\NAT@exlab}%
            \else\unskip\NAT@spacechar
              \NAT@hyper@{\NAT@date}%
            \fi
         \else
           \@citea\NAT@hyper@{%
             \NAT@nmfmt{\NAT@nm}%
             \hyper@natlinkbreak{%
               \NAT@aysep\NAT@spacechar}{\@citeb\@extra@b@citeb
             }%
             \NAT@date
           }%
         \fi
       \fi
     \or\@citea\NAT@hyper@{\NAT@nmfmt{\NAT@nm}}%
     \or\@citea\NAT@hyper@{\NAT@date}%
     \or\@citea\NAT@hyper@{\NAT@alias}%
     \fi \NAT@def@citea
     \else
       \ifcase\NAT@ctype
        \if\relax\NAT@date\relax
          \@citea\NAT@hyper@{\NAT@nmfmt{\NAT@nm}}%
        \else
         \ifx\NAT@last@nm\NAT@nm\NAT@yrsep
            \ifx\NAT@last@yr\NAT@year
              \def\NAT@temp{{?}}%
              \ifx\NAT@temp\NAT@exlab\PackageWarningNoLine{natbib}%
               {Multiple citation on page \thepage: same authors and
               year\MessageBreak without distinguishing extra
               letter,\MessageBreak appears as question mark}\fi
              \NAT@hyper@{\NAT@exlab}%
            \else
              \unskip\NAT@spacechar
              \NAT@hyper@{\NAT@date}%
            \fi
         \else
           \@citea\NAT@hyper@{%
             \NAT@nmfmt{\NAT@nm}%
             \hyper@natlinkbreak{\NAT@spacechar\NAT@@open\if*#1*\else#1\NAT@spacechar\fi}%
               {\@citeb\@extra@b@citeb}%
             \NAT@date
           }%
         \fi
        \fi
       \or\@citea\NAT@hyper@{\NAT@nmfmt{\NAT@nm}}%
       \or\@citea\NAT@hyper@{\NAT@date}%
       \or\@citea\NAT@hyper@{\NAT@alias}%
       \fi
       \if\relax\NAT@date\relax
         \NAT@def@citea
       \else
         \NAT@def@citea@close
       \fi
     \fi
     }}\ifNAT@swa\else\if*#2*\else\NAT@cmt#2\fi
     \if\relax\NAT@date\relax\else\NAT@@close\fi\fi}{#1}{#2}}
\makeatother

\newtheorem{prop}{Proposition}

\theoremstyle{remark}

\newcommand{\R}{\mathbb{R}}

\newcommand{\E}{\ensuremath{\mathbb{E}}}
\renewcommand{\Pr}{\ensuremath{\mathbb{P}}}

\renewcommand{\hat}{\widehat}
\renewcommand{\tilde}{\widetilde}

\def\bI{\mathbf{I}}

\def\det{\operatorname{det}}

\def\bzero{\boldsymbol 0}
\def\bone{\boldsymbol 1}
\def\bA{\boldsymbol A}

\def\bJ{\boldsymbol{J}}

\def\bX{\boldsymbol X}
\def\bW{\boldsymbol W}
\def\bY{\boldsymbol Y}
\def\bZ{\boldsymbol Z}

\def\bh{\boldsymbol h}

\def\bw{\boldsymbol w}
\def\bu{\boldsymbol u}
\def\bh{\boldsymbol h}

\def\bx{\boldsymbol x}
\def\by{\boldsymbol y}
\def\bz{\boldsymbol z}

\def\bzero{\boldsymbol 0}

\def\bSigma{\boldsymbol \Sigma}

\def\bmu{\boldsymbol \mu}

\DeclareMathOperator*{\argmin}{arg\,min}

\newcommand{\vague}{\stackrel{\lower0.2ex\hbox{$\scriptscriptstyle
                    \it{v} $}}{\to}}
\newcommand{\weak}{\stackrel{\lower0.2ex\hbox{$\scriptscriptstyle
                    \it{w} $}}{\to}}
\newcommand{\what}{\stackrel{\lower0.2ex\hbox{$\scriptscriptstyle
                    \it{\hat{w}} $}}{\to}}
\newcommand{\eqdis}{\stackrel{\lower0.2ex\hbox{$\scriptscriptstyle
                    \mathrm{d}$}}{=}}
\newcommand{\distr}{\stackrel{\lower0.2ex\hbox{$\scriptscriptstyle
                    \it{d} $}}{\to}}
\journal{}
\makeatletter
\def\ps@pprintTitle{%
  \let\@oddhead\@empty
  \let\@evenhead\@empty
  \let\@oddfoot\@empty
  \let\@evenfoot\@oddfoot
}
\makeatother

\NewDocumentCommand{\MakeTitleInner}{ +m +m +m }{
    \newpage%
    \null%
    \vskip 2em%
    \begin{center}%
        \let \footnote \thanks
        {\LARGE #1 \par}%  title
        \vskip 1.5em%
        {%
            \large
            \lineskip .5em%
            \begin{tabular}[t]{c}%
                #2% author
            \end{tabular}\par%
        }%
        \vskip 1em%
        {\large #3}%  date
    \end{center}%
    \par
    \vskip 1.5em%
}
\NewDocumentCommand{\MakeTitle}{ +m +m +m }{%
    \begingroup
        \renewcommand\thefootnote{\@fnsymbol\c@footnote}%
        \def\@makefnmark{\rlap{\@textsuperscript{\normalfont\@thefnmark}}}%
        \long\def\@makefntext##1{\parindent 1em\noindent
            \hb@xt@1.8em{%
                \hss\@textsuperscript{\normalfont\@thefnmark}%
            }##1%
        }%
        \if@twocolumn
            \ifnum \col@number=\@ne
                \MakeTitleInner{#1}{#2}{#3}
            \else
                \twocolumn[\MakeTitleInner{#1}{#2}{#3}]%
            \fi
        \else
            \newpage
            \global\@topnum\z@   % Prevents figures from going at top of page.
            \MakeTitleInner{#1}{#2}{#3}
        \fi
        \thispagestyle{plain}\@thanks
    \endgroup
    \setcounter{footnote}{0}%
    \setcounter{section}{0}%
    \setcounter{figure}{0}%
    \setcounter{table}{0}%
}
\makeatother

\usepackage{listings}
\definecolor{codegreen}{rgb}{0,0.6,0}
\definecolor{codegray}{rgb}{0.5,0.5,0.5}
\definecolor{codepurple}{rgb}{0.58,0,0.82}
\definecolor{backcolour}{rgb}{0.95,0.95,0.92}
\lstdefinestyle{mystyle}{
    backgroundcolor=\color{backcolour},   
    commentstyle=\color{codegreen},
    numberstyle=\tiny\color{codegray},
    basicstyle=\ttfamily\footnotesize,
    breakatwhitespace=false,         
    breaklines=true,                 
    captionpos=b,                    
    keepspaces=true,                 
    showspaces=false,                
    showstringspaces=false,
    showtabs=false,                  
    tabsize=2,
    keepspaces=true,
    columns=fullflexible
}
\begin{document}

\begin{frontmatter}



\title{
%\textcolor{red}{T1:} Spatial extremes: Gaussian location-scale mixture models at work\\
%\textcolor{red}{T2:} Extended modelling of spatial extremes with Gaussian location-scale mixtures\\
%\textcolor{red}{T3:} 
Continuous mixtures of Gaussian processes as models for spatial extremes
}

\author[inst4,inst5,inst6]{Lorenzo Dell'Oro}

%\affiliation[inst1]{organization={Dipartimento di Scienze Statistiche, Università di Padova},%Department and Organization
%            city={Padua},
%            country={Italy}}
\affiliation[inst4]{organization={IMAG, Université de Montpellier},%Department and Organization
            city={Montpellier},
            country={France}}
\affiliation[inst5]{organization={LEMON, Inria},%Department and Organization
            city={Montpellier},
            country={France}}
\affiliation[inst6]{organization={LJAD, Université Côte d'Azur},%Department and Organization
            city={Nice},
            country={France}}

\author[inst2]{Carlo Gaetan}

%\affiliation[inst2]{organization={Dipartimento di Scienze Ambientali, Informatica e Statistica, Università Ca' Foscari di Venezia},%Department and Organization
\affiliation[inst2]{organization={DAIS, Università Ca' Foscari di Venezia},
            city={Venice},
            country={Italy}}

\author[inst3]{Thomas Opitz}

%\affiliation[inst3]{organization={Biostatistics and Spatial Processes, INRAE},%Department and Organization
\affiliation[inst3]{organization={BioSP, INRAE},           % addressline={}, 
            city={Avignon},
            country={France}}

%% Abstract
\begin{abstract}
Spatial modelling of extreme values allows studying the risk of joint occurrence of extreme events at different locations and is of significant interest in climatic and other environmental sciences.
%A popular class of models, due to its flexibility and importance for asymptotic representations, is given by random location-scale mixtures of Gaussian processes. The general principle of a location or scale mixture is that a spatial ``baseline'' process is multiplied or shifted by a random variable, potentially altering its extremal dependence behaviour. 
A popular class of dependence models for spatial extremes is that of random location-scale mixtures, in which a spatial ``baseline'' process is multiplied or shifted by a random variable, potentially altering its extremal dependence behaviour.
Gaussian location-scale mixtures retain benefits of their Gaussian baseline processes while overcoming some of their limitations, such as symmetry, light tails and weak tail dependence. We review properties of Gaussian location-scale mixtures and develop novel constructions with interesting features, together with a general algorithm for conditional simulation from these models.
We leverage their flexibility to propose extended extreme-value models, that allow for appropriately modelling not only the tails but also the bulk of the data. This is important in many applications and  avoids the need to explicitly select the events considered as extreme. We propose new solutions for likelihood inference in parametric models of Gaussian location-scale mixtures, in order to avoid the numerical bottleneck given by the latent location and scale variables that can lead to high computational cost of standard likelihood evaluations.
%We leverage their flexibility to propose extended extreme-value models that allow for appropriately modelling not only the tails but also the bulk of the data. This avoids the uncertainty brought by the need to select an artificial threshold to define extreme events and separate the bulk from the tail.
%We also propose new solutions for likelihood inference in parametric models of Gaussian location-scale mixtures, in order to bypass the numerical bottleneck given by the latent location and scale variables that lead to high computational cost of standard likelihood evaluations.
The effectiveness of the models and of the inference methods is confirmed with simulated data examples, and we present an application to wildfire-related weather variables in Portugal. Although not detailed here, the  approaches would also be straightforward to use for modelling multivariate (non spatial) data. 
\end{abstract}



%% Keywords

\end{frontmatter}

\section{Introduction}

Gaussian processes  play a central role in spatial and spatio-temporal statistics by providing a flexible and tractable framework for modelling dependence in discrete and continuous domains. Extreme-value processes, by contrast, arise as asymptotic limits of suitably normalised pointwise maxima or threshold exceedances of stochastic processes and underpin modern theory for spatial extremes through max-stable and generalized Pareto processes \citep{Brown:Resnick:1977,de_Haan:1984,ferreira2014generalized}. The connection between these two classes of models is both theoretical and practical. Large families of max-stable processes can be constructed as limits of transformed Gaussian processes, for example via spectral representations in which exponentiated or scaled Gaussian processes serve as building blocks for max-stable dependence structures \citep{smith1990max,Kabluchko:Schlather:de_Haan:2009,Opitz2013}. These constructions establish a direct link between Gaussian dependence and asymptotic extreme-value behaviour. 

%From a modelling perspective, Gaussian processes are frequently used as latent components in hierarchical extreme-value models, governing spatial variation in marginal parameters such as location, scale, or tail index \citep{casson1999spatial,cooley2007bayesian}.

Despite these theoretical strengths, growing empirical and practical evidence suggests that max-stable processes are often ill-suited to the demands of modern  data analysis especially in the environmental context. Their asymptotic nature implies that they are primarily designed for modelling block maxima or exceedances above very high thresholds, whereas many contemporary applications require inference and stochastic simulation also at moderate levels and across a wide range of the distribution. Moreover, max-stable models typically impose strong forms of tail dependence that may not be supported by observed data, and likelihood-based inference is computationally challenging, often relying on composite likelihoods that become inefficient as the number of spatial locations increases.

Recent work has therefore argued for a shift away from max-stable processes as default modelling tools for spatial extremes especially for  environmental data  \citep{Huser2025}. Alternative approaches aim to retain the key insights of extreme-value theory while relaxing rigid asymptotic assumptions and improving computational scalability. 
These include hybrid frameworks mixing max-stability with other dependence types \citep{wadsworth2012dependence,Bacro:Gaetan:Toulemonde:2016},
generalized Pareto processes for threshold exceedances \citep{defondeville2018high}, and random-scale constructions. The latter retain a structure similar to asymptotic models but capture tail dependence more flexibly, including cases that are degenerate in asymptotic frameworks since spatial dependence gradually weakens and ultimately vanishes when events become more extreme \citep{Opitz:2016,Huser:Opitz:Thibaud:2017,Huser:Wadsworth:2019}.

Classical finite  mixtures of multivariate Gaussian distributions provide a powerful framework for density approximation and clustering \citep{mclachlan2000finite}, while more structured constructions such as scale mixtures, location mixtures, and mean--variance mixtures naturally accommodate heavy tails and heteroskedasticity \citep{andrews1974scale,barndorff1977infinitely}. Further extensions incorporating skewness have been developed through skew-normal and skew-$t$ formulations, offering additional flexibility for asymmetric data \citep{azzalini1996multivariate,azzalini2014skew}; a unified framework is presented in \cite{arellano2021formulation}. 
In spatial and spatio-temporal settings, Gaussian mixture and mixture-type models have been shown to effectively capture nonstationarity and complex dependence patterns beyond classical Gaussian processes \citep{fuentes2002periodogram,reich2012flexible}.

From an extreme-value perspective, location and scale mixtures of Gaussian processes (\citealp{Huser:Opitz:Thibaud:2017}, \citealp{krupskii2018factor}, including $t$-type processes, \citealp{Morris.al.2017}, and mean--variance mixtures, \citealp{%Zhang:Shaby:Wadsworth:2022,
zhang2022modeling})
form a bridge between classical Gaussian dependence and asymptotic extreme-value behaviour and offer a complementary mechanism for tail thickening by introducing latent random scaling that induces heavy-tailed behaviour and more realistic clustering of large values.
Another recent approach to scale mixtures of spatial processes with asymmetry in the tail dependence is made by \cite{krupskii2026skewmultiv}.

Despite their appealing modelling properties, the practical deployment of such rich mixture-of-Gaussian models is often limited by computational considerations. Likelihood-based inference typically involves high-dimensional vectors of latent variables and can lead to multimodal likelihood surfaces and expensive numerical integration. This challenge is exacerbated in spatial or multivariate contexts, where dependence structures further increase computational complexity. As a result, there is a strong need for computationally feasible inferential strategies, including efficient EM-type algorithms, Bayesian hierarchical formulations with scalable Markov chain Monte Carlo schemes, or approximate methods such as variational inference \citep{dempster1977maximum,fruhwirth2006finite,blei2017variational}. 

When direct likelihood maximisation is infeasible, two main inference approaches are typically adopted for Gaussian location and scale mixtures: i) EM algorithms,  which have recently been reviewed by \citet{lee2019mean}. However, they need to be adapted to the specific distributions of the random scale or location variables, and have not yet been studied in the general case in which both variables are present. Moreover, they can usually only be applied to the original marginal scales, and not to copula models; ii) Bayesian methods such as that proposed by \citet{Zhang:Shaby:Wadsworth:2022} in the context of censored inference on the tails of a scale mixture of a transformed Gaussian process.
Hence, developing inference procedures that preserve model flexibility while remaining computationally tractable is a key methodological objective that we try to address with this work.

This paper has three main objectives. First, in Section \ref{section:gaussian_locationscale_mixtures} we review the key theoretical and practical properties of Gaussian location-scale mixture models and introduce new constructions with enhanced tail flexibility and rich dependence structures.
We also develop a general algorithm for conditional simulation from these models, detailed in Section \ref{section:conditional_simulation}. 

Second, we address the computational challenge of likelihood-based inference for Gaussian location-scale mixtures. In particular, we propose a new two-step inferential method that exploits two different data transformations: in the first step, we use multivariate information to estimate the dependence parameter of the baseline Gaussian process; in the second step, we use univariate information to estimate the parameter of the random scale and location variables. Details are provided in Section \ref{section:estimation_methods}. This strategy alleviates the numerical burden induced by the latent variables, enabling the scalable and practical implementation of this class of models.
The new inferential method is evaluated on simulated data in Section \ref{section:simulation_studies}.

Finally, we leverage this flexibility to develop extended extreme-value models that describe both the bulk and the extremes of the distribution, moving beyond purely extreme-centric frameworks. This is important for analysing compound events, where not all data contributing to an extreme aggregate are extreme individually, and for stochastic generation of complete datasets used as inputs in impact models. We illustrate the models by analysing fire weather data in Portugal, using daily data of the Fire Weather Index available for more than 20 years at around 500 pixels, see Section \ref{section:application}.

%\texttt{
%\begin{itemize}
%    \item Literature (flexibility and utility of multivariate and spatial Gaussian mixtures, including scale mixtures, location mixtures, mean-variance mixtures, skew models)
%    \item Need for computationally feasible solutions for inference
%    \item Idea of extended modelling: there is a need for tail-flexible models in margins and dependence structure. Asymptotic extreme-value models lack flexibility: POT stability is mathematically elegant but not very realistic, as confirmed by empirical evidence. Model not only extreme events but the full distribution: this is important for analysing compound events (where not all data contributing to an extreme aggregate are extreme individually) and for stochastic generation of complete datasets used as inputs in impact models. Not only marginal distributions (EGPDs) but also dependence structure (copula).   
%\end{itemize}
%}

\section{Notation and definitions}\label{section:definitions}
We first introduce various mathematical objects used repeatedly throughout this paper. Whenever a mathematical operation involves a vector, say $\mathbf{a}$, and a scalar, say $b$, such as $\mathbf{a}+b$ or $\mathbf{a}/b$, the operation must be interpreted component-wise.

\subsection{Univariate distributions}

Let $X$ a random variable with cumulative distribution function (cdf)
$F$,  and survival function  $\overline{F}$.  
In the special case that $X$ is a standard Gaussian or Student random variable with degrees of freedom equal to $\nu$, $F$ is indicated by either a $\Phi$ or a $T_\nu$. The density function of a $d$-dimensional multivariate Gaussian with mean vector $\bzero$ and variance-covariance matrix $\bSigma$ is denoted as $\phi_d(\cdot;\bSigma)$, while $\phi:=\phi_1$ refers to the univariate standard Gaussian density.

The notation $X\sim\text{E}(\lambda)$ means that $X$ follows an exponential distribution with rate parameter $\lambda$, while $X\sim\text{AL}(\lambda_1,\lambda_2)$ means that $X$ follows an asymmetric Laplace distribution \citep{kotz2001laplace,gong2022asymmetric}, with cdf
\begin{equation}\label{eq:asymm_laplace_distrib}
    \begin{array}{ll}
    F_X(x;\,\lambda_1,\lambda_2)=\begin{cases}
        1-\lambda_2(\lambda_1+\lambda_2)^{-1}\exp(-\lambda_1 x), &\quad \text{if}\; x\ge0 \\
        \lambda_1(\lambda_1+\lambda_2)^{-1}\exp(\lambda_2 x), &\quad \text{if}\; x<0.
    \end{cases}
    \end{array}
\end{equation}
In the sequel, we exploit the fact that if $X_i$, $i=1,2$, are two independent exponential random variables with rate $\lambda_i$ then $X_1-X_2\sim\text{AL}(\lambda_1,\lambda_2) $.

A cdf $F$ with upper endpoint $x^*=\infty$ is said to be \textit{regularly varying} with index $\alpha\ge 0$ if $\bar{F}(tx)/\bar{F}(x)\to x^\alpha$ as $t\to\infty$, for any $x>0$.

The random variable $X$ with cdf $F_X$ is heavier-tailed than the random variable  $Y$ with cdf $F_Y$ if 
$$
\lim_{t \to \infty} 
\frac{\overline{F}_X(t)}{\overline{F}_Y(t)} 
= \infty .
$$

\subsection{Tail dependence coefficients for spatial processes}

 A common way to quantify the strength of the dependence in the upper tail region of two random random variable $X$ and $Y$ is through the bivariate \textit{(upper) tail dependence coefficient} $\chi$. It is defined, for $0<p<1$, as the limit  of
\begin{equation*}%\label{eq:chi_u_bivariate}
\begin{split}
        \chi(p)=&\,\Pr\left(F_{Y}(Y)>p\mid F_{X}(X)>p\right)=\frac{\Pr\left(F_{X}(X)>p,F_{Y}(Y)>p \right)}{\Pr(F_{X}(X)>p)}
        %\\        =&\,\frac{\Pr\left(F_{X}(X)>p,F_{Y}(Y)>u \right)}{1-p}
\end{split}
\end{equation*}
when this limit exists \citep{Joe:1997}, i.e.
\begin{equation*}%\label{eq:chi_bivariate}
	\chi=\lim_{p\rightarrow 1^-} \chi(p) \in [0,1].
\end{equation*}
If $\chi>0$, $X$ and $Y$ are said to be \textit{asymptotically dependent} (AD), while if $\chi=0$ the two variables are \textit{asymptotically independent} (AI).
Therefore, this coefficient discriminates between the two extremal dependence classes, and provides a quantification of the extremal dependence within the class of asymptotically dependent bivariate distributions, but it does not discriminate among asymptotically independent distributions.
An alternative measure that overcomes this limitation is $\bar{\chi}$ \citep{coles1999dependence}. It is defined, for $0<p<1$, as
\begin{equation*}
\begin{split}
    \bar{\chi} =&\lim_{p\rightarrow 1^-} \frac{2 \log \Pr(F_{X}(X)>p)}{\log\Pr(F_{X}(X)>p, F_{Y}(Y)>p)}-1
    %\\    =&\lim_{p\rightarrow 1^-} \frac{2\log(1-p)}{\log\Pr(F_{X}(X)>p, F_{Y}(Y)>p)}-1,
\end{split}
\end{equation*}
with $-1<\bar{\chi}\le 1$. If $X$ and $Y$ are asymptotically dependent, $\bar{\chi}=1$. On the other hand, for asymptotically independent variables, the value of this coefficient increases with the strength of the extremal dependence. In the case of independent variables, $\bar{\chi}=0$.
It follows that the pair $(\chi,\bar{\chi})$ provides information about the extremal dependence of $(X,Y)$ both in case of asymptotic dependence and independence.
The two coefficients can also refer to the lower tails; their values can be computed by replacing $X$ and $Y$ with $-X$ and $-Y$, respectively. In the following, when necessary, $\chi_U$ and $\bar{\chi}_U$ refer to the upper tail and $\chi_L$ and $\bar{\chi}_L$ to the lower tail, while when not specified, $\chi$ and $\bar{\chi}$ implicitly refer to the upper tail.

A real-valued spatial  process defined on a  domain $\mathcal{S}$ will be shortly denoted by $\{X(\mathbf{s})\}$ where $\mathbf{s}$ is a generic spatial location in $\mathcal{S}$.

The  coefficients $\chi$ and $\bar{\chi}$ can  also be evaluated for any pair of random variables  $X(\mathbf{s}_1)$ and $X(\mathbf{s}_2)$ of $\{X(\mathbf{s})\}$,  obtaining the coefficients $\chi(\mathbf{s}_1,\mathbf{s}_2)$ and $\bar{\chi}(\mathbf{s}_1,\mathbf{s}_2)$.
For simplicity and when there is no ambiguity, we use  $\chi$ and $\bar{\chi}$ to refer to $\chi(\mathbf{s}_1,\mathbf{s}_2)$ and $\bar{\chi}(\mathbf{s}_1,\mathbf{s}_2)$.
Likewise, if there is a correlation function $\rho(\mathbf{s}_1,\mathbf{s}_2)$ for $\{X(\mathbf{s})\}$, it will be denoted $\rho$ shortly.

\section{Gaussian location-scale mixtures}\label{section:gaussian_locationscale_mixtures}

\subsection{Model definition}\label{section:general_model_definition}

We define the general Gaussian location-scale mixture model as:
\begin{equation}\label{eq:gaussian_locationscale_mixtures}
    X(\mathbf{s}) = S + R \; W(\mathbf{s}), \quad \mathbf{s}\in\mathcal{S}\subset\mathbb{R}^d,
\end{equation}
where $S\in\R$, $R\ge0$, $\{W(\mathbf{s})\}$ is a standardized Gaussian process with means $0$, variances $1$ and correlation function $\rho(\mathbf{s}_1,\mathbf{s}_2)$, for two locations $\mathbf{s}_1,\mathbf{s}_2\in\mathcal{S}$, and $(S,R)$ is independent of $\{W(\mathbf{s})\}$.
For notational simplicity, we also assume by default that $S$ and $R$ are independent of each other, although this is not strictly necessary; indeed, some model classes do not make this assumption: two examples are normal mean-variance mixtures \citep{barndorff1982normal,zhang2022modeling} where $S =c R^2$ with $c\in\R$, and the multivariate skew-\textit{t} distribution \citep{azzalini2003distributions}, that can be written as a Gaussian location-scale mixture with dependence between $S$ and $R$ \citep[see][]{arellano2021formulation}.
 
The general specification \eqref{eq:gaussian_locationscale_mixtures} encompasses the following three subclasses of models, for which we present in the next subsection some noteworthy specification:
\begin{enumerate}
\item if $R = c_1\in\R^+$, it corresponds to a Gaussian location mixture. Without loss of generality, we assume $c_1=1$;
\item if $S = c_2\in\R$, it corresponds to a Gaussian scale mixture. Without loss of generality, we assume $c_2=0$;
\item if no constraint is placed on the values of $S$ and $R$, we are in the general case of Gaussian location-scale mixtures.
\end{enumerate}

In the presentation of the extremal dependence properties of a particular formulation of the model, the focus is directed towards the upper tail region of the distributions.
 Lower-tail properties are the same as upper-tail properties when $S$ is a symmetric distribution; otherwise, we can study lower-tail properties by considering the upper-tail properties of the modified process $-S+RW(\mathbf{s})$. 

\begin{figure}[htbp]
	\centering
	\begin{subfigure}{\linewidth}
		\centering
		\includegraphics[width=0.85\linewidth]{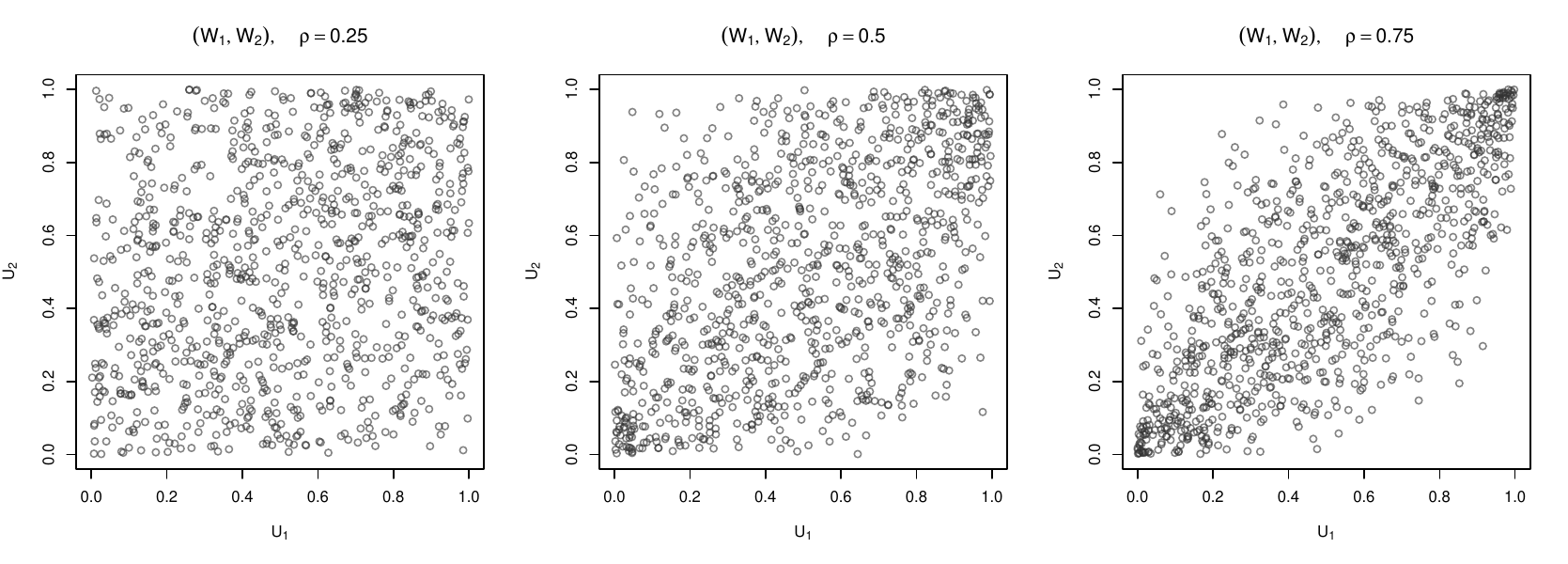}
	\end{subfigure}
	%\hfill
	\begin{subfigure}{\linewidth}
		\centering
		\includegraphics[width=0.85\linewidth]{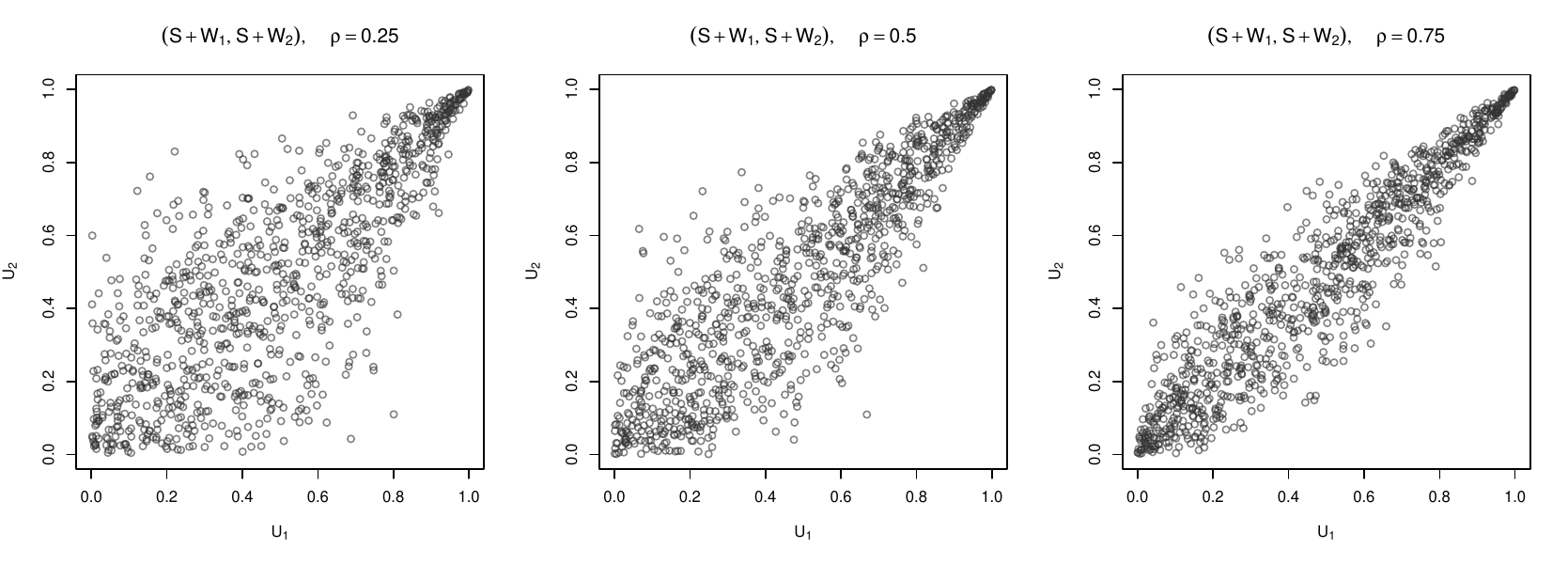}
		%    \caption{Simulation of 800 replications of model LM2, transformed in the uniform scale, with $S=S_1-S_2$, $S_1\sim\text{E}(\lambda_1)$, $S_2\sim\text{E}(\lambda_2)$, $\lambda_1=0.4$ and $\lambda_2=2.5$; $\text{cor}(W_1,W_2)=0.25$, $\text{cor}(W_3,W_4)=0.50$ and $\text{cor}(W_5,W_6)=0.75$.}
		%   \label{fig:biv_plot_2}
	\end{subfigure}
	%\hfill
	\begin{subfigure}{\linewidth}
		\centering
		\includegraphics[width=0.85\linewidth]{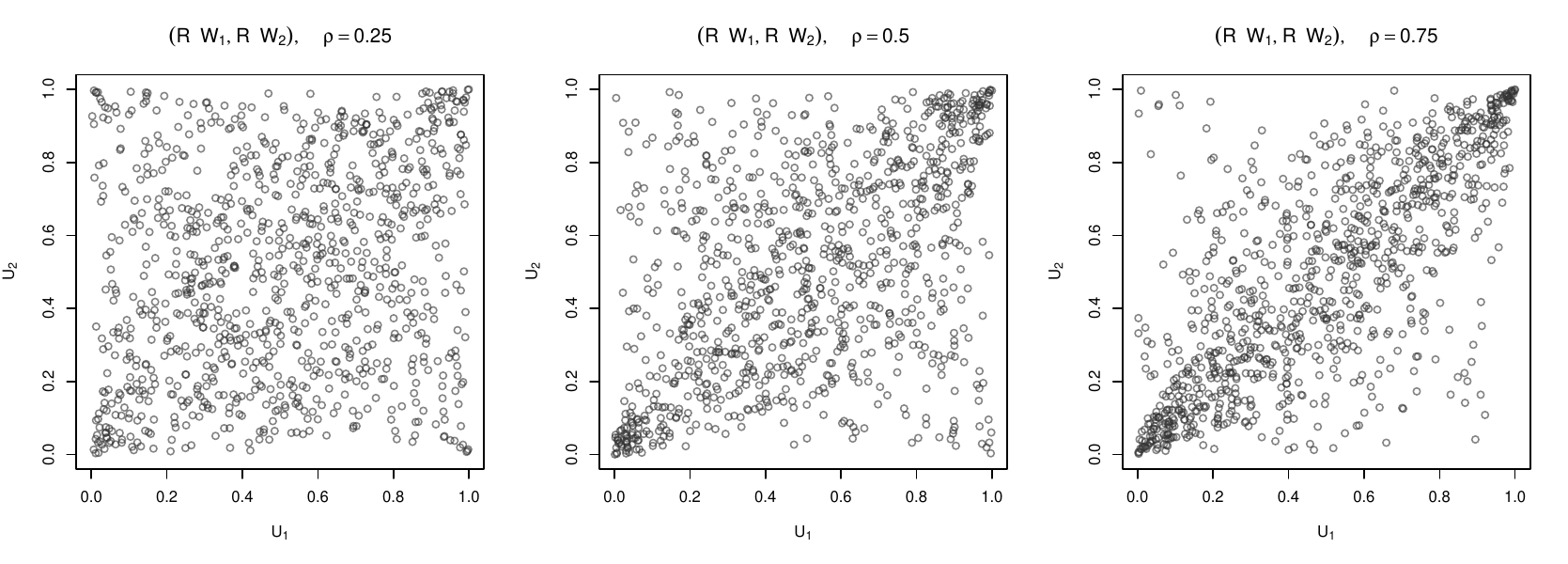}
		%    \caption{Simulation of 800 replications of model SM3, transformed in the uniform scale, with $R=1/\sqrt{G}$, $G\sim$ Gamma$(\nu/2,\nu/2)$ and $\nu=2$; $\text{cor}(W_1,W_2)=0.25$, $\text{cor}(W_3,W_4)=0.50$ and $\text{cor}(W_5,W_6)=0.75$.}
		%  \label{fig:biv_plot_3}
	\end{subfigure}
	%\hfill
	\begin{subfigure}{\linewidth}
		\centering
		\includegraphics[width=0.85\linewidth]{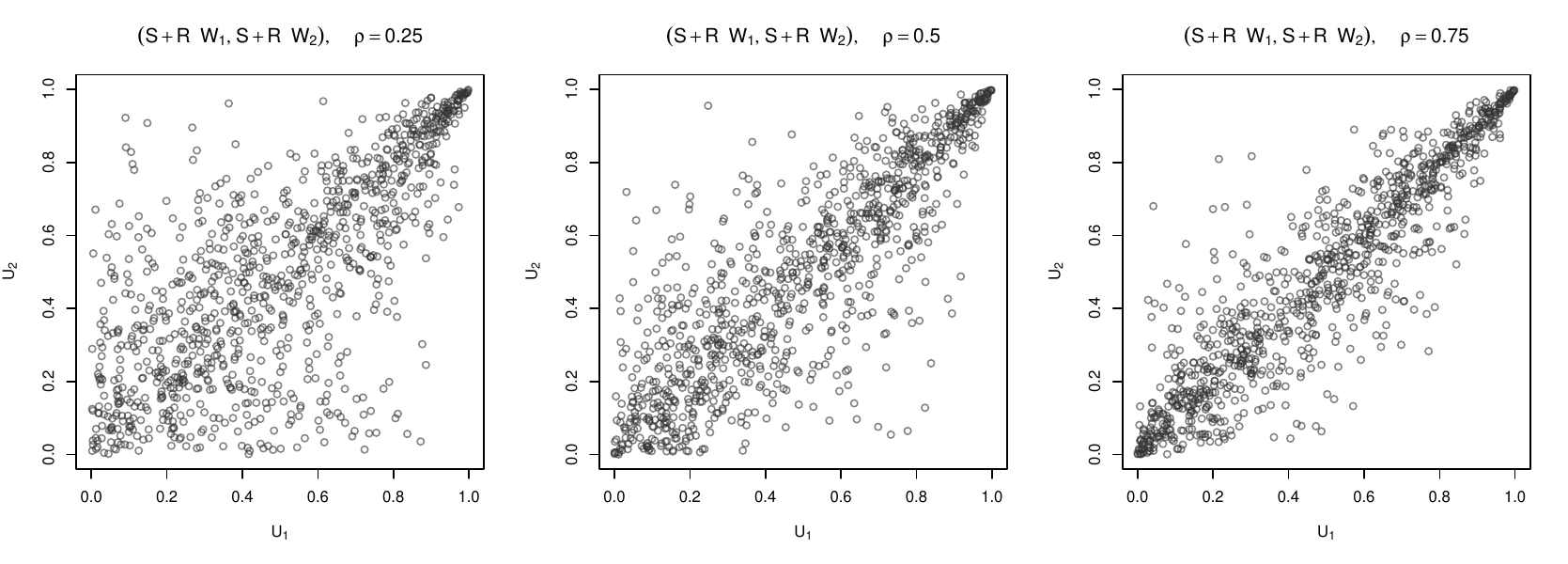}
		%    \caption{Simulation of 800 replications of model LSM2, transformed in the uniform scale, with $S=S_1-S_2$, $S_1\sim\text{E}(\lambda_1)$, $S_2\sim\text{E}(\lambda_2)$, $\lambda_1=0.4$, $\lambda_2=2.5$ and $R\sim\sqrt{E}$, $E\sim\text{E}(1/2)$; $\text{cor}(W_1,W_2)=0.25$, $\text{cor}(W_3,W_4)=0.50$ and $\text{cor}(W_5,W_6)=0.75$.}
		% \label{fig:biv_plot_4}
	\end{subfigure}
	\caption{Simulation of 800 replications of three bivariate Gaussian random variables, with $\rho=\text{cor}(W_1,W_2)=0.25,\, 0.5,\, 0.75$, respectively (top row), and of the corresponding models LM2, SM3 and LSM2 (second to fourth row). All variables were transformed to the uniform scale marginally.}
	\label{fig:biv_plots}
\end{figure}

To give a quick idea of the variety of bivariate dependence produced by different specifications  of  \eqref{eq:gaussian_locationscale_mixtures}, Figure \ref{fig:biv_plots} shows firstly three sets  of 800 simulated values from a two-dimensional  Gaussian vector $(W_1,W_2)^T$  with different correlation, namely $\rho=0.25,0.50,0.75$. 
Then each simulated set is modified by 800 simulated values of  $S$ and $R$  yielding realisations from $X_i$, $i=1,2$. All datasets are presented on uniform margins, i.e. as empirical copulas, to facilitate comparison of dependence properties.
\addtolength{\leftmargini}{1cm}
\begin{itemize}
	\item[(LM2)] $X_i=S+W_i$, with $S=E_1-E_2$, $E_1\sim\text{E}(0.4)$, $E_2\sim\text{E}(2.5)$;
	\item[(SM3)] $X_i= R \; W_i$  with $R=1/\sqrt{E_3}$, $E_3\sim$ E$(1)$; 
	\item[(LSM2)] $X_i=S+R \; W_i$ with $S=E_1-E_2$ and  $R=\sqrt{E_4}$, $E_4\sim$ E$(1/2)$.
\end{itemize} 
\addtolength{\leftmargini}{-1cm}
Section \ref{section:identifiability2_model} of the supplementary material provides an additional visualisation of the extremal dependence structure of these four models and a comparison of their distributions based on bivariate characteristic functions.

The Gaussian location-scale mixture model \eqref{eq:gaussian_locationscale_mixtures}
 can serve as a copula model  that provides only the dependence structure of the spatial process. In this case the  ``observed'' process $\{Y(\mathbf{s})\}$ has    specific marginal distributions.
Assuming for  notational simplicity that $\{X(\mathbf{s})\}$ and $\{Y(\mathbf{s})\}$ have space-invariant marginal distributions $F_X$ and $F_Y$, %the copula model
%Note that $F_\mathbf{s}$ can be non-stationary in space. %, while we generally assume the copula model \eqref{eq:gaussian_locationscale_mixtures} to be stationary.
it is possible to link the two processes with
%\begin{equation}\label{eq:margin_transf_YX}
$Y(\mathbf{s}) = F_Y^{-1}\left[F_X\{X(\mathbf{s})\}\right]$, for each $\mathbf{s}\in\mathcal{S}.$
%\end{equation}
%where $F_X$ is the cdf  of $X(\mathbf{s})$ and $F_Y$ is the cdf of $Y(\mathbf{s})$.
For an example of $F_Y$  we refer to Section \ref{section:application}.

\subsection{Examples }\label{section:example_models_gaussiansclocmixt}

We introduce examples of models from the three subclasses. Some of them were already discussed in previous literature, but we also propose new instances. 
 Table \ref{table:examples_gaussian_locscmix_models} condenses the assumptions about the distributions of $S$ and $R$, as well as  the resulting multivariate and marginal distributions when they correspond to a known family in the literature. It also includes their extremal dependence properties.

%\begin{landscape}
\begin{sidewaystable}%[h!]
\begin{center}
\caption{Examples of Gaussian location-scale mixture models detailed in Section \ref{section:example_models_gaussiansclocmixt}, with the names of the resulting multivariate or marginal distributions for $\bX$ when known, and their extremal dependence properties. The parameter $\rho\in(-1,1)$ is the bivariate correlation of the Gaussian process $\{W(\mathbf{s})\}$ for two generic spatial locations.
%, while $\Phi$ is the standard univariate Gaussian distribution function and $T_\alpha$ is the univariate Student's \textit{t} distribution function with $\alpha$ degrees of freedom.
}
%\scriptsize
\footnotesize
\begin{tabular}{|c|c|c|c|c|c|} 
 \hline
 Model & $F_S$, $F_R$  & Resulting distribution & Extremal dependence & Bivariate $\chi$ & Bivariate $\bar{\chi}$ \\ [0.5ex] 
 \hline\hline
 Gaussian & - & Multivariate Gaussian & AI & $0$ & $\rho$ \\ 
 \hline
 LM1 & $S\sim\text{E}(\lambda)$ & \makecell[c]{Closed form but\\not known family} & \makecell[c]{Upper tail: AD for $\lambda<\infty$\\Lower tail: AI} & $2[1-\Phi(\lambda\sqrt{1-\rho}/\sqrt{2})]$ & $1$ \\ 
 \hline
 LM2 & $S\sim\text{AL}(\lambda_1,\lambda_2)$ %\makecell[c]{$S=S_1-S_2\sim\text{AL}$\\ $S_1\sim\text{E}(\lambda_1$)\\ $S_2\sim\text{E}(\lambda_2$)} 
 & \makecell[c]{Closed form but\\not known family} & \makecell[c]{Upper tail: AD for $\lambda_1<\infty$\\Lower tail: AD for $\lambda_2<\infty$} & \makecell[c]{$\chi_U=2[1-\Phi(\lambda_1\sqrt{1-\rho}/\sqrt{2})]$\\$\chi_L=2[1-\Phi(\lambda_2\sqrt{1-\rho}/\sqrt{2})]$} & $1$ \\ \hline
 SM1 & \makecell[c]{$R=\sqrt{E}\sim\text{Rayleigh}(1)$\\ $E\sim\text{E}(1/2$)} & Multivariate Laplace & AI & $0$ & $\sqrt{2(1+\rho)}-1$ \\ \hline
 SM2 & \makecell[c]{$R=\sqrt{G}$ \\ $G\sim$ Gamma$(\alpha,1)$} & Mult.\ variance gamma & AI & $0$ & $\sqrt{2(1+\rho)}-1$ \\ \hline
 SM3 & \makecell[c]{$R=1/\sqrt{G}$ \\ $G\sim$ Gamma$(\nu/2,\nu/2)$} & \makecell[c]{Multivariate Student's \textit{t} \\ Gaussian limit as $\nu\rightarrow\infty$} & AD for $\nu<\infty$ & $2T_{\nu+1}\left(-\sqrt{\nu+1}\sqrt{\frac{1-\rho}{1+\rho}}\right)$ & 1 \\ \hline
 SM4 & \makecell[c]{$R=\sqrt{E}/G$ \\ $E\sim$ E$(1/2)$\\ $G\sim$ Gamma$(1/\gamma,1/\gamma)$} & \makecell[c]{Two-sided GPD (marg.)\\ $X\sim$ SGPD$(1,\gamma)$\\Laplace limit as $\gamma\rightarrow0^+$} & AD for $\gamma>0$ & $2T_{1/\gamma+1}\left(-\sqrt{1/\gamma+1}\sqrt{\frac{1-\rho}{1+\rho}}\right)$ & 1 \\ \hline
 SM5 & $R\sim$GPD($1,\gamma$) & No closed form & \makecell[c]{AI for $\gamma\le 0$\\ AD for $\gamma>0$} & \makecell[c]{$\gamma\le0:\; 0$ \\ $\gamma>0:\;2T_{\gamma+1}\left(-\sqrt{\gamma+1}\sqrt{\frac{1-\rho}{1+\rho}}\right)$}  & \makecell[c]{$\gamma<0:\rho$ \\ $\gamma=0:\;\sqrt[3]{4(1+\rho)}-1$ \\$\gamma>0:\;1$} \\ \hline
 LSM1 & \makecell[c]{$S\sim\text{E}(\lambda$)\\$R=\sqrt{E}$\\ $E\sim\text{E}(1/2$)} & \makecell[c]{Exponential ($S$)\\+ Laplace ($R\,W(\mathbf{s})$)} & \makecell[c]{Upper tail: AI if $\lambda\ge 1$\\Upper tail: AD if $\lambda<1$\\Lower tail: AI} & \makecell[c]{$\lambda\ge 1$: 0\\$\lambda<1$: $\mathbb{E}\left[\min\left\{\frac{Z_1^\lambda}{\mathbb{E}(Z_1^\lambda)},\frac{Z_2^\lambda}{\mathbb{E}(Z_2^\lambda)}\right\}\right]$,\\ with $Z_i=\exp\{RW_i\}$, $i=1,2$} & \makecell[c]{$\lambda>1:  \max\{2/\lambda-1,\rho\}$\\$\lambda\le 1: 1$} \\ \hline
 LSM2 & \makecell[c]{$S\sim\text{AL}(\lambda_1,\lambda_2)$\\ %$S_1\sim\text{E}(\lambda_1$)\\ $S_2\sim\text{E}(\lambda_2$)\\
 $R=\sqrt{E}$\\ $E\sim\text{E}(1/2$)} & \makecell[c]{Asymmetric Laplace ($S$)\\+ Laplace ($R\,W(\mathbf{s})$)} & \makecell[c]{Upper tail: AI if $\lambda_1\ge 1$\\Upper tail: AD if $\lambda_1<1$\\Lower tail: AI if $\lambda_2\ge 1$\\Lower tail: AD if $\lambda_2<1$} & \makecell[c]{Upper tail: as above with $\lambda=\lambda_1$\\ Lower tail: as above with $\lambda=\lambda_2$\\ and $Z_i=\exp\{-RW_i\}$, $i=1,2$} & \makecell[c]{As above with\\$\lambda=\lambda_1$ or $\lambda=\lambda_2$} \\ \hline
\end{tabular}
\label{table:examples_gaussian_locscmix_models}
\end{center}
\end{sidewaystable}
%\end{landscape}

\subsubsection{Gaussian location mixtures: \texorpdfstring{$X(\mathbf{s})=S+W(\mathbf{s})$}{X(s)=S+W(s)}}%: asymmetric upper and lower tails}

A key property of these models is that they allow for asymmetry in the upper and lower tail regions. The following  models  were studied in \cite{krupskii2018factor}.

\paragraph{Location Mixture 1 (LM1)} %(upper-tail dependence, closed-form marginal densities)}
Let $S\sim \mbox{E}(\lambda)$, where $\lambda$ is the rate parameter.
Then, $\{X(\mathbf{s})\}$ is AD with $\chi=\chi_L=\chi_U=2\Phi(-\lambda\sqrt{1-\rho}/\sqrt{2})$ and $\bar{\chi}=\bar{\chi}_U=\bar{\chi}_L=1$. %\citep[see][]{krupskii2018factor}.
As $\lambda\to\infty$, the finite-dimensional distribution of $\{X(\mathbf{s})\}$ converges in distribution to a multivariate Gaussian distribution and we get $\chi\rightarrow0$.
Note that finite-dimensional distributions of $\{X(\mathbf{s})\}$ can be expressed in closed form for any dimension \citep[see][]{krupskii2018factor}. % as $G(x)=\Phi(x)-\exp(\lambda^2/2-\lambda x)\Phi(x-\lambda)$.
%Key properties:
%\begin{itemize}
%    \item Asymptotic dependence in the upper tail, and asymptotic independence in the lower tail;
%    \item Closed-form marginal and multivariate densities.
%\end{itemize}

\paragraph{Location Mixture 2 (LM2)} %(upper- and lower-tail dependence; closed-form marginal densities)}
Let %$S=S_1-S_2$, with $S_1\sim\text{E}(\lambda_1)$ and $S_2\sim\text{E}(\lambda_2)$. Then $S$ is distributed as asymmetric Laplace distribution, 
$S\sim\mbox{AL}(\lambda_1,\lambda_2)$ (see Section \ref{section:definitions}). %\citep{kotz2001laplace,gong2022asymmetric}. %, denoted as $S\sim\text{Asym.Lapl.}(\lambda_1,\lambda_2)$, s<<uch that
%\begin{equation}\label{eq:asymm_laplace_distrib}
%    \begin{array}{ll}
%    F_S(x;\,\lambda_1,\lambda_2)=\begin{cases}
%        1-\lambda_2(\lambda_1+\lambda_2)^{-1}\exp(-\lambda_1 x), &\quad \text{if}\; x\ge0 \\
%        \lambda_1(\lambda_1+\lambda_2)^{-1}\exp(\lambda_2 x), &\quad \text{if}\; x<0.
%    \end{cases}
%    \end{array}
%\end{equation}
%From \eqref{eq:asymm_laplace_distrib}, it is clear that $F_S$ is exponential-tailed in the upper tail. Moreover, by studying the random variable $-S$, for which the distribution of $S$ is reflected at zero, it is possible to see that $F_S$ is exponential-tailed also in the lower tail. Note that the density of $\{W(\mathbf{s})\}$ is symmetric around $\mathbf{0}=(0,\dots,0)^\top$.
%Therefore, 
$\{X(\mathbf{s})\}$ is AD in both tails, with $\chi_U=2\Phi(-\lambda_1\sqrt{1-\rho}/\sqrt{2})$, $\chi_L=2\Phi(-\lambda_2\sqrt{1-\rho}/\sqrt{2})$ and $\bar{\chi}_U=\bar{\chi}_L=1$.
The finite-dimensional distributions of $\{X(\mathbf{s})\}$ can be expressed in closed form \citep[see][]{krupskii2018factor}. % as $$G(x)=\Phi(x)-\frac{\lambda_2}{\lambda_1+\lambda_2}\exp\left(\frac{\lambda_1^2}{2}-\lambda_1 x\right)\Phi(x-\lambda_1)+\frac{\lambda_1}{\lambda_1+\lambda_2}\exp\left(\frac{\lambda_2^2}{2}+\lambda_2 x\right)\Phi(-x-\lambda_2).$$
Three bivariate simulations from this model are shown in Figure~\ref{fig:biv_plots}.
%Key properties:
%\begin{itemize}
%    \item Asymptotic dependence in the upper and in the lower tail (asymmetric);
%    \item Closed-form marginal and multivariate densities.
%\end{itemize}

\subsubsection{Gaussian scale mixtures: \texorpdfstring{$X(\mathbf{s})=RW(\mathbf{s})$}{X(s)=RW(s)}}

Given the symmetric nature of the standard Gaussian distribution, which carries over to Gaussian scale mixture models,  we do not distinguish between their upper and lower tail regions.
We present five examples of models. Models SM1, SM2, and SM3 have been extensively studied in the literature; see, for example, the articles cited in the respective paragraphs. To the best of our knowledge, model SM4 has not yet been studied as a dependence model.

\paragraph{Scale Mixture 1 (SM1)  or Laplace process} %(asymptotic independence; closed-form marginal densities)} 
Let $R=\sqrt{E}$, with $E\sim\text{E}(1/2)$.
Then %$R\sim\text{Rayleigh}(1)$, i.e. $F_R(x)=1-\exp(-x^2/2)$, i.e.\ it is Weibull-tailed with index $\beta=2$ (see Definition \ref{def:weibull_tailed}), and 
$\{X(\mathbf{s})\}$ is a Laplace process \citep{Opitz:2016}. It is asymptotically independent with $\chi=0$ and $\bar{\chi}=\sqrt{2(1+\rho)}-1$, %(see Section \ref{section:ch1_gaussian_scale_location_mixtures}).
and its finite-dimensional distribution is available in closed form.
%Key properties:
%\begin{itemize}
%    \item Asymptotic independence in both tails;
%    \item Closed-form marginal and multivariate densities.
%\end{itemize}

\paragraph{Scale Mixture 2 (SM2)} %(Interpolation between Gaussian and Laplace process)}
Let $R=\sqrt{G}$, with $G\sim\text{Gamma}(\alpha,1)$, where $\alpha$ is the shape parameter. Then %$F_R$ is Weibull-tailed with index $\beta=2$ (see Definition \ref{def:weibull_tailed}), and 
$\{X(\mathbf{s})\}$ is asymptotically independent with $\chi=0$ and $\bar{\chi}=\sqrt{2(1+\rho)}-1$. 
The finite-dimensional distribution of $\{X(\mathbf{s})\}$ is a multivariate extension of the variance gamma \citep{madan1990variance}, which has been widely studied in mathematical finance \citep{fischer2025variance}.
It can be seen as an interpolation between Gaussian and Laplace processes.
%Key properties:
%\begin{itemize}
%    \item Asymptotic independence in both tails;
%    \item Interpolation between Gaussian and Laplace process.
%\end{itemize}

\paragraph{Scale Mixture 3 (SM3) or Student's \textit{t} process} %(asymptotic dependence; closed-form densities)}
Let $R=1/\sqrt{G}$, with $G\sim\text{Gamma}(\nu/2,\nu/2)$, where $\nu/2$ is both the shape and the rate parameter. Then $\{X(\mathbf{s})\}$ is a Student's \textit{t} process \citep{Roislien:Omre:2006}.
The finite-dimensional distribution of $\{X(\mathbf{s})\}$ is available in closed form
%, since $F_R$ is regularly varying with index $\nu$ \citep[see Definition \ref{def:regularly_varying} and][]{chan:li:2008}, 
and it is asymptotically dependent with $\chi=2T_{\nu+1}\left[-\sqrt{(\nu+1)(1-\rho)/(1+\rho)}\right]$ and $\bar{\chi}=1$ \citep[see][]{demarta2005}.
However, as $\nu\to\infty$, $F_R$ converges to a Dirac distribution at $1$, so $\{X(\mathbf{s})\}$ converges to the Gaussian process $\{W(\mathbf{s})\}$ and $\chi\rightarrow0$. 
Therefore, this model allows for asymptotic independence as a limit, at the boundary of the parameter space.
%Therefore, this mixture is flexible for modelling asymptotic dependence, but is quite rigid for asymptotic independence, since the only possible values of its coefficient $\bar{\chi}$ are $\bar{\chi}=1$, for $\nu<\infty$, and $\bar{\chi}=\rho$, for $\nu=\infty$ \citep{Huser:Opitz:Thibaud:2017}.
Three bivariate simulations from this model are shown in Figure \ref{fig:biv_plots}.
%Key properties:
%\begin{itemize}
%    \item Asymptotic dependence in both tails; asymptotic independence limit as $\nu\to\infty$;
%    \item Closed-form marginal and multivariate densities.
%\end{itemize}

\paragraph{Scale Mixture 4 (SM4) or elliptical generalized Pareto process} %(asymptotic dependence; generalized double Pareto margins)}
Let $R=\sqrt{E}/G$, with $E\sim\text{E}(1/2)$ and $G\sim\text{Gamma}(1/\gamma,1/\gamma)$, where $1/\gamma$ is both the shape and the rate parameter. Then, the unidimensional distribution of  $\{X(\mathbf{s})\}$ is a generalized double Pareto distribution \citep{armagan2013generalized}, also referred to as symmetric GPD (SGPD), two-sided GPD or type II compound Laplace distribution \citep{kotz2001laplace}, with density function $$f(x;\sigma,\gamma)=\frac{1}{2\sigma}\left(1+\frac{\gamma|x|}{\sigma}\right)^{-1/\gamma-1},\quad x\in\R,$$
denoted as $\text{SGPD}(\sigma, \gamma)$, with $\sigma=1$ and $\gamma>0$ by construction.
The process  $\{X(\mathbf{s})\}$ is AD with  $$\chi=2T_{1/\gamma+1}\left[-\sqrt{(1/\gamma+1)(1-\rho)/(1+\rho)}\right];$$
see  \ref{section:appendix_extremal_behaviour} for the proof.

As $\gamma\to0$, $G$ converges to a Dirac distribution at $1$, so $\{X(\mathbf{s})\}$ converges to a Laplace process. 
Therefore, this model also allows for asymptotic independence as a limit, at the boundary of the parameter space. %Key properties:
%\begin{itemize}
%    \item Asymptotic dependence in both tails; asymptotic independence limit as $\gamma\to0$;
%    \item Closed-form marginal densities (generalized double Pareto).
%\end{itemize}

\paragraph{Scale Mixture 5 (SM5)} %(transition between asymptotic dependence regimes)}
Let $R\sim\text{GPD}(1,\gamma)$, where $\gamma\in\R$ is the shape parameter. 
%\underline{Note}: This model was proposed in \cite{Huser:Opitz:Thibaud:2017}.\\
%\underline{Results and proofs}: 
When $\gamma>0$, %$F_R$ is regularly varying with index $\gamma$, so 
$\{X(\mathbf{s})\}$ is asymptotically dependent with $\chi=2T_{\gamma+1}[-\sqrt{(\gamma+1)(1-\rho)/(1+\rho)}]$ and $\bar{\chi}=1$.
When $\gamma=0$, %$F_R$ is exponential-tailed, i.e.\ Weibull-tailed with index $\beta=1$, so 
$\{X(\mathbf{s})\}$ is asymptotically independent with $\chi=0$ and $\bar{\chi}=\sqrt[3]{4(1-\rho)}-1$. %(see Section \ref{section:ch1_gaussian_scale_location_mixtures} and Definition \ref{def:weibull_tailed}).
When $\gamma<0$, $\{X(\mathbf{s})\}$ is asymptotically independent with $\chi=0$ and $\bar{\chi}=\rho$ \cite[see][for these results]{Huser:Opitz:Thibaud:2017}. %, by Proposition 9.2 in \cite{engelke2019extremal}.
Therefore, this model allows for a transition between the two extremal dependence classes in the interior of the parameter space.
%Key properties:
%\begin{itemize}
%    \item Transition between asymptotic dependence and asymptotic independence, in both tails, in the interior of the parameter space.
%\end{itemize}

\subsubsection{Gaussian location-scale mixtures}%: asymmetric tails with dependence transitions}

Let $\{X(\mathbf{s})\}=\{X(\mathbf{s}),\;\mathbf{s}\in\mathcal{S}\}$, with $X(\mathbf{s})=S+R\;W(\mathbf{s})$, $S\in\R$, $R > 0$, be a generic Gaussian location-scale mixture model.
While independence between $S$ and $R$ is not required in general, we assume it in the following two models.
A key property of these models is that they allow for asymmetry in the upper and lower tails, possibly with separate transition between extremal dependence regimes in the two tails.

To the best of our knowledge, models LSM1 and LSM2 are new proposals, although similar models have been studied by \cite{gong2022asymmetric} in the bivariate case and by \cite{gong2022flexible} in the spatial case.
%In particular, model LSM2 bears some similarity to those proposed in the aforementioned papers.
In those cases, the structure of the model was  $S+g(W(\mathbf{s}))$, with $g$ a marginal transformation of a standard Gaussian vector or process $\{W(\mathbf{s})\}$ to (asymmetric) Laplace distribution. This ensures that marginal parameters can be used to make either $S$ or $g(W(\mathbf{s}))$ relatively heavier-tailed than the other term.  
%,   a bivariate and a multivariate Gaussian were first transformed to marginally follow asymmetric Laplace and Laplace distributions, respectively, making them comparable with the respective random locations. 
Here,
similar marginal structure is obtained directly  by a proper Gaussian location-scale mixture process %with proper random location and scale variables, and 
without the need to apply any marginal transformation. %This means that the models proposed here are proper Gaussian location-scale mixtures, which 
This allows efficient conditional simulation following the approach in Section~\ref{section:conditional_simulation} and  estimation with the method proposed in Section \ref{section:estimation_methods}.
%the random location $S$ follows an asymmetric Laplace distribution, while $\{R\;W(\mathbf{s})\}$ is a Laplace process, naturally extending the construction of model SM1.

\paragraph{Location-Scale Mixture 1 (LSM1)} %(Upper tail-dependence transition)}
Let $S\sim \text{E}(\lambda)$, %, where $\lambda$ is the rate parameter, 
and let $R=\sqrt{E}$, with $E\sim\text{E}(1/2)$. Note that $\{R\,W(\mathbf{s})%,\;\mathbf{s}\in\mathcal{S}
\}$ is a Laplace process, with unidimensional $\text{Laplace}(1)$ distribution, whose tail heaviness is exactly the same as that of $S$ when $\lambda=1$.
Three cases can be distinguished; see \ref{section:appendix_extremal_behaviour} for the proof.
\begin{description}
	\item[$\lambda<1$:] $\{X(\mathbf{s})\}$ is AD, with
	\begin{equation}\label{eq:locscalemixture1_chi}
		\chi=\E\left[\min\left\{\frac{Z_1^\lambda}{\E(Z_1^\lambda)},\frac{Z_2^\lambda}{\E(Z_2^\lambda)}\right\}\right] \quad \text{and}\quad \bar{\chi}=1,
	\end{equation}
	where $Z_i=\exp\{R\;W(\mathbf{s}_i)\}$, $i=1,2$.
\item[$\lambda>1$:] $\{X(\mathbf{s})\}$ is AI with
\begin{equation}\label{eq:locscalemixture1_chibar}
	\chi=0 \quad \text{and}\quad \bar{\chi}=\max\left\{\frac{2}{\lambda}-1,\; \rho \right\},
	%    \bar{\chi}(\mathbf{s}_1,\mathbf{s}_2)=\begin{cases}\begin{array}{ll}
			%            2/\lambda-1 &\qquad\text{if}\; \lambda<2/(1+\rho)\\
			%            \rho &\qquad\text{if}\; \lambda\ge 2/(1+\rho),
			%        \end{array}
		%        \end{cases}
\end{equation}
where $\rho=\mbox{cor}(W(\mathbf{s}_1),W(\mathbf{s}_2))$.
\item[$\lambda=1$:] $\{X(\mathbf{s})\}$ is AI, with $\chi=0$ and $\bar{\chi}=1$
	\end{description}

Therefore, this model allows for a transition between the two extremal dependence classes for the upper tail, in the interior of the parameter space $\mathbb{R}^+$ of $\lambda$.
%Key properties:
%\begin{itemize}
%    \item Transition between asymptotic dependence and asymptotic independence, in the upper tail, in the interior of the parameter space.
%\end{itemize}

\paragraph{Location-Scale Mixture 2 (LSM2)} %(Separate upper-, lower-tail dependence transitions)} 
Let $S\sim AL(\lambda_1,\lambda_2)$ (see Section \ref{section:definitions}),  and let $R=\sqrt{E}$, with $E\sim\text{E}(1/2)$.
As $\lambda_2\rightarrow\infty$, the finite-dimensional distribution of this model converges to that of the LSM1 model with $\lambda = \lambda_1$.

Since the tail heaviness of the asymmetric Laplace distribution $F_S$ is governed by $\lambda_1$ in the upper tail and by $\lambda_2$ in the lower tail, %see \eqref{eq:asymm_laplace_distrib}, 
and is the same as that of an exponential distribution with rate parameter $\lambda_1$ or $\lambda_2$, respectively, we can extend the asymptotic properties of model LSM1 to both tails of this model. 
In the upper tail, $\{X(\mathbf{s})\}$ is asymptotically independent if $\lambda_1\ge1$ and asymptotically dependent if $\lambda_1<1$, where  $\chi_U$ and $\bar{\chi}_U$ are $\chi$ and $\bar{\chi}$ in \eqref{eq:locscalemixture1_chi} and \eqref{eq:locscalemixture1_chibar}, replacing $\lambda$ with $\lambda_1$.
In the lower tail, $\{X(\mathbf{s})\}$ is asymptotically independent if $\lambda_2\ge1$ and asymptotically dependent if $\lambda_2<1$, again replacing $\lambda$ in \eqref{eq:locscalemixture1_chi} and \eqref{eq:locscalemixture1_chibar} with $\lambda_2$ to obtain $\chi_L$ and $\bar{\chi}_L$; the only other difference from \eqref{eq:locscalemixture1_chi} is that here $Z_i=\exp\{-R\;W(\mathbf{s}_i)\}$, $i=1,2$.%, and $\lambda$ is replaced by $\lambda_2$, while the other values of $\chi_L$ and $\bar{\chi}_L$ are analogous to those of the upper tail.

Therefore, this model allows for a transition between the two extremal dependence classes in the interior of the parameter space, for both tails, and the extremal dependence in the two tails is governed by two different parameters, $\lambda_1$ for the upper tail and $\lambda_2$ for the lower tail.
%Key properties:
%\begin{itemize}
%    \item Separate transition between asymptotic dependence and asymptotic independence in the upper tail and in the lower tail, in the interior of the parameter space.
%\end{itemize}

\section{Conditional simulation}\label{section:conditional_simulation}

In Gaussian location-scale mixture models, we can exploit the Gaussian structure  $\{W(\mathbf{s})\}$ in Equation \eqref{eq:gaussian_locationscale_mixtures} to easily perform unconditional or conditional spatial simulations.
Here, we present a general conditional simulation algorithm that can be used with any distribution of $S$ and $R$. It generalizes the one proposed \citep{Huser:Opitz:Thibaud:2017} for Gaussian scale mixtures. 

The vector $\bX$ of  $m+h$ random variables  $\bX=(X(\mathbf{s}_1),\ldots,X(\mathbf{s}_{m+h}))^\top$
of the process \eqref{eq:gaussian_locationscale_mixtures}  is   partitioned into the vectors $\bX_1=(X(\mathbf{s}_1),\dots,X(\mathbf{s}_m))^\top$ and $\bX_2=(X(\mathbf{s}_{m+1}),\dots,X(\mathbf{s}_{m+h}))^\top$. By analogy, we partition $\bW=(W(\mathbf{s}_1),\ldots,W(\mathbf{s}_{m+h}))^\top$ 
into  $\bW_1$ and $\bW_2$.
%, such that $\bX_1$ and $\bW_1$ are defined on $\mathcal{S}_1\subset\mathcal{S}$, while $\bX_2$ and $\bW_2$ are defined on $\mathcal{S}_2\subset\mathcal{S}$, with $\mathcal{S}_1\cap\mathcal{S}_2 =\emptyset$.
Let $\bSigma_{i;j}$, $i,j=1,2$, be the corresponding block matrices stemming from $\bSigma$, the correlation matrix of $\bW=(\bW_1^\top,\bW_2^\top)^\top$.
Suppose that we want to simulate from $\bX_2$ conditional on the observed values $\bX_1=\bx_1$.
We can do so using Algorithm \ref{alg:conditional_simulation}.
For Monte--Carlo estimation of properties of the conditional distribution, such as point predictions at unobserved locations, multiple pairs $(s,r)$ can be simulated in the Markov chain in Step $1$, and Steps~$2$--$4$ can be run for each pair to obtain multiple realizations of $\bX_{2}\mid\bX_{1}$. %; 
%their simulated distribution can be summarized, e.g., by the median, if a point prediction is of interest.
In point $1$, independence between $S$ and $R$, as in our case,  can simplify the simulation procedure. 
Moreover, if interest is in simulating the corresponding vector $\mathbf{Y}_2$ on the scale of the observed process $Y(\mathbf{s})$ (see Section \ref{section:general_model_definition}), this can be done via marginal transformation of the simulated conditional vector $\bX_2\mid \bX_1$ using the unconditional marginal distributions $F_Y$ and $F_X$ of $\mathbf{Y}_2$ and $\bX_2$, respectively.
The conditional simulation algorithm is applied in Section \ref{section:application_conditional_simulation} to real data, with an illustration in Figure \ref{fig:pt_conditional_simulation_Viseu}.
The median of the simulated data is used as point prediction on a hold-out region, and its values are compared to the observed ones in that region.
We note that the potential applications of the proposed simulation method extend beyond point prediction. It generates the full conditional distributions at unobserved locations, enabling the calculation of threshold exceedance probabilities based on the flexible extremal dependence structure of the model class presented in this paper.

\begin{algorithm}[
t]
\caption{Conditional simulation of Gaussian location-scale mixtures}\label{alg:conditional_simulation}
\begin{algorithmic}
\State
{\bf Goal:} Simulate $\bX_{2} \mid \bX_{1}= \mathbf{x_1}$ in a Gaussian location-scale mixture vector $\bX=(\bX_{1},\bX_{2})$\\
{\bf Inputs:} Conditioning components $\mathbf{x}_1$; Gaussian variance-covariance matrix $\bSigma$
\begin{enumerate}
    \item Simulate $(S,R)$ from the conditional distribution of $(S,R)\mid \bX_1$
    \begin{equation*}
    \begin{split}
        f_{(S,R)|\bX_1=\bx_1}(r,s)=&\frac{f_{(S,R,\bX_1)}(s,r,\bx_1)}{f_{\bX_1}(\bx_1)}=\frac{f_{(S,R,\bW_1)}(S,R,(\bx_1-s)/r)r^{-m}}{f_{\bX_1}(\bx_1)}\\
        =&\frac{f_{S,R}(s,r)\phi_m((\bx_1-s)/r;\,\bSigma_{1;1})r^{-m}}{f_{\bX_1}(\bx_1)}
    \end{split}
    \end{equation*}
    %here we assume independence between $R$ and $S$. This can be done
    through a MCMC Metropolis-Hastings algorithm;
%    \item Compute $\bw_{1,i} = (\bx_1 - s_i) / r_i$;
%    \item Simulate $\bw_{2,i}$ from $\bW_{2}\mid\bW_1=\bw_{1,i} \sim \mathcal{N}(\cdots)$;
%    \item Compute $\bx_{2,i} = s_i + r_i \bw_{2,i}$;
%    \item If necessary, $\mathbf{y}_{2,i} | \mathbf{y}_{1,i} = F^{-1}(G(\bx_{2,i}))$.
    \item Compute $\bW_{1} = (\bX_1 - S) / R$;
    \item Simulate $\bW_{2}$ from $\bW_{2}\mid\bW_1 \sim \mathcal{N}(\bmu_{2|1},\bSigma_{2|1})$, with $\bmu_{2|1}=\bSigma_{2;1}\bSigma_{1;1}^{-1}\bW_1$ and $\bSigma_{2|1}=\bSigma_{2;2}- \bSigma_{2;1}\bSigma_{1;1}^{-1}\bSigma_{1;2}$;
    \item Compute and return $\bX_{2} = S + R\; \bW_{2}$.
 %   \item If necessary, $\mathbf{y}_{2} \mid \mathbf{y}_{1} = F^{-1}(G(\bx_{2}))$.
\end{enumerate}
\end{algorithmic}
\end{algorithm}

%Point $5.$ requires preliminary estimation of the marginal distributions $F$ and $G$ of the processes $Y(s)$ and $X(\mathbf{s})$, respectively.

The MCMC Metropolis-Hastings algorithm \citep{robert2004monte}
 inside of Step $1$ of Algorithm \ref{alg:conditional_simulation} creates a homogenous Markov chain, $(S^{(k)},R^{(k)})$, $k=1,2,\ldots$, whose distribution converges to the  distribution of  $(S,R)\,|\,\bX_1$. % $K$ (e.g., set to $K=50\,000$ in the data application) of simulations from $(S,R)\mid\bX_1$ to ensure convergence to the stationary distribution, i.e., 
%$$(S^{(k)},R^{(k)})\mid\bX_1 \;\xrightarrow{d}\; (S,R)\mid\bX_1,\qquad \mbox{for } k\rightarrow\infty.$$
The transition from $(S^{(k)},R^{(k)})$ to $(S^{(k+1)},R^{(k+1)})$ is described in the following.
First, propose $(S',R')$ such that   $S'=S^{(k)}+Z_S$ and 
$\log(R')=\log(R^{(k)})+Z_R$, where  $Z_S,Z_R\stackrel{\text{ind.}}{\sim} \mathcal{N}(0,1)$. A logarithmic transformation could also be applied to $S$ when its support is $\R^+$. Then, we set
\begin{equation}\label{eq:metropolis_hastings_update}
    \left(S^{(k+1)},R^{(k+1)}\right)=\begin{cases}\begin{array}{ll}
        %(R^{'}_{(k+1)},S^{'}_{(k+1)}) &\text{with probability}\;\rho(r',s',r,s)\\
        (S',R') &\text{with probability}\;\rho(S',R',S^{(k)},R^{(k)}),\\
        (S^{(k)},R^{(k)}) & \text{otherwise,}
    \end{array}
    \end{cases}
\end{equation}
where the acceptance ratio is (see \ref{section:appendix_metropolis_hastings} for the derivation)
%    $$\rho(r',s',r^{(k)},s^{(k)})=\min\left\{\frac{f_{R}(r')f_{S}(s')f_{\bW_1}((\bx_1-s')/r')r'^{-m} g(r^{(k)},s^{(k)}|r',s')}{f_{R}(r^{(k)})f_{S}(s^{(k)})f_{\bW_1}((\bx_1-s^{(k)})/r^{(k)})r^{(k)-m}g(r',s'|r^{(k)},s^{(k)})},1\right\}$$
%and $g(a,b\,|\,c,d)$ is the sampling density from $(c,d)$ to $(a,b)$; note that
%$$g(r^{(k)},s^{(k)}|r',s')=g(r^{(k)}|r')g(s^{(k)}|s')=\phi(\log(r^{(k)})-\log(r'))\frac{1}{r^{(k)}}\phi(s^{(k)}-s'),$$
%$$g(r',s'|r^{(k)},s^{(k)})=g(r'|r^{(k)})g(s'|s^{(k)})=\phi(\log(r')-\log(r^{(k)}))\frac{1}{r'}\phi(s'-s^{(k)}),$$
%and, due to the symmetry of the Gaussian density function of the random walk proposal, all terms but the Jacobian $1/r^{(k)}$ and $1/r'$ simplify, leading to:
$$\rho(s',r',s^{(k)},r^{(k)})=\min\left\{\frac{f_{S,R}(s',r')\phi_m((\bx_1-s')/r';\bSigma_{1;1})r'^{-m+1}}{f_{S,R}(s^{(k)},r^{(k)}))\phi_m((\bx_1-s^{(k)})/r^{(k)};\bSigma_{1;1})r^{(k-m+1)}},1\right\}.$$
In practice, the distribution of the initial states of the Markov chain may
be quite different to the distribution of $(S,R)\mid\bX_1$. To mitigate this, it is desirable to identify a ``burn-in" $(S^{(k)},R^{(k)})$, $k=1,\ldots K$, which is then discarded. %We set $K=50\,000$ in the data application.
After the burn-in period, the values of the Markov chain still have autocorrelation. A thinning procedure can be applied to generate approximately independent realizations of the conditional distribution of $\bX_{2}\mid \bX_{1}$.

\section{Inference}\label{section:estimation_methods}

Two cases can be distinguished in real-data applications: when \eqref{eq:gaussian_locationscale_mixtures} is used to directly model observed spatial data, and when \eqref{eq:gaussian_locationscale_mixtures} provides a copula model describing the dependence structure of an underlying spatial process.
 The latter is a common framework when we want to specify and fit sub-asymptotic models for extremes 
\citep{Huser2025}. 	
In such a case, the model for  the ``observed'' process $\{Y(\mathbf{s})\}$ is the model induced by the transformation 
$Y(\mathbf{s}) = F^{-1}_Y\left(F_X (X(\mathbf{s}))\right)$. Note that, for notational simplicity,
we assume that the cdfs of $X(\mathbf{s})$ and $Y(\mathbf{s})$ ($F_X$ and $F_Y$, resp.) are invariant with respect to $\mathbf{s}$.

A likelihood-based approach requires the evaluation  of  the 
partial derivatives of the cdf of  
$\bX=(X(\mathbf{s}_1),\ldots,X(\mathbf{s}_m))^\top$
for given $m$ (distinct) spatial locations $\mathbf{s}_1,\dots,\mathbf{s}_m$.
We denote by $F_{\bX}(\cdot;\theta)$ the cdf of $\bX$. The unknown parameter $\theta=(\theta_W,\theta_{S,R})$, of the distribution of $\bX$,  is split into two parameters: $\theta_W$, the parameter(s) in the covariance matrix $\bSigma_{\theta_W}$ of the standard Gaussian vector $\bW=(W(\mathbf{s}_1),\ldots,W(\mathbf{s}_m))^\top$;  $\theta_{S,R}$, the parameter(s) of the distribution of $S$ and $R$ (if any). 
We obtain the cdf of $\bX$ by
\begin{equation}\label{eq:chapter2_integral_density}
	F_{\bX}(\bx;\,\theta)=\int_{-\infty}^{+\infty}\int_{0}^{+\infty} \Phi_m\left(\frac{x_1-s}{r},\dots,\frac{x_m-s}{r}; \bSigma_{\theta_W} \right) f_{S,R}(s,r;\theta_{S,R}) \,ds\,dr,
\end{equation}
where $\bx=(x_1,\dots,x_m)^\top$,  $f_{S,R}(\,\cdot\,;\,\theta_{S,R})$ is the joint density of $R$ and $S$ and 
$\Phi_m\left(\,\cdot\,;\bSigma \right)$ is the m-dimensional cdf of a standard Gaussian vector with correlation matrix $\bSigma$. 
In few cases, such  as the Student's \textit{t} stochastic process and the Laplace stochastic process (see Section \ref{section:example_models_gaussiansclocmixt}), expression \eqref{eq:chapter2_integral_density}  is available in analytical form. However, this is not true in general, so the evaluation of the likelihood functions is usually performed by numerical integration \citep{Huser:Opitz:Thibaud:2017,krupskii2018factor}.
This leads to a high numerical cost, especially when the number of spatial locations $m$ is large, since an $m$-dimensional Gaussian density function has to be evaluated for each value of $r$ and $s$ within the integrals. The computational cost is even higher  when the model is fitted as copula model.

In the following, we propose a two-step estimation procedure that does not rely on specific assumptions about the distributions of $R$ or $S$, with the aim of reducing the computational burden.

First, we transform vector $\bX$ to obtain a vector whose distribution depends only on the distribution of $\bW$ and for which we know the analytical expression.
This allows us to avoid numerical integration. 
We can then use this distribution to estimate the parameter $\theta_W$, denoted by $\hat{\theta}_W$.
In the second step, we reduce the dimension of $\bX$ to  a random variable whose distribution depends on all the parameters, $\theta=(\theta_W,\theta_{S,R})$.
We plug the value $\hat{\theta}_W$ into the value of $\theta_W$  to obtain a profile distribution from which to make inferences about the value of $\theta_{S,R}$. 

We will begin by describing our proposal in the case that we observe $n$ independent value, $\bx_i$, $i = 1, \ldots, n$ of $\bX$, subsections \ref{section:new_method_step1} and \ref{section:new_method_step2}.
Then we describe how the estimation procedure can be modified when using the distribution of $\bX$ as a copula model (subsection \ref{section:new_method_copula}). 
We conclude this section with a discussion of how to assess estimation uncertainty.

\subsection{Step 1: estimation of \texorpdfstring{$\theta_W$}{theta W}}\label{section:new_method_step1}

The estimation method is inspired by the REstricted Maximum Likelihood (REML; \citealp{patterson1971recovery}).
We identify transformations of the vector $\bX$ that are independent of $S$ and $R$, whose analytical distributions can be employed for likelihood-based inference on $\theta_W$. The transformed vector, denoted by $\bZ$, will have different derivations in the three cases: location, scale, and location-scale mixtures.
In all three cases, the restricted likelihood function for $\theta_W$ is then constructed as
$$\mathcal{L}(\theta_W;\,\bz_1,\dots,\bz_n)=\prod_{i=1}^n f_{\bZ}\left(\bz_i;\,\theta_W\right),$$
where $\bz_i$ are i.i.d.\ observations of $\bZ$.

\subsubsection*{Gaussian location mixtures}

 We transform $\bX$ into 
\begin{equation}\label{eq:Z_tilde_1}
\bZ=\left(X(\mathbf{s}_2)-X(\mathbf{s}_1),\dots,X(\mathbf{s}_m)-X(\mathbf{s}_1)\right)^\top  =(W(\mathbf{s}_2)-W(\mathbf{s}_1),\dots,W(\mathbf{s}_m)-W(\mathbf{s}_1))^\top.
\end{equation}
Its density function (see \ref{app1:density_Z_case1}) is given by
$$f_{\bZ}(\bz;\,\theta_W) = (2\pi)^{-(m-1)/2} |\bA\bSigma_{\theta_W} \bA^\top|^{-{1}/{2}} \exp\left(-\frac{1}{2} \bz^\top (\bA\bSigma_{\theta_W} \bA^\top)^{-1}\bz \right),$$
where $\bz=(x(\mathbf{s}_2)-x(\mathbf{s}_1),\dots,x(\mathbf{s}_m)-x(\mathbf{s}_1))^\top$ and $\bA$ is the $(m-1)\times m$ matrix
\begin{equation}\label{eq:matrix_A}
    \bA = \left[ \begin{matrix} -1&1&0&\cdots&0\\ -1&0&1&\cdots&0 \\ \vdots&\vdots&&\ddots&\vdots\\ -1&0&0&\cdots&1 \ \end{matrix} \right].
\end{equation}
%and its density function, for $\bz\in\R^{m-1}$, is
%$$f_{\bZ}(\bz;\,\bSigma) = (2\pi)^{-(m-1)/2} |\bA\bSigma \bA^\top|^{-{1}/{2}} \exp\left(-\frac{1}{2} \bz^\top (\bA\bSigma \bA^\top)^{-1}\bz \right).$$
%Then, the restricted likelihood function for $\theta_W$ is
%$$\mathcal{L}(\theta_W;\,\bz_1,\dots,\bz_n)=\prod_{i=1}^n f_{\bZ}\left(\bz_i;\,\theta_W\right),$$
%where $\bz_i=(x_i(\mathbf{s}_2)-x_i(\mathbf{s}_1),\dots,x_i(\mathbf{s}_m)-x_i(\mathbf{s}_1))^\top$, $i=1,\dots,n$, are i.i.d. observations of $\bZ$.

\subsubsection*{Gaussian scale mixtures}
We transform $\bX$ into 
\begin{equation}\label{eq:Z_tilde_2}
    \bZ=\left(\frac{X(\mathbf{s}_2)}{X(\mathbf{s}_1)},\dots,\frac{X(\mathbf{s}_m)}{X(\mathbf{s}_1)}\right)^\top = \left(\frac{W(\mathbf{s}_2)}{W(\mathbf{s}_1)},\dots,\frac{W(\mathbf{s}_m)}{W(\mathbf{s}_1)}\right)^\top.
\end{equation}
Its density function (see \ref{app1:density_Z_case2}), is given by
\begin{equation*}
\begin{split}
    f_{\bZ}(\bz;\,\theta_W) %=& \int_0^\infty f_Z(z;\,\bSigma) d\tilde{z}_1 = \int_0^\infty 2\; \phi_\bSigma(\tilde{z}_1,\tilde{z}_1\tilde{z}_2,\dots,\tilde{z}_1\tilde{z}_m)\; \tilde{z}_1^{m-1} d\tilde{z}_1\\
    %=& \; 2 \int_0^\infty (2\pi)^{-{m}/{2}} |\bSigma|^{-{1}/{2}} \exp\left(-\frac{1}{2}\tilde{z}_1^2\; \dot{\bz}^\top \bSigma^{-1}\dot{\bz} \right) \tilde{z}_1^{m-1} d\tilde{z}_1 \qquad\qquad [y=\tilde{z}_1^2]\\
    %=& \; 2 \int_0^\infty (2\pi)^{-{m}/{2}} |\bSigma|^{-{1}/{2}} \exp\left(-\frac{1}{2}y\; \dot{\bz}^\top \bSigma^{-1}\dot{\bz} \right) y^{(m-1)/2} \frac{1}{2}y^{-{1}/{2}}dy\\
    %=& \; (2\pi)^{-{m}/{2}} |\bSigma|^{-{1}/{2}} \int_0^\infty \exp\left(-{1}/{2}y\; \dot{\bz}^\top \bSigma^{-1}\dot{\bz} \right) y^{{m}/{2}-1} dy\\
    %=& \; (2\pi)^{-{m}/{2}} |\bSigma|^{-{1}/{2}} \frac{\Gamma\left(m/2\right)}{\left(\frac{1}{2} \dot{\bz}^\top \bSigma^{-1}\dot{\bz} \right)^{m/2}} \int_0^\infty \frac{\left(\frac{1}{2} \dot{\bz}^\top \bSigma^{-1}\dot{\bz} \right)^{m/2}}{\Gamma\left(m/2\right)} \exp\left(-\frac{1}{2}y\; \dot{\bz}^\top \bSigma^{-1}\dot{\bz} \right) y^{{m}/{2}-1} dy\\
    =& \; \pi^{-m/2} |\bSigma_{\theta_W}|^{-{1}/{2}} \Gamma\left(m/2\right)\left(\dot{\bz}^\top \bSigma_{\theta_W}^{-1}\dot{\bz} \right)^{-m/2},
\end{split}
\end{equation*}
where $\dot{\bz}=(1,\bz^\top)^\top$ and $\bz
=(x(\mathbf{s}_2)/x(\mathbf{s}_1),\dots,x(\mathbf{s}_m)/x_i(\mathbf{s}_1))^\top$.
%where the last integration for $y$ comes from the gamma density.
%Note that, when $W(\mathbf{s}_1)$ is independent of $W(\mathbf{s}_2),\dots,W(\mathbf{s}_m)$, $\bZ$ follows a multivariate Cauchy distribution, i.e.\ a multivariate Student's \textit{t} with 1 degree of freedom.
%Then, the  restricted  likelihood function for $\theta_W$ %, on which $\bSigma$ is based, is
%$$\mathcal{L}(\theta_W;\bz_1,\dots,\bz_n)=\prod_{i=1}^n \Big[f_{\bZ}\left(\bz_i;\,\theta_W\right)\mathbb{1}\{x_i(\mathbf{s}_1)>0\} + f_{\bZ}\left(-\bz_i;\,\theta_W\right)\mathbb{1}\{x_i(\mathbf{s}_1)<0\}\Big],$$
%where $\mathbb{1}\{\cdot\}$ is the indicator function and 
%$\bz_i
%=\left({x_i(\mathbf{s}_2)}/{x_i(\mathbf{s}_1)},\dots,{x_i(\mathbf{s}_m)}/{x_i(\mathbf{s}_1)}\right)^\top$, $i=1,\dots,n$, are i.i.d. observations of $\bZ$.%, for $\quad i=1,\dots,n$.
%are $n$ i.i.d.  observations of $\bZ$.

\subsubsection*{Gaussian location-scale mixtures}
We transform $\bX$ into 
\begin{equation}\label{eq:Z_tilde_3}
\begin{split}
    \bZ=& \left(\frac{X(\mathbf{s}_3)-X(\mathbf{s}_1)}{X(\mathbf{s}_2)-X(\mathbf{s}_1)},\dots,\frac{X(\mathbf{s}_m)-X(\mathbf{s}_1)}{X(\mathbf{s}_2)-X(\mathbf{s}_1)}\right)^\top \\
    =& \left(\frac{W(\mathbf{s}_3)-W(\mathbf{s}_1)}{W(\mathbf{s}_2)-W(\mathbf{s}_1)},\dots,\frac{W(\mathbf{s}_m)-W(\mathbf{s}_1)}{W(\mathbf{s}_2)-W(\mathbf{s}_1)}\right)^\top.
\end{split}
\end{equation}
Its density function (see \ref{app1:density_Z_case3}) is
\begin{equation*}
    f_{\bZ}(\bz;\,\theta_W) = \pi^{-(m-1)/{2}} |\bA\bSigma_{\theta_W} \bA^\top|^{-{1}/{2}} \Gamma\left(\frac{m-1}{2}\right)\left[\dot{{\bz}}^\top (\bA\bSigma_{\theta_W} \bA^\top)^{-1}\dot{{\bz}} \right]^{-(m-1)/{2}},
\end{equation*}
where $\dot{{\bz}}=(1,\bz^\top)^\top$ and 
$\bz=\left([x(\mathbf{s}_3)-x(\mathbf{s}_1)]/[x(\mathbf{s}_2)-x(\mathbf{s}_1)],\ldots,[x(\mathbf{s}_m)-x(\mathbf{s}_1)]/[x(\mathbf{s}_2)-x(\mathbf{s}_1)]\right)^\top$.
%The  restricted likelihood function for $\theta_W$ is
%\begin{equation*}
%    \mathcal{L}(\theta_W;\bz_1,\dots,\bz_n)=\prod_{i=1}^n \Big[f_{\bZ}(\bz_i;\,\theta_W)\mathbb{1}\{x_i(\mathbf{s}_2)-x_i(\mathbf{s}_1)>0\} \; + f_{\bZ}(-\bz_i;\,\theta_W)\mathbb{1}\{x_i(\mathbf{s}_2)-x_i(\mathbf{s}_1)<0\}\Big],
%\end{equation*}
%where 
%$\bz_i=\left([x_i(\mathbf{s}_3)-x_i(\mathbf{s}_1)]/[x_i(\mathbf{s}_2)-x_i(\mathbf{s}_1)],\ldots,[x_i(\mathbf{s}_m)-x_i(\mathbf{s}_1)]/[x_i(\mathbf{s}_2)-x_i(\mathbf{s}_1)]\right)^\top$, $i=1,\dots,n$, are i.i.d. observations of $\bZ$.

\subsubsection*{Invariance of the estimators to the selection of the reference locations}
In the first step, we choose one or two reference locations for the transformed vector $\bZ$.
However,  the choice of the reference locations has no impact on the resulting estimates. Indeed, if another location $\mathbf{s}_k$, $k\in\{1,\dots,m\}$ (or a pair of locations $\mathbf{s}_j,\mathbf{s}_k$, $j,k\in\{1,\dots,m\}$) is chosen instead of $\mathbf{s}_1$ (or instead of $\mathbf{s}_1,\mathbf{s}_2$), the corresponding likelihood functions would be proportional to those shown here, leading to the same maximum. A proof is provided in the supplementary material, Section \ref{suppl:section:example_deference_locations}.

\subsection{Step 2: estimation of \texorpdfstring{$\theta_{S,R}%\mbox{ and } \theta_R
$}{theta S R}}\label{section:new_method_step2}

We consider parameter estimation in the case of a Gaussian location-scale mixture; the adaptation to simpler cases presents no difficulty. The vectors $\bX$ and $\bW$ are transformed into the (spatial) average of their $m$ respective elements,  $\bar{X}$ and $\bar{W}$, which are linked by the equation
\begin{equation}\label{eq:average}
 \bar{X}=S+R\;\bar{W}.
\end{equation}  Note that $    \bar{W}\sim \mathcal{N}\left(0,\sigma_{\bar{W}}^2\right)$, with $\sigma_{\bar{W}}^2=m^{-2}\;\mathbf{1}^\top \bSigma_{\theta_W} \mathbf{1}$, and that $\theta_W$ has been estimated in the first step.
Identifiability of $\theta_{S,R}$ from the distribution of $\bar{X}$, when both $S$ and $R$ are non-degenerate random variables, is discussed in Section \ref{section:identifiability1_parameter} in the supplementary material. A short proof is also provided for the two location-scale mixture models presented in this work: LSM1 and LSM2.

In general, the distribution of $\bar{X}$  is not known and
once again, it would be necessary to resort to numerical integration as in \eqref{eq:chapter2_integral_density} to evaluate the cdf 
$$
F_{\bar{X}}(x;\theta_W, \theta_{S,R})=\int_{-\infty}^{+\infty}\int_{0}^{+\infty} \phi\left(\frac{x-s}{r\,\sigma_{\bar{W}}} \right)  f_{S,R}(s,r;\,\theta_{S,R}) \,ds\,dr,
$$
or its density.
However, the simulation of $\bar{X}$ does not require the knowledge of $F_{\bar{X}}$ and, unlike the vector case $\bX$, the simulation  is fast. This opens up the possibility of deriving  estimates of $\theta_{S,R}$ based on a minimum distance method between distributions  (\citealp{wolfowitz1957minimum,bolthausen1977convergence})
by approximating $F_{\bar{X}}$ with a Monte Carlo method.

More precisely, we consider the averages $\bar{x}_i$ from each $\bx_i$, $i=1,\dots,n$, and we denote $F_{n}$ their empirical cdf.   A minimum distance estimator of $\theta_{S,R}$, pretending that  $\theta_W$ is known, minimizes a distance $\tau(\theta_{S,R})$ between the empirical cdf $F_n(x)$ and the theoretical cdf $F(x;\,\theta_{S,R})$.
In particular, minimization of the $L^2$   distance
$$
\tau(\theta_{S,R})=\int [F_{n}(x)-F_{\bar{X}}(x;\,\theta_W, \theta_{S,R})]^2 F_{\bar{X}}(dx;\,\theta_W, \theta_{S,R})
$$  
with respect to $\theta_{S,R}$ leads to a Cram\'er-von Mises estimator \citep{pollard1980minimum}.
Estimators minimizing the $L^2$ distance are typically strongly consistent and asymptotically normal; they are also more robust to model misspecification than maximum likelihood estimators \citep{parr1981minimum}. We consider the   sample version of the distance, i.e. %the Cram\'er-von Mises statistic \citep{anderson1962distribution}
\begin{equation}\label{eq:cramer_von_mises}
	T(\theta_{S,R}) = %\frac{1}{12n}+
    \sum_{i=1}^n \left[\frac{i-1/2}{n}-F_{B,\bar{X}}\left(\bar{x}_{(i)};\theta_W,\theta_{S,R}\right)\right]^2,
\end{equation}
where  $\bar{x}_{(i)}$, $i=1,\dots,n$, are the ordered  values of $\bar{x}_{i}$, $i=1,\dots,n$.
Note that we approximate $F_{\bar{X}}$ by  $F_{B,\bar{X}}$, the empirical cdf  of $B$  i.i.d. copies of \eqref{eq:average}, with
$(S,R)\sim F_{S,R}(\cdot;\,\theta_{S,R})$ and
$\bar{W}\sim \mathcal{N}(0,m^{-2}\;\mathbf{1}^\top \bSigma_{\hat\theta_W} \mathbf{1})$. A formal proof of consistency of the (simulated) estimator minimizing \eqref{eq:cramer_von_mises} is given in Section \ref{section:smde_supplementary} of the supplementary material. We typically set $B=10^6$  from each distribution, setting the same seeds at each optimization iteration, as shown in the supplementary code.
Although this intermediate Monte Carlo approximation must be repeated for each value of $\theta_{S,R}$ in the iterations of the minimization algorithm of \eqref{eq:cramer_von_mises}, in our experience the procedure remains %robust and 
quite fast because it only involves simulating from univariate distributions.
%The new inference procedure we propose in this work does not require assuming independence. 
%From \eqref{eq:gaussian_locationscale_mixtures}, we can distinguish three subclasses of models, leading to some differences in the inference and simulation approaches we propose.

\subsection{Copula models}\label{section:new_method_copula}
The two-step estimation method described above starts from the premise to observe data from $\bX$.
However, as explained in Section \ref{section:general_model_definition}, in real data applications model \eqref{eq:gaussian_locationscale_mixtures} may be used as a copula model for the dependence inside of $\mathbf{Y}=(Y(\mathbf{s}_1),\dots,Y(\mathbf{s}_m))^\top$. According to this modelling approach each element of $\bY$ is linked to each element of $\bX$ by
$
X(\mathbf{s}_i) =F_X^{-1}\left(F_Y(Y(\mathbf{s}_i))\right),\quad i=1,\ldots,m,
$
where $F_X$ and $F_Y$ are the cdf of $X(\mathbf{s}_i)$ and $Y(\mathbf{s}_i)$, respectively. 

We assume that the analytical expression of  $F_Y$ is either known or can be consistently estimated, see  Section \ref{section:application} for an example.
In this way, we get 
\begin{equation}\label{eq:tranform_copula_UY}
	\mathbf{U}=(U(\mathbf{s}_1),\dots,U(\mathbf{s}_m))^\top=\left(F_Y(Y(\mathbf{s}_1)),\dots,F_Y(Y(\mathbf{s}_m))\right)^\top.
\end{equation}

Then, to transform $\mathbf{U}$ into $\bX$, one would need to know the marginal distributions of the latter, $F_X$. 
As mentioned above, however, analytical expressions for 
$F_X$ and $F_X^{-1}$, the quantile function, are rarely available. 
On the other hand, for a fixed value of $\theta_{S,R}$ an inexpensive Monte Carlo approximation of the quantile function $F^{-1}_X$  can be obtained by generating a sufficiently large  number of i.i.d.\ copies of $X(\mathbf{s})=S+R W(\mathbf{s})$, with
$(S,R)\sim F_{S,R}(\cdot;\,\theta_{S,R})$ and
$W\sim \mathcal{N}(0,1)$.
We denote this approximation by $F^{-1}_X(\cdot;\,\theta_{S,R})$, with a slight abuse of notation, and we set
\begin{equation}\label{eq:tranform_copula_XU}
	\bX(\theta_{S,R})=\left(F_X^{-1}(U(\mathbf{s}_1);\,\theta_{S,R}),\dots,F_X^{-1}(U(\mathbf{s}_m);\,\theta_{S,R})\right)^\top.
\end{equation}
We now incorporate this additional Monte Carlo approximation into the two-step procedure for estimating the parameters  $\theta_W$ and $\theta_{S,R}$, as described above, to obtain Algorithm \ref{alg:estimation_copula}. The input of the algorithm are the vectors $\bu_i$, $i = 1, \ldots, n$, obtained by transforming the $n$ independent observed values of $\bY$, $\by_i$, using \eqref{eq:tranform_copula_UY}.

\begin{algorithm}[H]
\caption{Parameter estimation in the copula case}\label{alg:estimation_copula}
\begin{algorithmic}
\State
{\bf Input:} $\mathbf{u}_i$, $i=1,\ldots,n$.
\State {\bf Output:} Estimated values    $\hat{\theta}_W$ and $\hat\theta_{S,R}$
\begin{enumerate}
    \item For a current value of $\theta_{S,R}$% and $\theta_R$
    , compute $\bx_i(\theta_{S,R})$, $i=1,\ldots,n$ as in \eqref{eq:tranform_copula_XU};
    \item Estimate $\theta_W$ with the method described in Section \ref{section:new_method_step1}, obtaining $\hat{\theta}_W$;
    \item Compute the Cram\'er-von Mises statistic $T(\theta_{S,R})$ as in \eqref{eq:cramer_von_mises};
    \item Iterate points $1$--$3$ by changing the fixed values of $\theta_{S,R}$% and $\theta_R$
    , to minimize $T(\theta_{S,R})$.
\end{enumerate}

\end{algorithmic}
\end{algorithm}
The iterative estimation of  ${\hat{\theta}}_{W}$ and  ${\hat{\theta}}_{S,R}$ increases the computational cost for the computational case.  A comparison between the computation time in the copula and non-copula cases can be found in Section \ref{section:simul_computation}, and in particular in Table \ref{table:simul_computation_time}.

\subsection{Confidence intervals}\label{section:new_method_interval}
One option for obtaining confidence intervals for the estimates of $\theta_W$ and $\theta_{S,R}$ would be to compute the asymptotic variances of the two-step estimators \citep{gong1981pseudo,zhao2005composite} and use them to construct the confidence intervals.

Here, however, a parametric bootstrap procedure can be employed to approximate these confidence intervals thanks to the ease of simulating from the models. In the non-copula case, once point estimates $\hat{\theta}_W$ and $\hat{\theta}_{S,R}$ are obtained as described in Sections \ref{section:new_method_step1} and \ref{section:new_method_step2}, it is sufficient to simulate, from the Gaussian location-scale mixture model \eqref{eq:gaussian_locationscale_mixtures}, $B$ dataset analogues to the observed one for spatial locations and number of replications. Then, for $b=1,\dots,B$, estimates $\hat\theta_W^b$ and $\hat\theta_{S,R}^b$ can be obtained following the same procedures; their distribution can be used to build confidence intervals. Since the estimator $\hat{\theta}_W$ is obtained by maximizing a correctly specified likelihood, we could alternatively assess its uncertainty by using standard tools for maximum likelihood estimators.

Interval estimation becomes slightly more complicated in the copula case (Section \ref{section:new_method_copula}). 
%since the vectors on the Gaussian location-scale mixture marginal scale are constructed from vectors of uniforms, as shown in \eqref{eq:tranform_copula_XU}, with a transformation $F_X^{-1}(\cdot;\theta_{S,R})$ that depends on a parameter.
In this case, after $B$ datasets from model \eqref{eq:gaussian_locationscale_mixtures} are simulated using the estimated parameters $\hat{\theta}_W$ and $\hat{\theta}_{S,R}$, the data must be transformed by applying the function $F_X(\cdot;\hat{\theta}_{S,R})$, which leads to $B$ datasets on the uniform marginal scale.
Then, two options are available to obtain estimates $\hat{\theta}_W^b$ and $\hat{\theta}_{S,R}^b$, for $b=1,\dots,B$:
\begin{itemize}
    \item The standard option is to apply to each dataset the procedure of Section \ref{section:new_method_copula}
    %, that is, estimating $\theta_{S,R}$ by minimizing \eqref{eq:cramer_von_mises_v2} to obtain $\hat{\theta}_{S,R}^b$ and then computing $\hat{\theta}^b_W\{\hat{\theta}^b_{S;R}\}$
    ; in some cases, this may be computationally demanding since it involves repeating $B$ times the recursive estimation described in Algorithm \ref{alg:estimation_copula}.
    \item A faster option is to obtain each estimate $\hat{\theta}_{S,R}^b$ by fixing $\theta_W$ to the value $\hat{\theta}_W$ estimated on the observed dataset and minimizing \eqref{eq:cramer_von_mises}; then, each $\hat{\theta}_W^b$ can be estimated, with the method described in Section \ref{section:new_method_step1}, on the dataset transformed by $F_X^{-1}(\cdot\;;\hat{\theta}^b_{S,R})$. Since uncertainty is not propagated between the estimators of $\theta_{S,R}$ and $\theta_W$,  the width of confidence intervals could be  (slightly) underestimated.   
\end{itemize}

%\section{Numerical examples}
%\subsection{Simulated data example}
%\subsection{Real data example}
\section{Simulation study}\label{section:simulation_studies}

In this section (see also the Supplementary Material), the proposed inferential method is evaluated on simulated data, and its performance is compared to that of some of the existing ones.
In particular, data are simulated from Gaussian location-scale mixtures at locations sampled randomly in a square  $[0,200]\times [0,200]$, which is similar to the dimension (in km) of the spatial domain of the real data application (see Section \ref{section:application}).
The  correlation function of the Gaussian process $\{W(\mathbf{s})\}$ is  a Mat\'ern correlation function with range parameter $\varphi>0$ and smoothness parameter $\eta>0$, such that $\theta_W=(\varphi,\eta)$ and, for $\mathbf{s}_1,\mathbf{s}_2\in\mathcal{S}\subset \mathbb{R}^2$,
\begin{equation*}
    \rho(\mathbf{s}_1,\mathbf{s}_2)=\frac{1}{2^{\eta-1}\Gamma(\eta)}\left(\frac{2\sqrt{\eta}\|\mathbf{s}_1-\mathbf{s}_2\|}{\varphi}\right)^\eta K_\eta\left(\frac{2\sqrt{\eta}\|\mathbf{s}_1-\mathbf{s}_2\|}{\varphi}\right),
\end{equation*}
where $\|\cdot\|$ is the Euclidean distance and  $K_\eta$ is the modified Bessel function of the second kind of order $\eta$.
In the simulations, the true values of these parameters are fixed at $\varphi=50$ and $\eta=0.5$. Note that the value $\eta=0.5$ defines the exponential correlation function; other simulations with $\eta=1$ were also conducted, leading to very similar results in terms of bias and variance of the estimators, and are not shown here to limit the amount of results presented.
The number of spatial locations, $m$, and the number of independent replications in each spatial location, $n$, are considered according the following   data-poor, data-rich and intermediate configurations:
%\begin{description}
%    \item[] A: $m=50$, $n=100$;
%    \item[]  B: $m=100$, $n=500$;
%    \item[]  C: $m=200$, $n=1000$;
%    \item[]  D: $m=400$, $n=2000$.
%\end{description}
A) $m=50$, $n=100$; B) $m=100$, $n=500$; C) $m=200$, $n=1000$; D) $m=400$, $n=2000$.
For each replication, the $m$ locations are sampled uniformly within $[0,200]\times[0,200]$, increasing the uncertainty in the estimation.

Section~\ref{section:simul_comparison} compares the proposed method with the maximization of the full likelihood, for two models for which the latter is available in closed form; Section \ref{section:simul_consistency} (in the supplementary material) numerically illustrates the consistency of the proposed method for models belonging to the three classes of Gaussian location, scale and location-scale mixtures; Section \ref{section:simul_integration} (in the supplementary material) compares the proposed method with the integration-based one from equation~\eqref{eq:chapter2_integral_density}; Section \ref{section:simul_copula} (in the supplementary material) reproduces the estimations of Sections \ref{section:simul_comparison} and \ref{section:simul_consistency} adopting a copula approach; finally, Section \ref{section:simul_computation} includes comments on the computation time necessary for parameter estimation in all the above cases.

\subsection{Comparison with full likelihood}\label{section:simul_comparison}
In general, the finite-dimensional distribution of Gaussian location-scale mixtures \eqref{eq:gaussian_locationscale_mixtures} is not known in closed form.
However, in few cases, such as Laplace and Student's \textit{t} processes (see Section \ref{section:example_models_gaussiansclocmixt}) it is available in closed form. In these cases, we can perform inference on the model parameter by maximizing the full likelihood and use it as a benchmark to quantify the loss of statistical efficiency of the  proposed inferential solutions.
In the following cases, inference is performed on 100 simulated datasets for each of the configurations A, B, C and D (see above).

\begin{figure}[t!]
    \centering
    \includegraphics[width=1\linewidth]{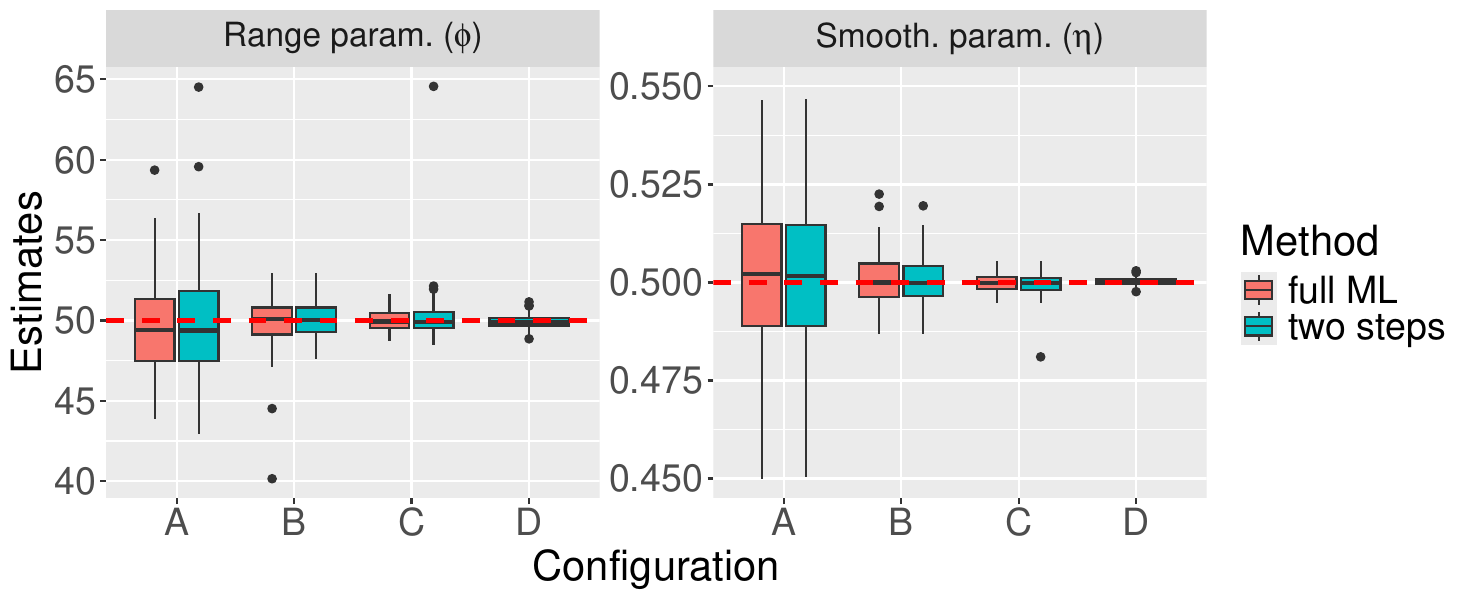}
    \caption{Estimation of the parameters of a Laplace process on 100 simulated datasets, for each configuration, with true values $\varphi=50$ and $\eta=0.5$ (red lines). Orange boxplots report full maximum likelihood estimates (not computed for Configuration D for computational reasons), while blue boxplots show the estimates based on the likelihood of ratios explained in Section \ref{section:new_method_step1}.}
    \label{fig:boxplots_sim_laplace}
\end{figure}

Figure \ref{fig:boxplots_sim_laplace} shows results for the estimation of $\theta_W=(\varphi,\eta)$ for a Laplace process (model SM1 in Table \ref{table:examples_gaussian_locscmix_models}). The scaling random variable $R$ has no parameter to estimate in this case, since it has standard Rayleigh distribution. For the parameter $\theta_W$, the proposed estimation method appears unbiased and does not lose much efficiency, compared to the full maximum likelihood.
\begin{figure}[t!]
    \centering
    \begin{subfigure}{\linewidth}
           \includegraphics[width=1\linewidth]{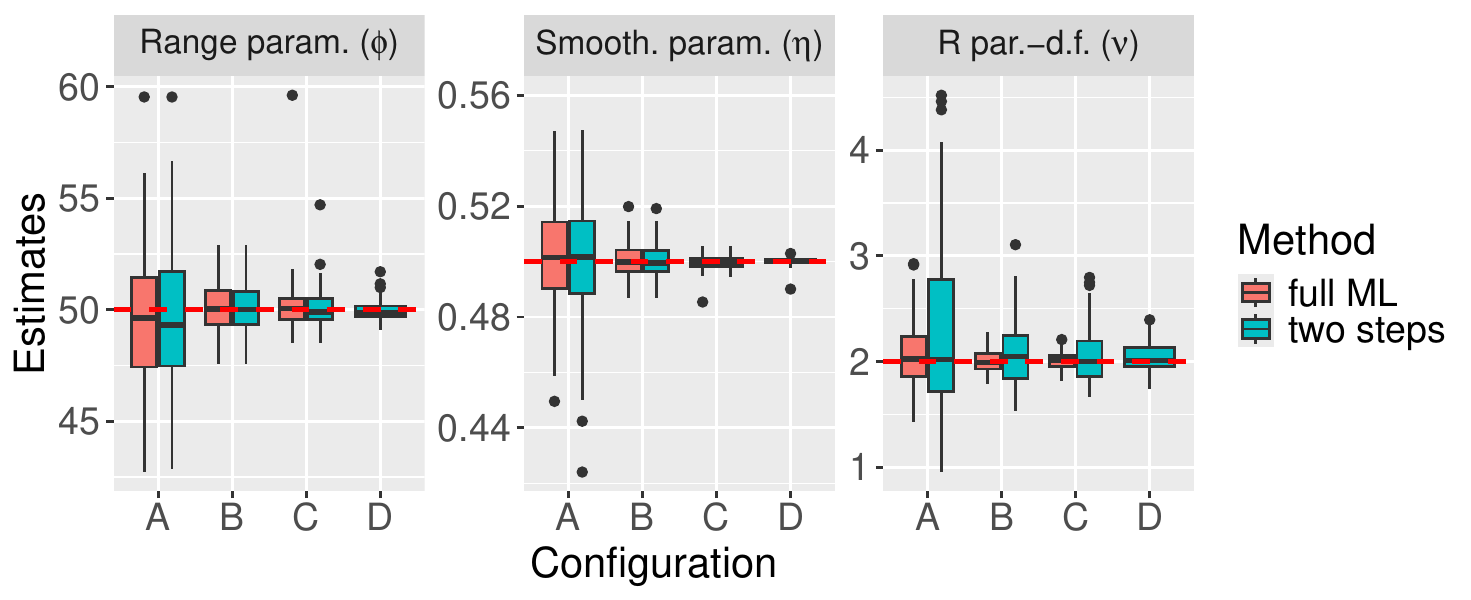} 
    \end{subfigure}
  %  \begin{subfigure}{\linewidth}
  %         \includegraphics[width=1\linewidth]
  %  \end{subfigure}
    \caption{Estimation of the parameters of a Student's \textit{t} process on 100 simulated datasets for each configuration, with true values (red lines) $\varphi=50$, $\eta=0.5$ and $\nu=2$. %(top) and $\nu=8$ (bottom). 
    Orange boxplots report full maximum likelihood estimates (not computed for Configuration D for computational reasons), while blue boxplots show the two-step solution for inference explained in Section \ref{section:estimation_methods}.}
    \label{fig:boxplots_sim_t_2_8}
\end{figure}
%\begin{figure}[htpb!]
%    \centering
%    \includegraphics[width=1\linewidth]
%    \caption{Estimation of the parameters of a Student's \textit{t} process on 100 simulated datasets for each configuration, with true values $\varphi=50$, $\eta=0.5$ and $\nu=8$ (red lines). Orange boxplots refer to full maximum likelihood estimates, while blue boxplots refer to the two-step solution for inference explained in Section \ref{section:estimation_methods}.}
%    \label{fig:boxplots_sim_t_8}
%\end{figure}

Figure \ref{fig:boxplots_sim_t_2_8} displays analogue estimation results on a Student's \textit{t} process (model SM3 in Table \ref{table:examples_gaussian_locscmix_models}) with $\nu=2$ %and $\nu=8$ 
degrees of freedom. %For $\nu$ increasing to infinity, the process tends to a Gaussian process, and $\nu$ generally becomes more difficult to identify from data. 
The estimators of $\theta_W=(\varphi,\eta)$ show performances similar to the maximum likelihood estimators, as in the previous case. However, the estimator of $\theta_R=\nu$ exhibits clearly higher variability than the one based on the full likelihood, especially in the lower-dimensional configurations.
This is due to the relative statistical inefficiency of the minimum distance estimator used in Step 2 (see Section \ref{section:new_method_step2}), compared to the maximum likelihood estimator. The efficiency of the two-step estimator could be improved by replacing the second step with the maximization of the likelihood based on the distribution of $\bar{X}$ in \eqref{eq:average}, in particular when it is available in closed form and in the non-copula cases. However, for the cases presented here, we opted to use the same estimation method, which is valid for every distribution of $S$ and $R>0$, and for all copula cases.

To further confirm consistent behaviour of the two-step estimation procedure, in particular for the parameter $\theta_{S,R}$ of the random scale and location of the mixture models, another simulation study is provided in Section \ref{section:simul_consistency} in the supplementary material, for some of the models presented in Section \ref{section:example_models_gaussiansclocmixt}.

\subsection{Computation time}\label{section:simul_computation}
\noindent
In addition to consistency and statistical efficiency, an important feature of estimation methods for Gaussian location-scale mixtures is computational efficiency. In many cases, the computation times may be inflated by the presence of latent variables in the models, with additional parameters to estimate. Table~\ref{table:simul_computation_time} reports the average computation time, in seconds, for the simulations of Section \ref{section:simulation_studies} and Section \ref{section:simul_supplementary} in the supplementary material.
These averages are computed over the 100 datasets of each simulation study and over all the models belonging to each setting, with the exception of the integration-based method, for which 50 datasets are simulated. In particular, the case with 2 parameters refers to the Laplace process (model SM1, Figures \ref{fig:boxplots_sim_laplace}, \ref{fig:boxplots_sim_laplace_int} and \ref{fig:boxplots_sim_laplace_copula});
the case with 3 parameters covers the Student's \textit{t} process (model SM3, Figures \ref{fig:boxplots_sim_t_2_8} and \ref{fig:boxplots_sim_t_2_copula}), as well as model LM1 (Figures \ref{fig:boxplots_sim_exp}, \ref{fig:boxplots_sim_exp_int} and \ref{fig:boxplots_sim_exp_copula}), model SM4 (Figures \ref{fig:boxplots_sim_sgpd} and \ref{fig:boxplots_sim_sgpd_copula}) and model LSM1 (Figures \ref{fig:boxplots_sim_exp_lapl} and \ref{fig:boxplots_sim_exp_lapl_copula}); finally, the case with 4 parameters refers to model LM2 (Figure \ref{fig:boxplots_sim_aslapl}).
The code for the simulation studies is written in the R programming language and executed on a 2.3 GHz machine with eight cores and 12 GB of memory. We use the Nelder-Mead method for multivariate optimization and the Brent method for univariate optimization.

\begin{table}[t!]
\begin{center}
\caption{Average computation time (rounded to at most 2 significant digits), in seconds, for parameter estimation on a single dataset in the simulation studies of Section \ref{section:simulation_studies} and Section \ref{section:simul_supplementary}, for the cases with 2, 3 and 4 parameters, without and with copula, and for configurations A ($m=50, n=100$), B ($m=100, n=500$), C ($m=200, n=1000$) and D ($m=400, n=2000$).}
\begin{tabular}{| c | c | c | c | c | c | c | c | c | c |} 
 \hline
  \multirow{2}{*}{\makecell[c]{Number of\\parameters}} & \multirow{2}{*}{Method} & \multicolumn{4}{c|}{No copula} & \multicolumn{4}{c|}{Copula}  \\
 \cline{3-10} & & A & B & C & D & A & B & C & D \\
 \hline\hline
 \multirow{3}{*}{2}& two steps & 0.3 & 1 & 5 & 32 & 1 & 2 & 8 & - \\
 \cline{2-10}
 & full likelihood & 2.6 & 70 & 550 & - & 3 & 50 & 540 & - \\
 \cline{2-10}
 & integration & 450 & 8600 & - & - & - & - & - & - \\
 \hline\hline
 \multirow{3}{*}{3} & two steps & 45 & 50 & 55 & 85 & 70 & 110 & 250 & - \\
 \cline{2-10}
 & full likelihood & 15 & 130 & 800 & - & 130 & 300 & 1200 & - \\ 
 \cline{2-10}
 & integration & 470 & 5300 & - & - & - & - & - & - \\
 \hline\hline
 4 & two steps & 290 & 310 & 320 & 370 & - & - & - & - \\ \hline
\end{tabular}
\label{table:simul_computation_time}
\end{center}
\end{table}

The computation times shown in Table \ref{table:simul_computation_time} indicate that the two-step approach is much faster than the maximization of the closed-form full likelihood, with much better scalability as the number of locations $m$ and replications $n$ is increased, such as in Configurations C and D.
This may be due to the separation of the functions to be optimized in two steps, and to the simpler form of these functions compared to that of some likelihood functions.
The third method, based on the numerical integration of \eqref{eq:chapter2_integral_density}, is much more computationally demanding and becomes prohibitive for Configurations C and D. The cases reported here refer only to single integrals, as explained in Section \ref{section:simul_integration}. A general case of Gaussian location-scale mixture (with both random location $S$ and random scale $R$) would involve a double integral, making the  computation time even longer.
On the other hand, the  proposed method requires similar computation time for the three cases of Gaussian location, scale and location-scale mixtures.

\section{Real data application to fire weather data in Portugal}\label{section:application}

\subsection{Data description}\label{section:application_datadescription}

ERA5-Land is a large hourly global dataset of various weather variables on a longitude-latitude grid of size $0.1^{\circ}\times 0.1^{\circ}$. It is obtained by assimilating observation data (e.g., weather stations, radar, satellite) into a physical climate model.  Data are available for the period from 1940 to today, and are routinely used as a realistic approximation of the true weather of the past. Following common practice, we here refer to them as ``observations", even though producing them has required using a model.   
\begin{figure}[t]
    \centering
    \includegraphics[width=0.85\linewidth]{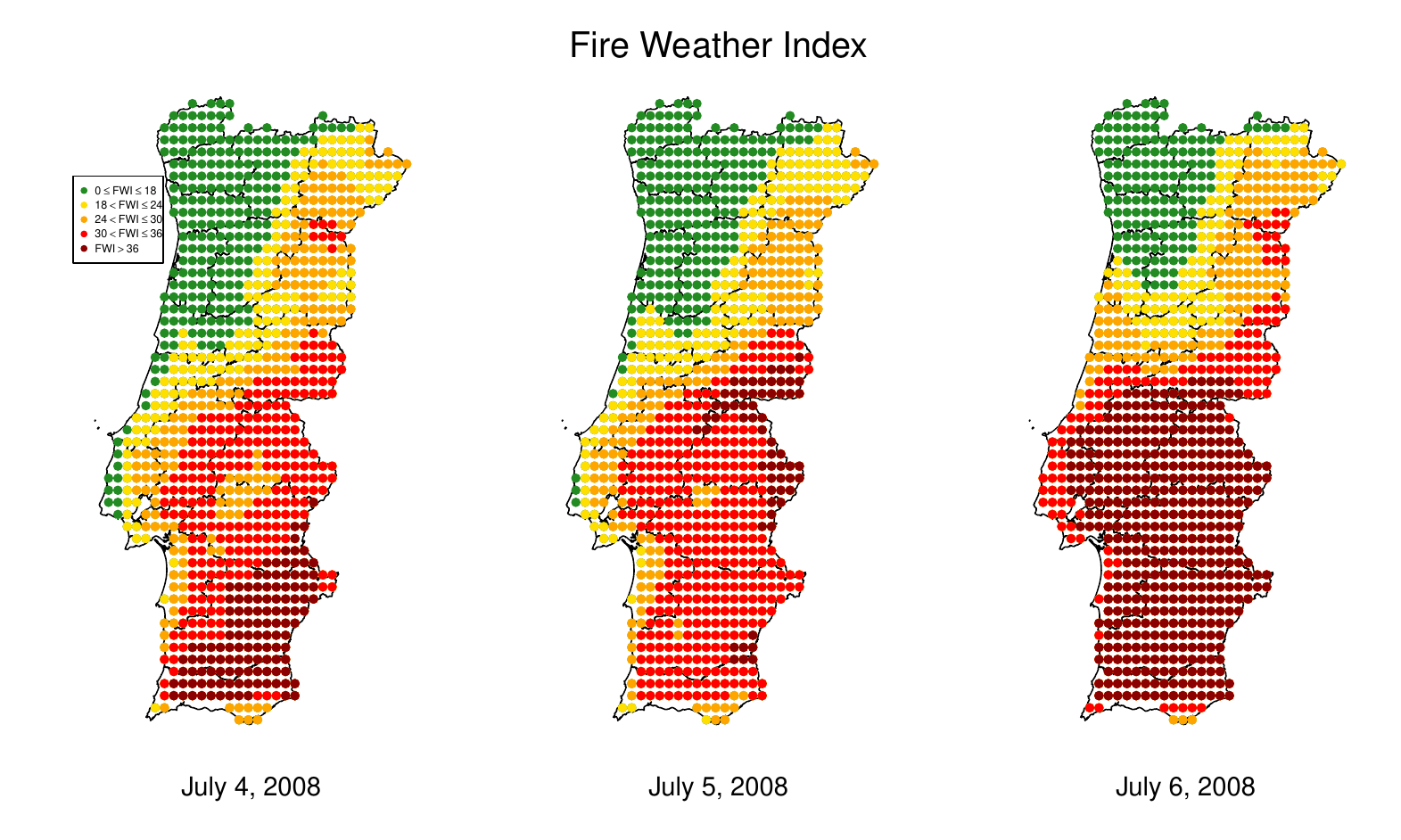}
    \caption{Three consecutive days of observed FWI over Portugal.}
    \label{fig:portugal3days}
\end{figure}
We apply the models we propose to a dataset of daily Fire Weather Index (FWI), a continuous random variable measuring the weather-related risk of wildfires, computed from ERA5-Land data on the whole domain of continental Portugal for recent years. The original dataset includes $23$ years (from 2000 to 2022) of daily observations on $947$ spatial locations. For illustration of the data, the maps in Figure~\ref{fig:portugal3days} show FWI values observed for three consecutive days in summer 2008.
Although the FWI, which is based on weather information, appears to take higher values in the south of Portugal, the area with the highest historical susceptibility to wildfires is the north of the country, one of Europe's wildfire hotspots, due to the land-use and land-cover configurations(\citealp{verde2010assessment}; \citealp{neves2023landsat}; see also \citealp{moghli2025fireres}).
Therefore, we split the geographical domain into two subsets, as shown in Figure \ref{fig:portugal_northsouth} in the supplementary material, and  our analysis focuses only on the northern region.
This reduces the dimensionality of the dataset to $m=491$ spatial locations.
Moreover, to ensure approximate temporal stationarity, we restrict the analysis to the two main wildfire months of July and August, leading to $62$ values at each location for each of the $23$ years, for a total of $n=1426$ replications.

\subsection{Marginal modelling}\label{section:application_marginal_modelling}

As explained in Section \ref{section:general_model_definition}, Gaussian location-scale mixtures \eqref{eq:gaussian_locationscale_mixtures} may be used as copula models for the dependence structure of the data and specifying a marginal distribution of the observed data. 
%which are observed on another marginal scale $\{Y(\mathbf{s})\}$, with marginal distribution $F_Y^\mathbf{s}$ that may depend on the spatial location $\mathbf{s}$, for $\mathbf{s}\in\mathcal{S}$, as explained below.
A typical approach in extreme value analysis, based on the peaks over thresholds theory, is to fix some high threshold, which can also depend on the spatial location $\mathbf{s}$, and model the positive excesses over the threshold  with a generalized Pareto distribution (GPD).

Various  extensions of the GPD have been proposed to avoid the explicit selection of a threshold by modelling instead all the data, while trying to maintain a good fit on the tails \citep{naveau2024multivariate}.
These extensions reduce the uncertainty brought into the model by the selection of the threshold, and allow to model the entire marginal distribution of the data. %, not only the threshold exceedances.

We use the Extended Generalized Pareto Distribution (EGPD) \cite{naveau2016modeling} as a model for $F_Y$ that complies with extreme value theory in both the upper and lower tails.   
 Its cdf is defined as
\begin{equation}\label{eq:egpd_def}
    F_Y(y)=B\left(H_\xi\left({y}/{\sigma}\right)\right),
\end{equation}
where $H_\xi(\cdot)=1-(1+\xi x)^{-1/\xi}$, for $\xi\ne0$, is the cdf of a GPD, with shape parameter $\xi$.
Here  $\sigma>0$ is a scale parameter and $B(\cdot)$ is  a continuous cdf on $[0, 1]$ satisfying the following three conditions for three positive constants $a$, $b$ and $c$: $\lim_{u\to0}\bar{B}(1-u)/u=a$; $\lim_{u\to0}B(u\,w(u))/B(u)=b$  where $w(\cdot)$ is any positive function such that $w(u)=1+o(u)$ as $u\to0$; $\lim_{u\to0}B(u)/u^\kappa=c$ with $\kappa\ge0$.

\begin{figure}[t]
\begin{subfigure}{0.44\linewidth}
\includegraphics[width=\linewidth]{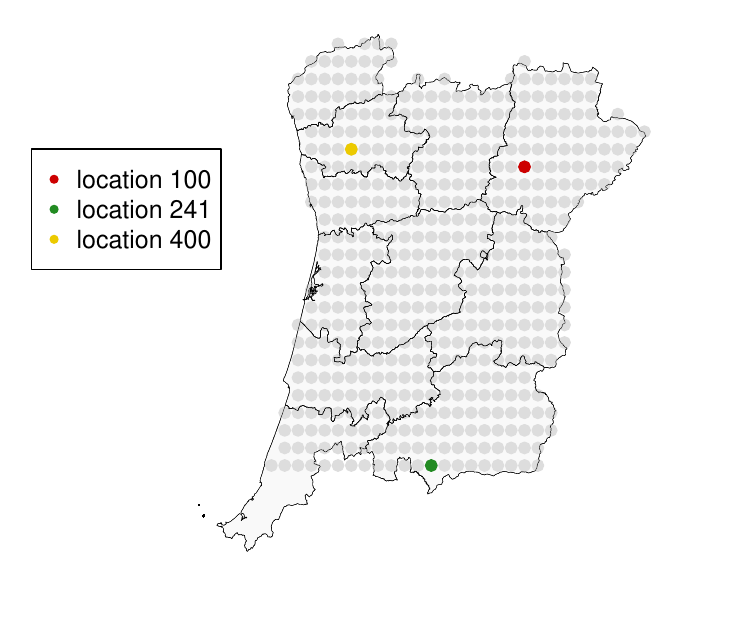}
\end{subfigure}
\hfill
\begin{subfigure}{0.44\linewidth}
\includegraphics[width=\linewidth]{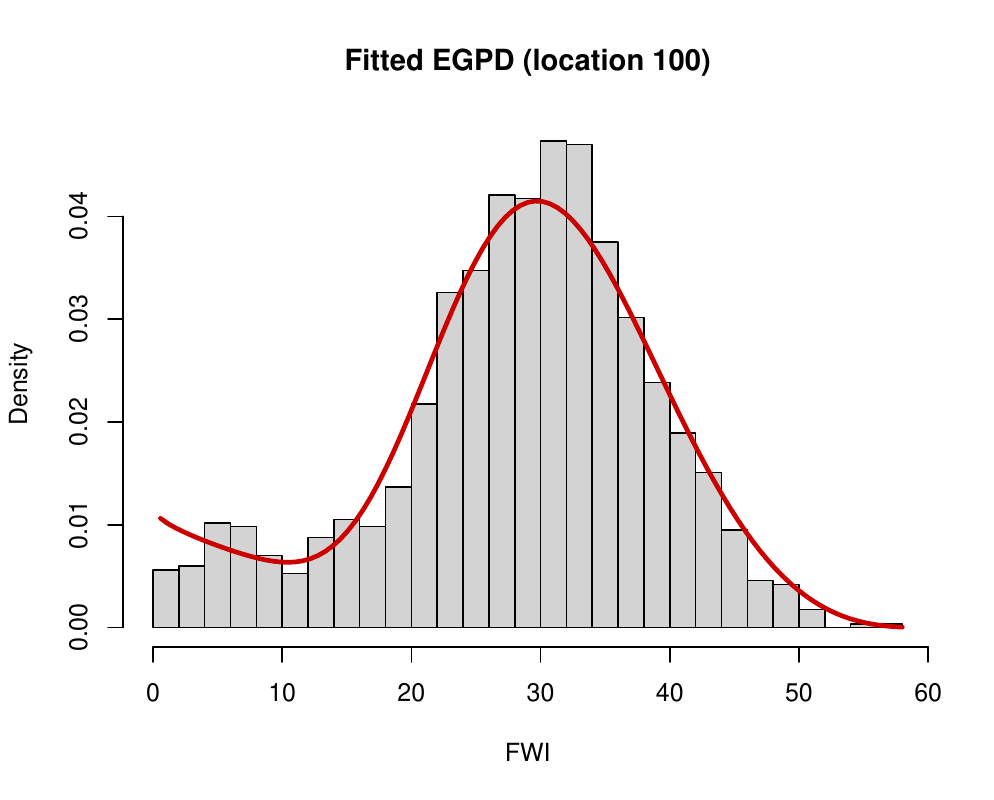}
\end{subfigure}
\begin{subfigure}{0.44\linewidth}
\includegraphics[width=\linewidth]{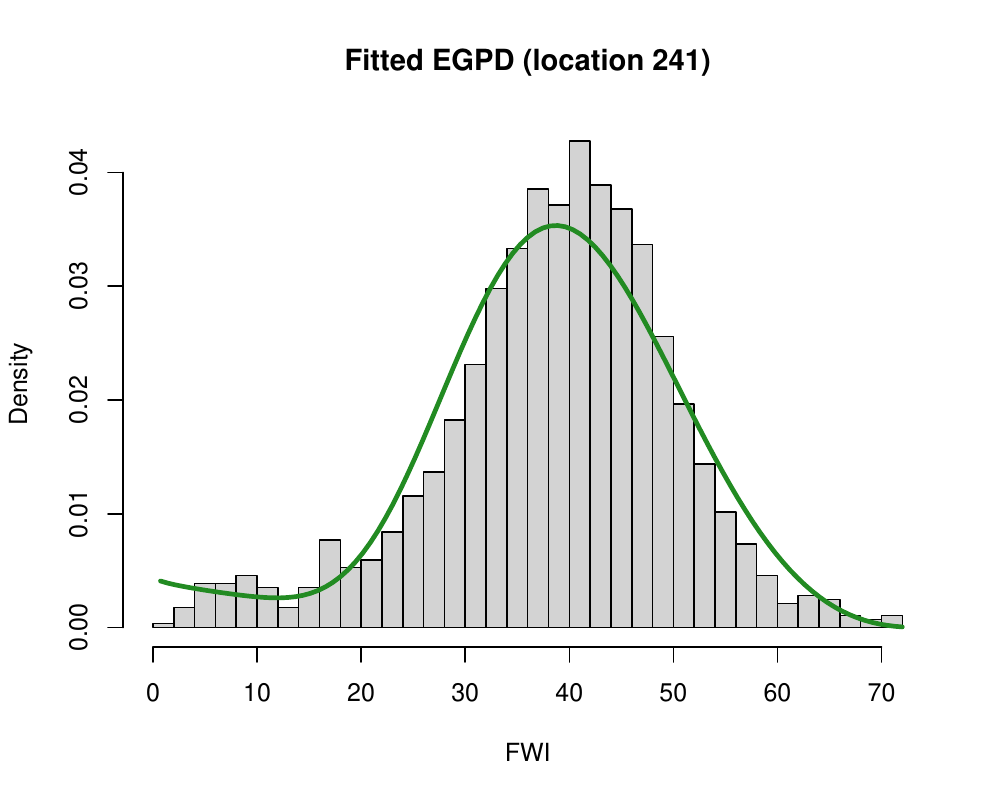}
\end{subfigure}
\hfill
\begin{subfigure}{0.44\linewidth}
\includegraphics[width=\linewidth]{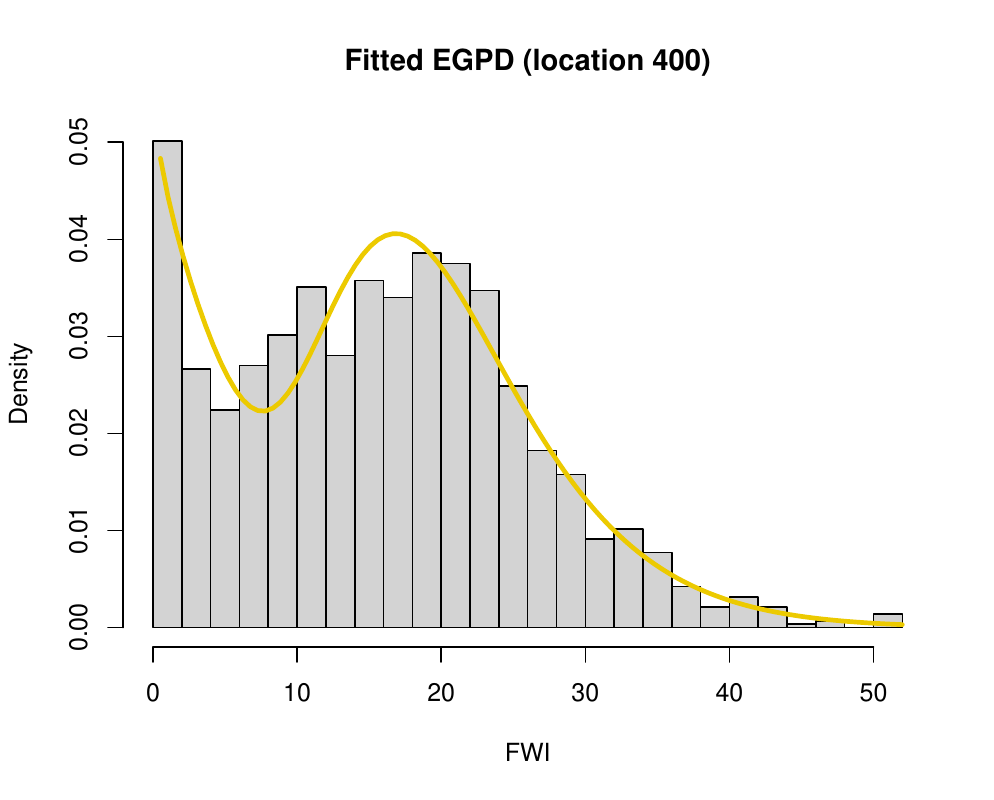}
\end{subfigure}
\caption{Three locations in the north of Portugal and the corresponding FWI histograms and fitted EGPD densities, with spatially-varying parameters $\sigma$, $\xi$ and $p$.}
\label{fig:pt_3locations}
\end{figure}

We choose the  parametric specification  
 $B(u)=p u^{\kappa_1} + (1-p) u^{\kappa_2}$, with $p\in[0,1]$ and $\kappa_2\ge\kappa_1>0$ \citep{naveau2016modeling}. In this case, the lower tail behaviour is controlled by $\kappa_1$, while $\kappa_2$ modifies the shape of the density in its central part. 
 This specification permits possible bimodality in the marginal distribution of the data, as illustrated in Figure \ref{fig:pt_3locations}. This bimodality is likely due to the way the FWI is constructed; high-intensity rainfall events cause the index value to approach zero.
 
 %This allows  for possible bimodality in the marginal distribution of the data (see Figure \ref{fig:pt_fitted_egpd_hist} in the supplementary material). 
%This bimodality is probably a consequence of how the fire weather index is constructed: high-intensity rainfall events push the value of the index very close to zero.

The marginal model presented above has five parameters, $\sigma$, $\xi$, $p$, $\kappa_1$ and $\kappa_2$.
%A first approach is to assume that these parameters are constant over the north of Portugal and estimate them by maximizing an independence likelihood \citep{Varin:Reid:Firth:2011}.
%This leads to the following estimated values: $\hat{\sigma}=16.44$, $\hat{\xi}=-0.22$, $\hat{p}=0.22$, $\hat{\kappa}_1=0.74$ and $\hat{\kappa}_2=5.81$.
%The corresponding EGPD density is shown in Figure \ref{fig:pt_fitted_egpd_hist} in the supplementary material, over the histogram of all the observed FWI values in the north of Portugal.
%A different model specification 
We account for the spatial non-stationarity of the data, visible in Figure \ref{fig:portugal3days},
by letting parameters $\sigma$, $\xi$ and $p$ in \eqref{eq:egpd_def} vary according to smooth functions of the coordinate $\mathbf{s}$, leading to a cdf $F_Y^{\mathbf{s}}$ of $Y(\mathbf{s})$ varying with $\mathbf{s}$.
The estimation of these functions using a penalized likelihood approach based on regression splines is implemented in the experimental version of the R package \texttt{evgam} \citep{Youngman:2020}, % (downloaded in December 2025) 
available at \href{https://github.com/byoungman/evgam}{github.com/byoungman/evgam}. 
The extra flexibility seems able to capture the non-stationary behaviour of the data, as shown in Figure \ref{fig:pt_3locations} for three spatial locations that are far from each other, and also in Figure \ref{fig:pt_quantile_maps} in the supplementary material, where we compare empirical and fitted quantiles of order $u\in\{0.5,0.75,0.9,0.95\}$ for all the locations in the north of Portugal.
Therefore we use the estimates of $F_Y^s$ to transform the data as in
\eqref{eq:tranform_copula_UY}, and then use the resulting vectors $\bu_i$, $i=1,\dots,n$, in Algorithm \ref{alg:estimation_copula} to fit the dependence model parameters.
\subsection{Dependence modelling}\label{section:application_dependence_modelling}

We consider several candidate models in the class of Gaussian location-scale mixtures: the Gaussian location mixture 1 (LM1), the Gaussian scale mixture 5 (SM5), and the Gaussian location-scale mixture 2 (LSM2). 
The parameter estimation process took approximately 15 minutes for the three-parameter models LM1 and SM5, and three hours for the four-parameter model LSM2. This can be compared with the copula cases in Table \ref{table:simul_computation_time}).
\begin{table}[t]
    \centering
    \caption{Estimated parameter values (and corresponding $0.95$ bootstrap confidence intervals) for a Gaussian process and models LM1, SM5, LSM2.}
    \small
\begin{tabular}{|c|c|c|}
        \hline
        Model & $\hat{\theta}_W$ & $\hat{\theta}_{S,R}$ \\ \hline\hline
        %Gaussian & {\small $\hat{\varphi}=117.509$ $(116.866,118.267)$, $\hat{\eta}=1.202$ $(1.198,1.205)$} & {\small -} \\ \hline
        %LM1 & {\small $\hat{\varphi}=106.457$ $(105.896,107.492)$, $\hat{\eta}=1.198$ $(1.193,1.202)$} & {\small $\hat{\lambda}=2.343$ $(2.136,2.644)$} \\ \hline
        %SM5 & {\small $\hat{\varphi}=86.219$ $(85.239,87.348)$, $\hat{\eta}=1.313$ $(1.309,1.317)$} & {\small $\hat{\gamma}=0.015$ $(-0.127,0.172)$} \\ \hline
        %LSM2 & {\small $\hat{\varphi}=94.731$ $(93.947,107.212)$, $\hat{\eta}=1.277$ $(1.259,1.280)$} & {\small $\hat{\lambda}_1=1.102$ $(0.990,1.240)$, $\hat{\lambda}_2=0.848$ $(0.787,0.937)$} \\ \hline
        Gaussian & {\small $\hat{\varphi}=117.51$ $(116.87,118.27)$, $\hat{\eta}=1.20$ $(1.20,1.21)$} & {\small -} \\ \hline
        LM1 & {\small $\hat{\varphi}=106.46$ $(105.90,107.49)$, $\hat{\eta}=1.20$ $(1.19,1.20)$} & {\small $\hat{\lambda}=2.34$ $(2.14,2.64)$} \\ \hline
        SM5 & {\small $\hat{\varphi}=86.22$ $(85.24,87.35)$, $\hat{\eta}=1.31$ $(1.31,1.32)$} & {\small $\hat{\gamma}=0.02$ $(-0.13,0.17)$} \\ \hline
        LSM2 & {\small $\hat{\varphi}=94.73$ $(93.95,107.21)$, $\hat{\eta}=1.28$ $(1.26,1.28)$} & {\small $\hat{\lambda}_1=1.10$ $(0.99,1.24)$, $\hat{\lambda}_2=0.85$ $(0.79,0.94)$} \\ \hline
    \end{tabular}
    \label{tab:estimated_param_copula}
\end{table}
\begin{figure}[htbp]
\begin{subfigure}{0.48\linewidth}
\includegraphics[width=\linewidth]{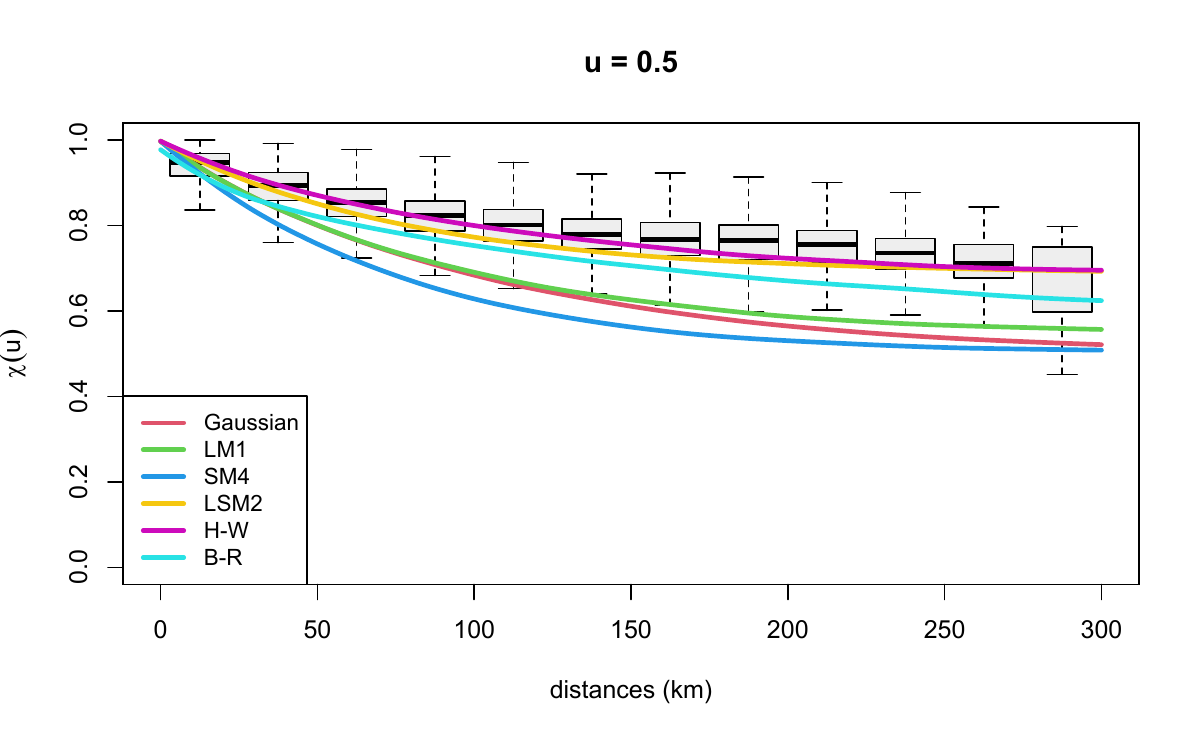}
\end{subfigure}
\hfill
\begin{subfigure}{0.48\linewidth}
\includegraphics[width=\linewidth]{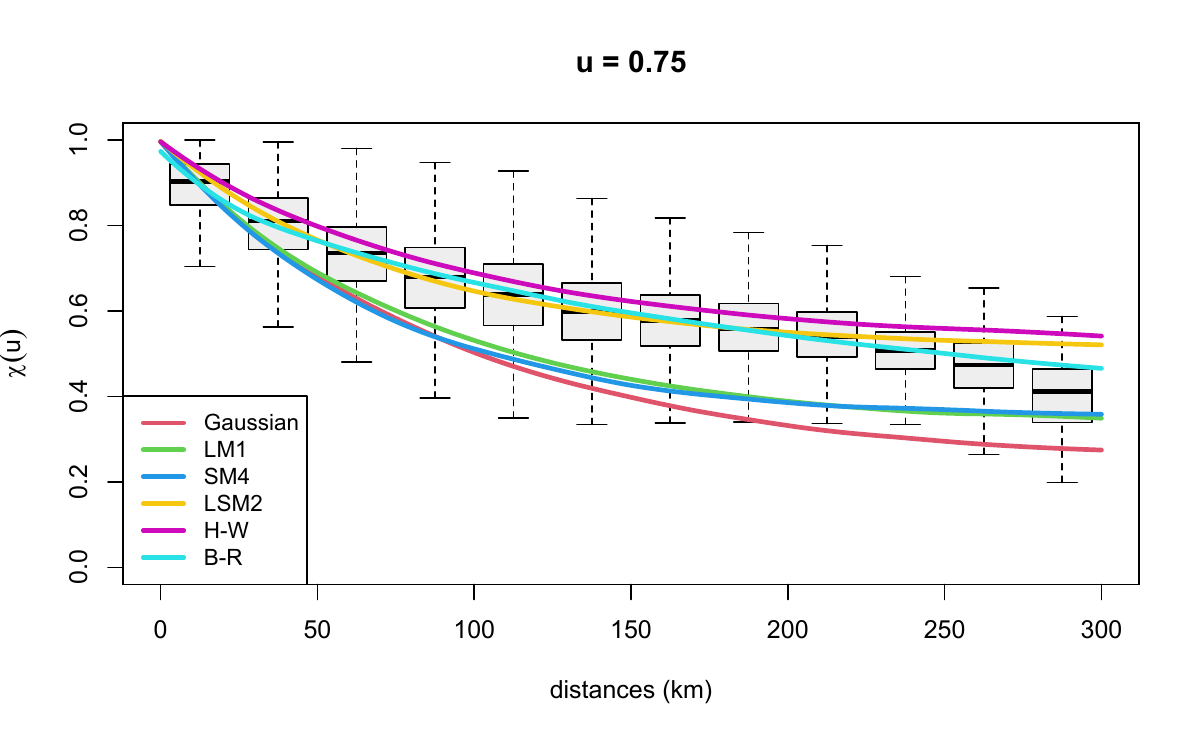}
\end{subfigure}
\begin{subfigure}{0.48\linewidth}
\includegraphics[width=\linewidth]{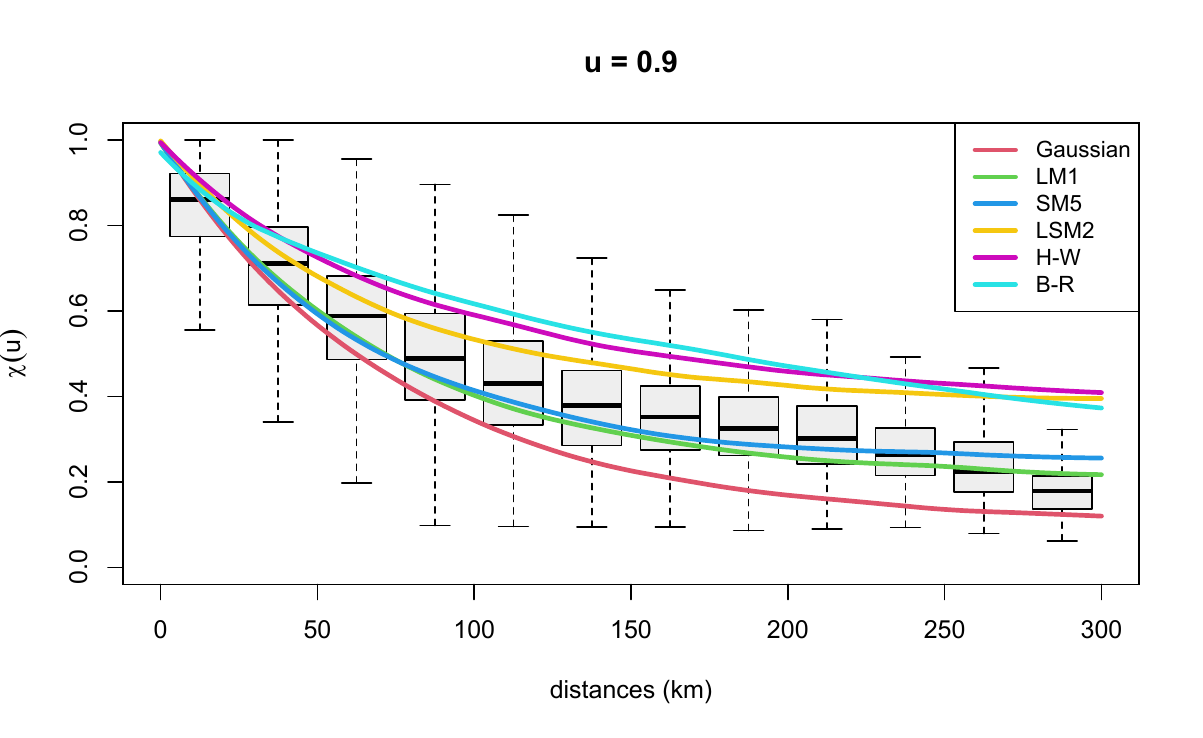}
\end{subfigure}
\hfill
\begin{subfigure}{0.48\linewidth}
\includegraphics[width=\linewidth]{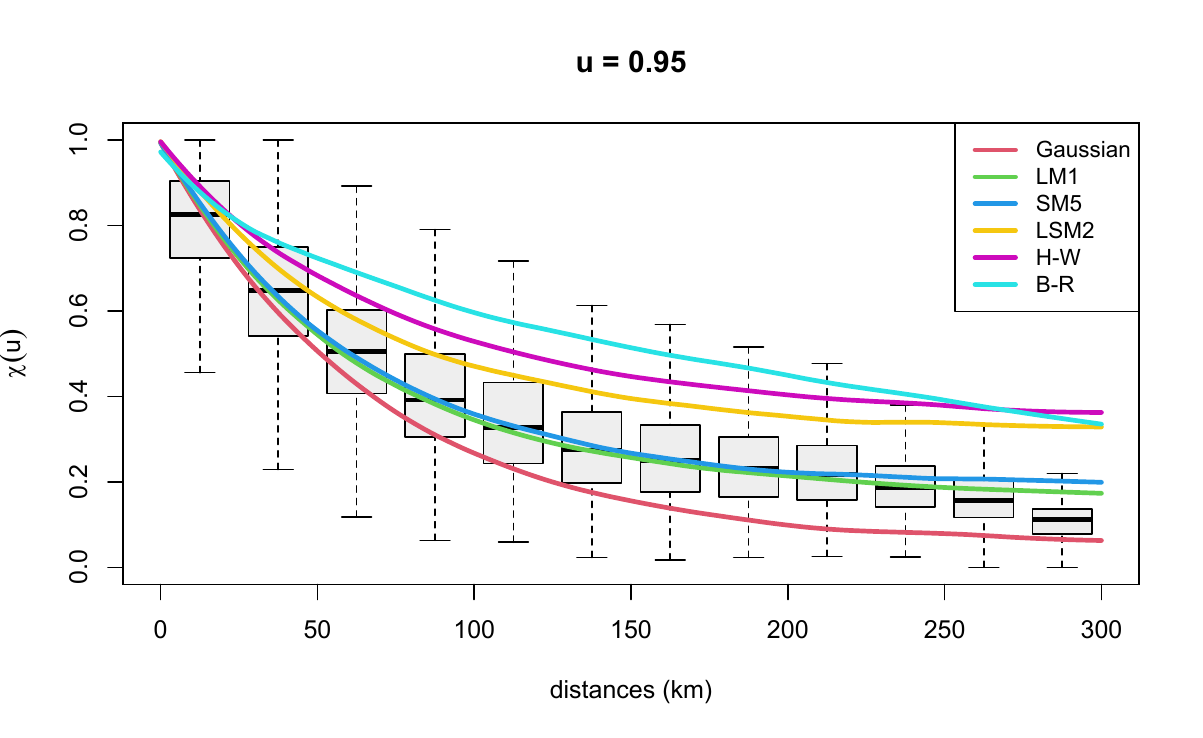}
\end{subfigure}
\begin{subfigure}{0.48\linewidth}
\includegraphics[width=\linewidth]{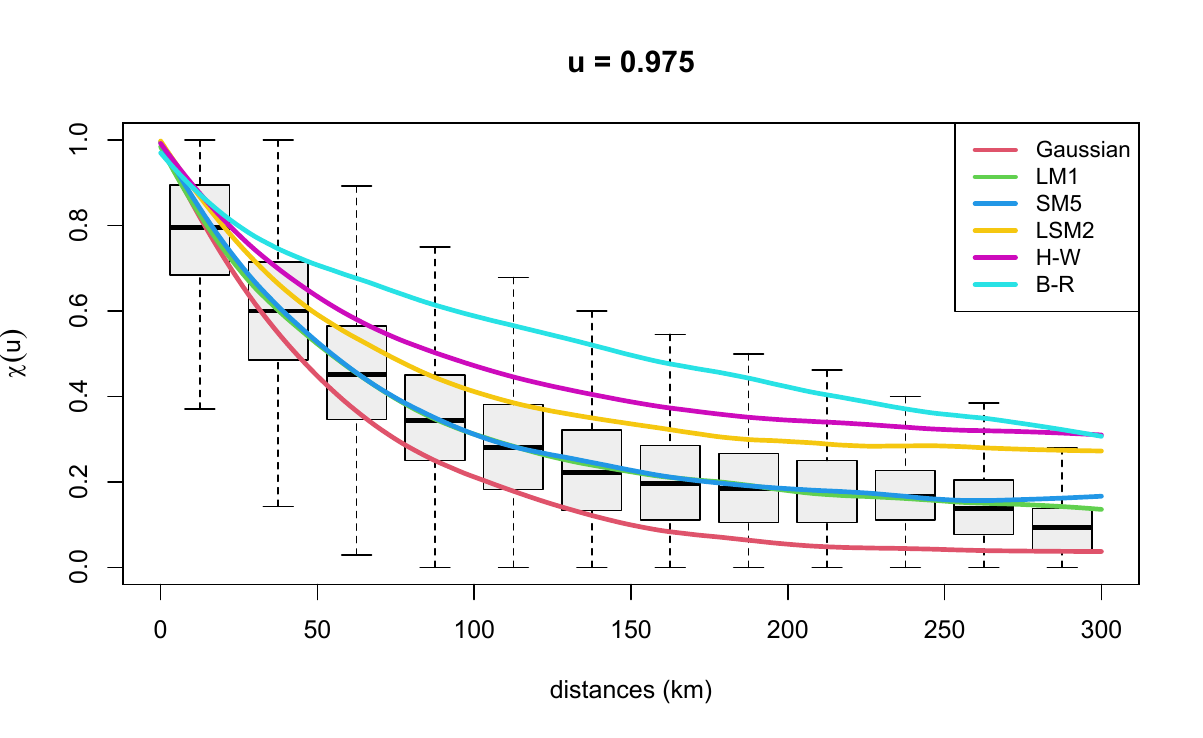}
\end{subfigure}
\hfill
\begin{subfigure}{0.48\linewidth}
\includegraphics[width=\linewidth]{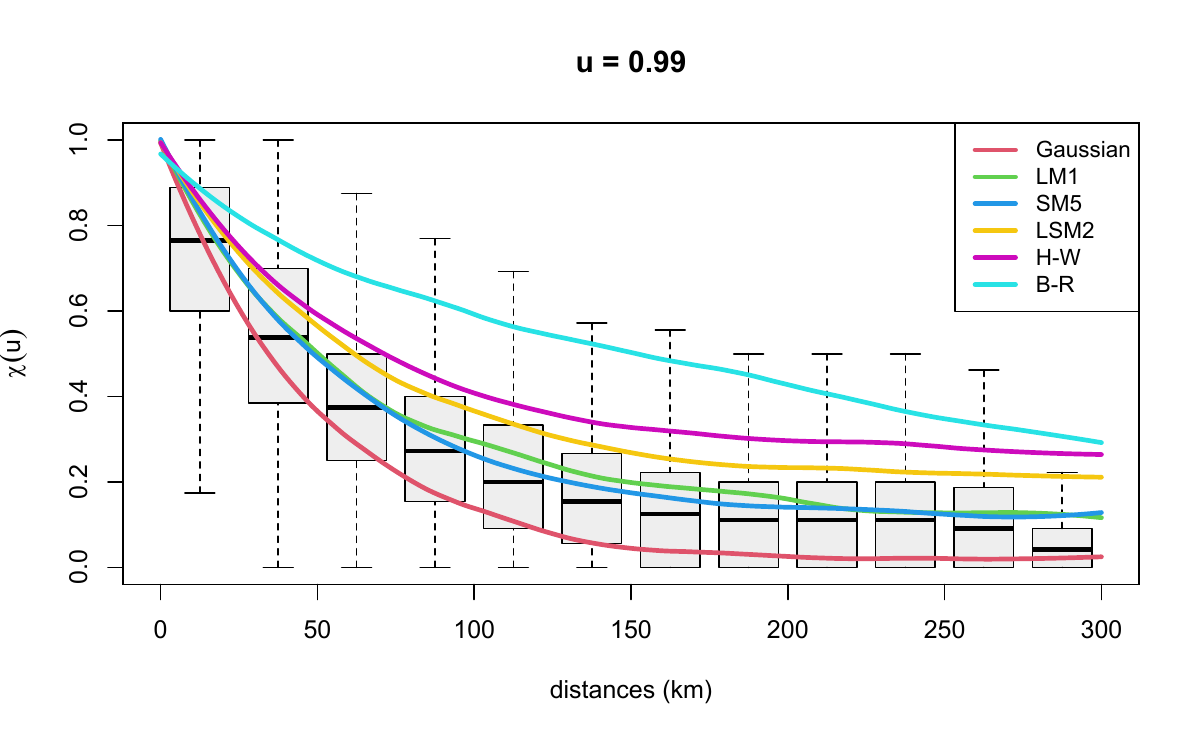}
\end{subfigure}
\caption{Empirical bivariate $ \chi(u)$ for pairs at different distances and thresholds $u\in\{0.5,0.75,0.9,0.95,0.975,0.99\}$ and fitted models: Gaussian process (red lines), Gaussian location mixture 1 (green lines), Gaussian scale mixture 5 (blue lines), Gaussian location-scale mixture 2 (yellow lines), Huser-Wadsworth model (purple lines) and Brown-Resnick process (light blue lines).}
\label{fig:portugal_chi_dist}
\end{figure}
The estimated parameter values for these models and for a Gaussian process are reported in Table \ref{tab:estimated_param_copula}. %as follows: $\hat{\varphi}=117.509$ $(116.866,118.267)$ and $\hat{\eta}=1.202$ $(1.198,1.205)$ for the Gaussian process; $\hat{\varphi}=106.457$ $(105.896,107.492)$, $\hat{\eta}=1.198$ $(1.193,1.202)$ and $\hat{\lambda}=2.343$ $(2.136,2.644)$ for model LM1; $\hat{\varphi}=86.219$ $(85.239,87.348)$, $\hat{\eta}=1.313$ $(1.309,1.317)$ and $\hat{\gamma}=0.015$ $(-0.127,0.172)$ for model SM5; $\hat{\varphi}=94.731$ $(93.947,107.212)$, $\hat{\eta}=1.277$ $(1.259,1.280)$, $\hat{\lambda}_1=1.102$ $(0.990,1.240)$ and $\hat{\lambda}_2=0.848$ $(0.787,0.937)$ for model LSM2.
Note that these results indicate, for the upper tail of the data distribution, compatibility with both asymptotic dependence and asymptotic independence. Indeed, the confidence interval for $\gamma$ in model SM5, which includes $0$, and that for $\lambda_1$ in model LSM2, which includes $1$, do not show a significant prevalence of one extremal dependence class.

Figure \ref{fig:portugal_chi_dist} shows the empirical coefficients $ \chi(u)$ for $u=0.5,0.75,0.9,0.95,0.975,0.99$ and for different distances in space (data are summarized by boxplots for bins of size 25 km), and compares them with the values of the coefficient obtained by simulation from the fitted models. The standard Gaussian process underestimates the risk of bivariate occurrences of extreme values, while models LM1 and SM5 seem to be more appropriate for describing the behaviour of data at high quantiles, i.e.\ for $u\ge0.9$. Model LSM2, which shows the best fit for central quantiles, corresponding to $u=0.5$ and $u=0.75$, overestimates the coefficient $\chi$ for higher quantiles.
If  the analysis focuses on wildfire-prone conditions  corresponding to relatively extreme fire-weather index values, LM1 and SM5 are the most appropriate models for representing spatial clustering of extreme events.
In this context, model selection should depend on the applied purpose of the analysis, as well as considerations about data features and the size of the dataset. For example, if the data exhibits tail asymmetry with slower-than-Gaussian joint tail decay, this can only be captured by Gaussian location-scale mixtures such as LSM2, rather than LM1 or SM5. Conversely, if only upper-tail dependence is of interest and/or a smaller dataset is available, more parsimonious models such as LM1 and SM5 may be preferred.

For comparison, two additional copula models were fitted to the data and the resulting $\chi(u)$ displayed in Figure \ref{fig:portugal_chi_dist}. These models are a Huser-Wadsworth random-scale model (H-W; \citealp{Huser:Wadsworth:2019}) and a max-stable Brown-Resnick process (B-R; \citealp{Brown:Resnick:1977}). Both models overestimate the  dependence at high quantiles, and the latter seems too rigid to capture the decay of $\chi(u)$ values at increasing spatial distances. Section \ref{section_alternative_dep_models} of the supplementary material provides details on the adopted inferential methods and the estimated parameter values.

In addition to the bivariate dependence diagnostics shown in Figure~\ref{fig:portugal_chi_dist}, we can evaluate how well  the models  reproduce the spatial extent of extreme events. Given that we have $n=1426$ daily observations across $m=491$ spatial locations, we compute  $N_i(p)$,  the number of locations at which values exceeding the  $p^{(th)}$ quantile are observed on day $i$, for $i=1,\dots,n$. We assume that the $N_i(p)$ are i.i.d. copies of the  $N(p)$  a discrete random variable  taking values in $\{0,1,\dots,m\}$. The distribution of this variable differs for the estimated models and for $p$ and it can be  approximated by Monte Carlo.  Table~\ref{tab:nll_exceed_count} shows the values of the negative log-likelihood values for different model specification  for $p\in\{0.90,0.95,0.99\}$. Models LM1 and SM5 produce the best results, confirming their ability to capture the spatial extent of extremal dependence, as shown in Figure~\ref{fig:portugal_chi_dist}.

\begin{table}[t]
    \centering
    \caption{Negative log-likelihood of the observed threshold exceedance counts $N_i(p)$, for the six models shown in Figure~\ref{fig:portugal_chi_dist}.}
    \begin{tabular}{|c|c|c|c|c|c|c|}
    \hline
    Model & Gaussian & LM1 & SM5 & LSM2 & H-W & B-R \\ \hline
    nll ($p=0.90$)     & 4546.3 & 4498.7 & \textbf{4420.0} & 4432.6 & 4457.9 & 4455.3 \\ \hline
    nll ($p=0.95$)   & 3154.2 & 3114.9 & \textbf{3087.8} & 3099.4 & 3101.5 & 3129.4\\ \hline
    nll ($p=0.99$)     & 1184.3 & \textbf{1168.7} & 1170.9 & 1175.3 & 1171.1 & 1217.1\\ \hline
    \end{tabular}
    \label{tab:nll_exceed_count}
\end{table}

\subsection{Conditional simulation}\label{section:application_conditional_simulation}

\begin{figure}[htbp!]
\centering
\begin{subfigure}{0.738\linewidth}
\centering
\includegraphics[width=\linewidth]{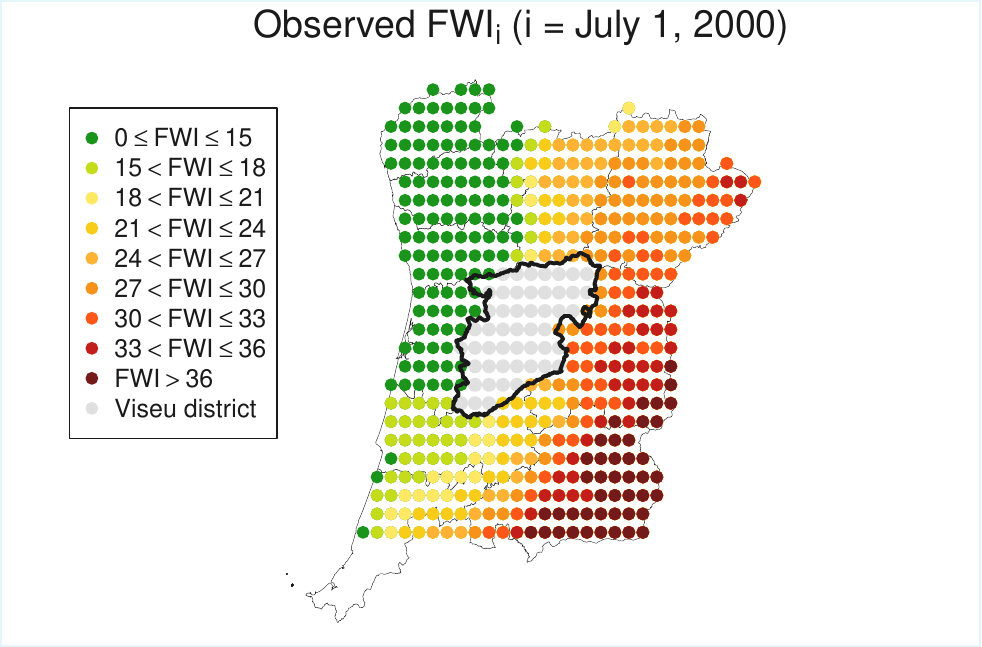}
%\label{fig:pt_observed_noViseu}
\end{subfigure}

%\end{figure}
%\begin{figure}[t]
\begin{subfigure}{0.171\linewidth}
\includegraphics[width=\linewidth]{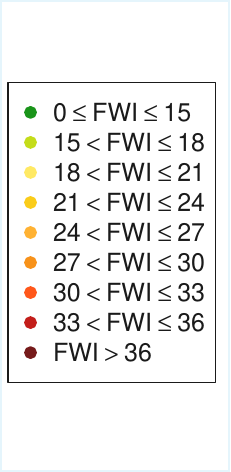}
\end{subfigure}
\begin{subfigure}{0.279\linewidth}
\includegraphics[width=\linewidth]{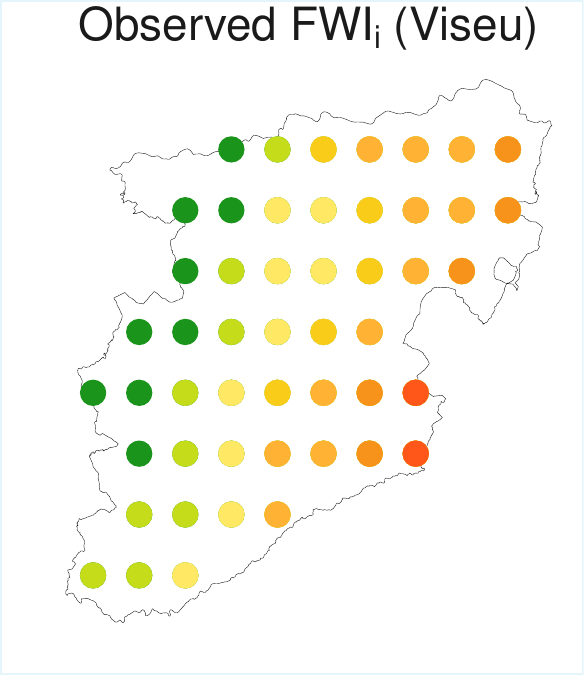}
\end{subfigure}
\begin{subfigure}{0.279\linewidth}
\includegraphics[width=\linewidth]{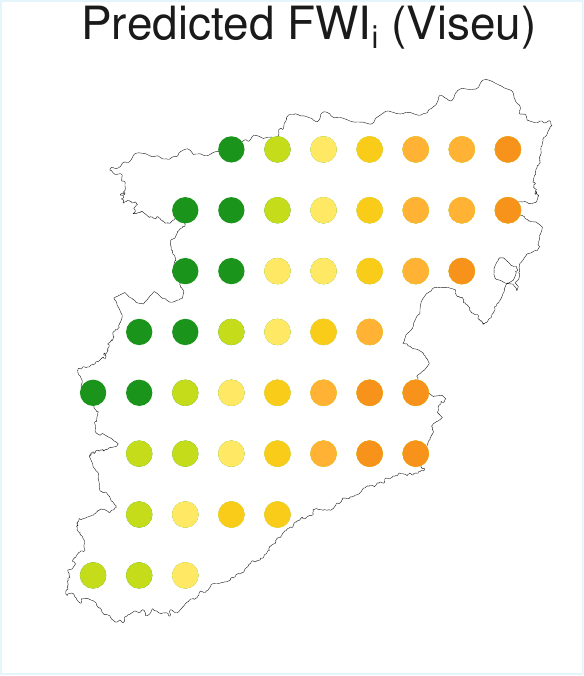}
\end{subfigure}

\begin{subfigure}{0.171\linewidth}
\includegraphics[width=\linewidth]{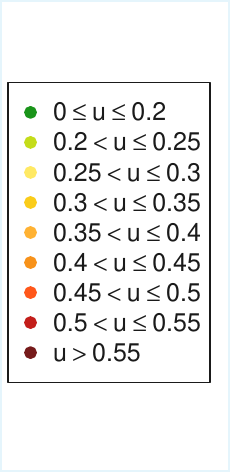}
\end{subfigure}
\begin{subfigure}{0.279\linewidth}
\includegraphics[width=\linewidth]{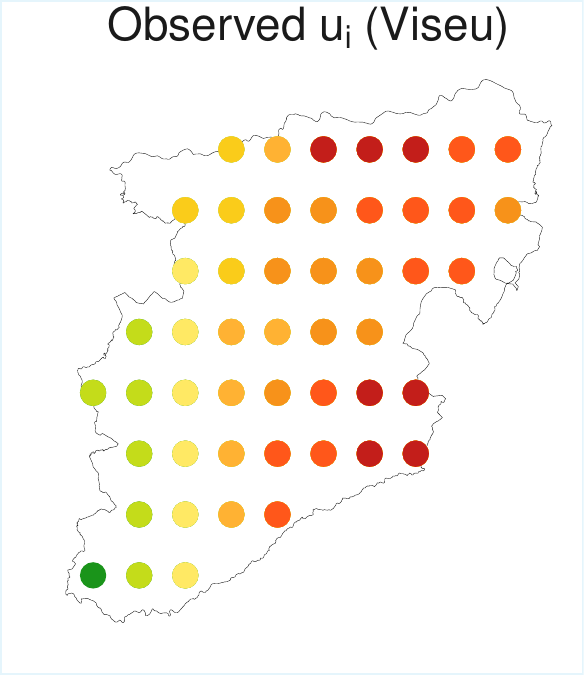}
\end{subfigure}
\begin{subfigure}{0.279\linewidth}
\includegraphics[width=\linewidth]{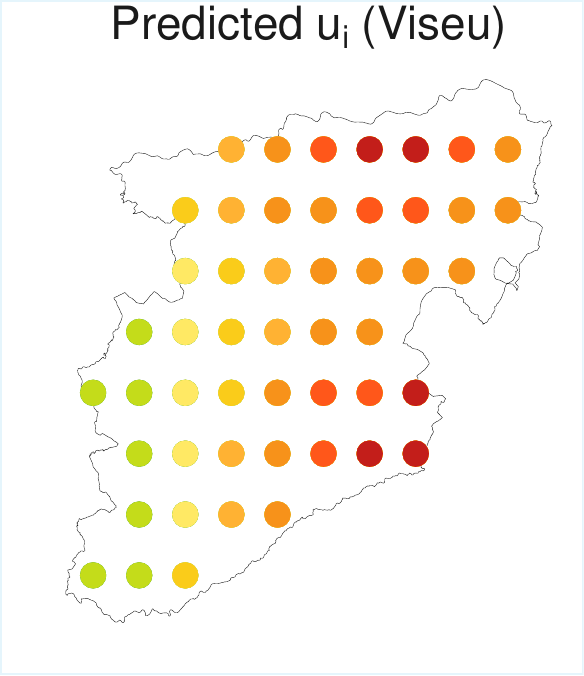}
\end{subfigure}
\caption{Conditional simulation described in Section \ref{section:application_conditional_simulation}. The top figure displays the observed FWI values on July 1st, 2000, in the north of Portugal, excluding the Viseu district. The central and bottom figures show the observed and predicted values in the Viseu district on the FWI and uniform scale, respectively.}
\label{fig:pt_conditional_simulation_Viseu}
\end{figure}

The conditional simulation algorithm presented in Section \ref{section:conditional_simulation} is applied to the FWI data in the north of Portugal, as illustrated in Figure \ref{fig:pt_conditional_simulation_Viseu}.
In particular, data from the 1st of July, 2000, are analysed. The Viseu district is assumed to be unobserved, and its observations are left out from the dataset; therefore, in the following $\bX_2$ and $\mathbf{Y}_2$ refer to the Viseu district, while $\bX_1$ and $\mathbf{Y}_1$ refer to the other districts in the north of Portugal.
The model assumed for the dependence structure of the data is Gaussian location mixture 1, with the parameter values estimated in Section \ref{section:application_dependence_modelling}.
The Metropolis-Hastings simulation scheme (see 1.\ in Algorithm \ref{alg:conditional_simulation}) involved $k=3\cdot 10^5$ simulated values of $S\mid\bX_1$, with a thinning rate of 100.
Since multiple values of $S$ are generated in this way, the entire conditional distribution of $\bX_2\mid\bX_1$ is simulated. The corresponding distribution of $\mathbf{Y}_2\mid\mathbf{Y}_1$, on the FWI scale, is obtained by applying $\hat{F}_Y^{\mathbf{s}\;-1}\{\hat{F}_X(\cdot)\}$ to it, with the spatially-varying $\hat{F}_Y^{\mathbf{s}}$ estimated in Section \ref{section:application_marginal_modelling} and $\hat{F}_X$ estimated in Section \ref{section:application_dependence_modelling}. To obtain point predictions, these distributions are summarized by their medians.
In particular, the predicted values of $\mathbf{Y}_{2,i}$, where $i$ refers to the day (1st of July, 2000), are compared to the observed ones in the Viseu district, showing a high degree of similarity.
Since this may be due to the spatially varying marginal distribution, Figure \ref{fig:pt_conditional_simulation_Viseu} also shows the same comparison on the uniform scale, that is, on the scale of $\hat{F}_X(\bX_2)$. The simulated data on this scale are independent of the marginal distribution, but the predicted values are still very close to the observed ones.

By repeating the above procedure  for each of the $n=1426$ daily observations, and taking the median value of the conditionally simulated distribution of $\mathbf{Y}_2\mid\mathbf{Y}_1$ as the point prediction at each location in the Viseu district, we obtained an RMSE of $1.09$. For comparison, predictions obtained using ordinary Kriging based on an estimated variogram fitted with a Mat\'ern covariance model yielded an RMSE of $1.74$. The use of an intermediate method, which involves the estimated marginal EGPD being used to transform the data into a uniform scale and then into a standard Gaussian scale, and the application of ordinary Kriging to the transformed data, results in an RMSE of 1.54.

\section{Conclusions}

The objective of this work was threefold. First, a review of the key properties of Gaussian location-scale mixtures was conducted. These mixtures constitute a general class that includes the known subclasses of Gaussian location mixtures and Gaussian scale mixtures.
The capacity to define both a random scale and a random location introduces enhanced flexibility into the model's formulation, particularly concerning extremal dependence in both the upper and lower tail. In this framework, 13 models are proposed and studied with respect to their extremal dependence structure and, when available, their marginal and multivariate distributions.
A subset of these models have been previously examined in the existing literature; however, others constitute novel propositions. In the latter case, the presented results are supported by analytical proofs.
For this class of models, a conditional simulation algorithm is proposed.

Secondly, we propose a method for inference that is both general, being valid for any specified distribution of the random scale and the random location, and fast and stable, not involving numerical integrations.
Our inferential approach could also be useful for multivariate (not spatial) models that make use of location and scale mixtures of multivariate Gaussian distributions.

Thirdly, we leverage the flexibility of these models, in conjunction with an inferential method that does not involve any censoring, to propose an extended modelling of both the tails and the bulk of the data distribution. Consequently, an extended modelling approach is adopted for marginal distributions as well, based on the extended generalized Pareto distribution.

A possible extension would be to model also temporal dependence, defining Gaussian location-scale mixtures as space-time models.
However, the construction of these models would not be straightforward.
A promising idea is to adapt the proposed inference technique to space-time location-mixture models as $X(\mathbf{s},t)=S(t)+ W(\mathbf{s},t)$, for $\mathbf{s}\in\mathcal{S},\; t\in\mathcal{T}$, where $S(t)$ is a time process independent of the space-time Gaussian process $W(\mathbf{s},t)$.
Existing inference approaches are either prohibitive for this setting or come with an extremely high computational cost. By adapting our transformation trick, we can consider $\{X(\mathbf{s},t)-X(\mathbf{s}_1,t)\}=\{W(\mathbf{s},t)-W(\mathbf{s}_1,t)\}$, for $\mathbf{s}_1\in\mathcal{S}$, which is a Gaussian process.
The specification of the process $S(t)$ is not obvious because it must consider inferential aspects while maintaining the realism of the model when applied to real data. Moreover, it would be necessary to develop different techniques for the cases of Gaussian scale and location-scale mixtures, since the corresponding transformations would not lead to  processes with closed-form likelihoods.
%To avoid issues in high-dimensional data settings with many spatial locations and time steps, one could adopt an approach based on Gaussian Markov random fields and latent-variable representations to estimate the parameters in $\theta_W$, for example in the INLA-SPDE framework \citep{rue2009approximate,lindgren2011explicit}. %However, a modelling effort has to be made to avoid that these models would have too rigid or not realistic dependence structures in space and time and to combine short-range dependence with long-range independence (see e.g.\ \citealp{hazra2025efficient}).

\section*{Acknowledgments}
The authors thank Rapha{\"e}l Huser and Likun Zhang for their comments on a previous version of this work.

%\newpage
\appendix

\section{Extremal behaviour of SM4, LSM1} \label{section:appendix_extremal_behaviour} 

\subsection{Proof for SM4 model}
Let $X(\mathbf{s})=RW(\mathbf{s})$ with $R=\sqrt{E}/G$, where $E\sim\text{E}(1/2)$ and $G\sim\text{Gamma}(1/\gamma,1/\gamma)$.
The density function of the inverse Gamma random variable $1/G$ is regularly varying with index $-(1/\gamma+1)$, so, by Karamata's theorem \citep[see][]{bingham1989regular}, its distribution function is regularly varying with index $-1/\gamma$ (see Section \ref{section:definitions}).
Then, since $1/G$ is heavier tailed than $\sqrt{E}$, which follows a Rayleigh distribution, by Breiman's lemma \citep{breiman1965some} $R$ is also regularly varying with index $1/\gamma$. Following Section 3.1 of \cite{Huser:Opitz:Thibaud:2017}, we have that 
$\{X(\mathbf{s})\}$ is asymptotically dependent with $$\chi=2T_{1/\gamma+1}\left[-\sqrt{(1/\gamma+1)(1-\rho)/(1+\rho)}\right].$$ 
\qed

\subsection{Proof for LSM1 model}

Let $X(\mathbf{s})=S+RW(\mathbf{s})$  with
$S\sim \text{E}(\lambda)$, and $R=\sqrt{E}$, with $E\sim\text{E}(1/2)$.

\noindent For $\lambda<1$, we need to prove that 
$\{X(\mathbf{s})\}$ is AD, with
\begin{equation}\label{eq:locscalemixture1_chi_app}
	\chi=\E\left[\min\left\{\frac{Z_1^\lambda}{\E(Z_1^\lambda)},\frac{Z_2^\lambda}{\E(Z_2^\lambda)}\right\}\right],
\end{equation}
where $Z_i=\exp\{R\;W(\mathbf{s}_i)\}$, $i=1,2$ and $\bar{\chi}=1$. 
To prove this result, consider the exponential  transformation 
$$
\exp\{X(\mathbf{s})\}=\exp\{S\}\exp\{R\;W(\mathbf{s})\},\;\mathbf{s}\in\mathcal{S},
$$
that defines a scale mixture with scaling variable $\exp\{S\}\sim\text{Pareto}(\lambda)$ and $\exp\{R W(\mathbf{s})\}\sim\text{Pareto}(1)$. 
Therefore, Proposition 1 of \cite{engelke2019extremal} provides the expression \eqref{eq:locscalemixture1_chi_app}, since the $\chi$ index is invariant to monotonically increasing transformations of the random variable, such as the logarithmic one.
%Also note that $\exp\{S\}\sim\text{Pareto}(\lambda)$ and the consitional distribution of $\exp\{R\;W(\mathbf{s})\}$ given that $R\;W(\mathbf{s})>0$ is a Pareto($1$) distribution. This leads to a reparametrized form of the model of \cite{Huser:Wadsworth:2019}, which is coherent with result \eqref{eq:locscalemixture1_chi_app}.

\noindent  For $\lambda>1$, we need to prove that $\{X(\mathbf{s})\}$ is AI, i.e. $\chi=0$, with
\begin{equation*}%\label{eq:locscalemixture1_chibar}
	\bar{\chi}=\max\left\{\frac{2}{\lambda}-1,\; \rho \right\},
	%    \bar{\chi}(\mathbf{s}_1,\mathbf{s}_2)=\begin{cases}\begin{array}{ll}
			%            2/\lambda-1 &\qquad\text{if}\; \lambda<2/(1+\rho)\\
			%            \rho &\qquad\text{if}\; \lambda\ge 2/(1+\rho),
			%        \end{array}
		%        \end{cases}
\end{equation*}
where $\rho=\mbox{cor}(W(\mathbf{s}_1),W(\mathbf{s}_2))$.
To prove this result, consider again the multiplicative representation $\exp\{X(\mathbf{s})\}$ and see \cite{engelke2019extremal}, Proposition 5.

\noindent For $\lambda=1$, $\{X(\mathbf{s})\}$ is asymptotically independent, with $\chi=0$ and $\bar{\chi}=1$, as is possible to prove by considering the same multiplicative representation and following \cite{engelke2019extremal}, Proposition 6, case 3(c).
%We can further exploit this analogy to say that, when $\lambda=1$, $\{X(\mathbf{s})\}$ is asymptotically independent, with $\chi=0$ and $\bar{\chi}=1$, as in the Huser-Wadsworth model when $\delta=0.5$.
\qed

\section{Details of the Metropolis-Hastings algorithm for conditional simulation}\label{section:appendix_metropolis_hastings}

The acceptance probability in \eqref{eq:metropolis_hastings_update}, for each step of the Metropolis-Hastings algorithm, is
$$\rho(s',r',s^{(k)},r^{(k)})=\min\left\{\frac{f_{S,R}(s',r')\phi_m((\bx_1-s')/r';\bSigma_{1;1})r'^{-m} g(r^{(k)},s^{(k)}|r',s')}{f_{S,R}(s^{(k)},r^{(k)})\phi_m((\bx_1-s^{(k)})/r^{(k)};\bSigma_{1;1})r^{(k)-m}g(r',s'|r^{(k)},s^{(k)})},1\right\},$$
where $g(a,b\,|\,c,d)$ is the sampling density from $(c,d)$ to $(a,b)$; note that
$$g(r^{(k)},s^{(k)}|r',s')=g(r^{(k)}|r')g(s^{(k)}|s')=\phi(\log(r^{(k)})-\log(r'))\frac{1}{r^{(k)}}\phi(s^{(k)}-s'),$$
$$g(r',s'|r^{(k)},s^{(k)})=g(r'|r^{(k)})g(s'|s^{(k)})=\phi(\log(r')-\log(r^{(k)}))\frac{1}{r'}\phi(s'-s^{(k)}),$$
and, due to the symmetry of the Gaussian density function of the random walk proposal, all terms but the Jacobian $1/r^{(k)}$ and $1/r'$ simplify, leading to:
$$\rho(s',r',s^{(k)},r^{(k)})=\min\left\{\frac{f_{S,R}(s',r')\phi_m((\bx_1-s')/r';\bSigma_{1;1})r'^{-m+1}}{f_{S,R}(s^{(k)},r^{(k)})\phi_m((\bx_1-s^{(k)})/r^{(k)};\bSigma_{1;1})r^{(k)-m+1}},1\right\}.$$

\section{Derivation of density functions for the vectors \texorpdfstring{$\bZ$}{Z}}\label{app1:density_Z}

\subsection{Gaussian location mixtures}\label{app1:density_Z_case1}
The vector \eqref{eq:Z_tilde_1} can be written as
$$
    \bZ=\left(X(\mathbf{s}_2)-X(\mathbf{s}_1),\dots,X(\mathbf{s}_m)-X(\mathbf{s}_1)\right)^\top 
    =\left(W(\mathbf{s}_2)-W(\mathbf{s}_1),\dots,W(\mathbf{s}_m)-W(\mathbf{s}_1)\right)^\top.
$$
Note that this is a linear combination of the components of a multivariate Gaussian, so $\bZ\sim \mathcal{N}_{m-1}(\bzero,\bA\bSigma_{\theta_W} \bA^\top)$, where $\bA$ is the $(m-1)\times m$ matrix in \eqref{eq:matrix_A},
%$$\bA = \left[ \begin{matrix} -1&1&0&\cdots&0\\ -1&0&1&\cdots&0 \\ \vdots&\vdots&&\ddots&\vdots\\ -1&0&0&\cdots&1 \ \end{matrix} \right]$$
and its density function, for $\bz\in\R^{m-1}$, is
$$f_{\bZ}(\bz;\,\theta_W) =\phi_{m-1}(\bz;\bA\bSigma_{\theta_W} \bA^\top) = (2\pi)^{-(m-1)/2} |\bA\bSigma_{\theta_W} \bA^\top|^{-{1}/{2}} \exp\left(-\frac{1}{2} \bz^\top (\bA\bSigma_{\theta_W} \bA^\top)^{-1}\bz \right).$$

\subsection{Gaussian scale mixtures}\label{app1:density_Z_case2}
The vector \eqref{eq:Z_tilde_2} can be written as
\begin{equation*}
    \bZ=\left(\frac{X(\mathbf{s}_2)}{X(\mathbf{s}_1)},\dots,\frac{X(\mathbf{s}_m)}{X(\mathbf{s}_1)}\right)^\top = \left(\frac{W(\mathbf{s}_2)}{W(\mathbf{s}_1)},\dots,\frac{W(\mathbf{s}_m)}{W(\mathbf{s}_1)}\right)^\top.
\end{equation*}
To obtain the density function of $\bZ$, we can first define another vector
%$$\tilde{\bZ}=\left(W(\mathbf{s}_1),\frac{W(\mathbf{s}_2)}{W(\mathbf{s}_1)},\dots,\frac{W(\mathbf{s}_m)}{W(\mathbf{s}_1)}\right)^\top \;\Big|\; W(\mathbf{s}_1)>0.$$
$$\tilde{\bZ}=\left(W(\mathbf{s}_1),\frac{W(\mathbf{s}_2)}{W(\mathbf{s}_1)},\dots,\frac{W(\mathbf{s}_m)}{W(\mathbf{s}_1)}\right)^\top.$$
%Note that the definition of $\tilde{\bZ}$ is conditional but, in practical cases, all its components can be multiplied by $-1$ if $W(\mathbf{s}_1)<0$, given the symmetry of its distribution and the focus on the dependence structure only.
The density function of $\tilde{\bZ}$ is
$%$
f_{\tilde{\bZ}}(\tilde{\bz};\,\theta_W) = %2\;
\phi_m(\tilde{z}_1,\tilde{z}_1\tilde{z}_2,\dots,\tilde{z}_1\tilde{z}_m;\bSigma_{\theta_W})\, |\tilde{z}_1|^{m-1}, $ % $ % \;\mathbb{1}(\tilde{z}_1>0),$$
where $\tilde{z}_j$, $j=1,\dots,m$, is the $j$-th component of $\tilde{\bz}=(\tilde{z}_1,\dots,\tilde{z}_m)^\top$. % and $\phi_m(\cdot;\bmu,\bSigma)$ is the multivariate Gaussian density of $\mathcal{N}(\bmu,\bSigma)$.
Defining also $\bz=(\tilde{z}_2,\dots,\tilde{z}_m)^\top$ and $\dot{\bz}=(1,\tilde{z}_2,\dots,\tilde{z}_m)^\top=(1,\bz^\top)^\top$, we can obtain the density function of $\bZ$ as

\begin{equation*}
\begin{split}
%    f_{\bZ}(\bz;\,\theta_W) =& \int_0^\infty f_{\tilde{\bZ}}(\tilde{\bz};\,\theta_W) d\tilde{z}_1 = \int_0^\infty 2\; \phi_m(\tilde{z}_1,\tilde{z}_1\tilde{z}_2,\dots,\tilde{z}_1\tilde{z}_m;\bSigma_{\theta_W})\; \tilde{z}_1^{m-1} d\tilde{z}_1\\
    f_{\bZ}(\bz;\,\theta_W) =& \int_{-\infty}^\infty f_{\tilde{\bZ}}(\tilde{\bz};\,\theta_W) d\tilde{z}_1 = \int_{-\infty}^\infty  \phi_m(\tilde{z}_1,\tilde{z}_1\tilde{z}_2,\dots,\tilde{z}_1\tilde{z}_m;\bSigma_{\theta_W})\; |\tilde{z}_1|^{m-1} d\tilde{z}_1\\
%    =& \; 2 \int_0^\infty (2\pi)^{-{m}/{2}} |\bSigma_{\theta_W}|^{-{1}/{2}} \exp\left(-\frac{1}{2}\tilde{z}_1^2\; \dot{\bz}^\top \bSigma_{\theta_W}^{-1}\dot{\bz} \right) \tilde{z}_1^{m-1} d\tilde{z}_1 \qquad\qquad [\mbox{setting }y=\tilde{z}_1^2]\\
    =&  \int_{-\infty}^\infty (2\pi)^{-{m}/{2}} |\bSigma_{\theta_W}|^{-{1}/{2}} \exp\left(-\frac{1}{2}\tilde{z}_1^2\; \dot{\bz}^\top \bSigma_{\theta_W}^{-1}\dot{\bz} \right) |\tilde{z}_1|^{m-1} d\tilde{z}_1 \\
%    =& 2\; \int_0^\infty (2\pi)^{-{m}/{2}} |\bSigma_{\theta_W}|^{-{1}/{2}} \exp\left(-\frac{1}{2}\tilde{z}_1^2\; \dot{\bz}^\top \bSigma_{\theta_W}^{-1}\dot{\bz} \right) \tilde{z}_1^{m-1} d\tilde{z}_1 \\
    =& \; 2 \int_0^\infty (2\pi)^{-{m}/{2}} |\bSigma_{\theta_W}|^{-{1}/{2}} \exp\left(-\frac{1}{2}y\; \dot{\bz}^\top \bSigma_{\theta_W}^{-1}\dot{\bz} \right) y^{(m-1)/2} \frac{1}{2}y^{-{1}/{2}}dy \quad\qquad [\mbox{setting }y=\tilde{z}_1^2] \\
    =& \; (2\pi)^{-{m}/{2}} |\bSigma_{\theta_W}|^{-{1}/{2}} \int_0^\infty \exp\left(-y\; \dot{\bz}^\top \bSigma_{\theta_W}^{-1}\dot{\bz} /2\right) y^{{m}/{2}-1} dy\\
    =& \; (2\pi)^{-{m}/{2}} |\bSigma_{\theta_W}|^{-{1}/{2}} \frac{\Gamma\left(m/2\right)}{\left( \dot{\bz}^\top \bSigma_{\theta_W}^{-1}\dot{\bz} /2\right)^{m/2}} \int_0^\infty \frac{\left(\dot{\bz}^\top \bSigma_{\theta_W}^{-1}\dot{\bz} /2 \right)^{m/2}}{\Gamma\left(m/2\right)} \exp\left(-y\; \dot{\bz}^\top \bSigma_{\theta_W}^{-1}\dot{\bz}/2 \right) y^{{m}/{2}-1} dy\\
    =& \; \pi^{-m/2} |\bSigma_{\theta_W}|^{-{1}/{2}} \Gamma\left(m/2\right)\left(\dot{\bz}^\top \bSigma_{\theta_W}^{-1}\dot{\bz} \right)^{-m/2},
\end{split}
\end{equation*}
where the last integration for $y$ comes from the gamma density.
%Note that, when $W(\mathbf{s}_1)$ is independent of $W(\mathbf{s}_2),\dots,W(\mathbf{s}_m)$, $\bZ$ follows a multivariate Cauchy distribution, i.e.\ a multivariate Student's \textit{t} with 1 degree of freedom.

\subsection{Gaussian location-scale mixtures}\label{app1:density_Z_case3}
The vector \eqref{eq:Z_tilde_3} can be written as
\begin{equation*}
\begin{split}
    \bZ= \left(\frac{X(\mathbf{s}_3)-X(\mathbf{s}_1)}{X(\mathbf{s}_2)-X(\mathbf{s}_1)},\dots,\frac{X(\mathbf{s}_m)-X(\mathbf{s}_1)}{X(\mathbf{s}_2)-X(\mathbf{s}_1)}\right)^\top 
    = \left(\frac{W(\mathbf{s}_3)-W(\mathbf{s}_1)}{W(\mathbf{s}_2)-W(\mathbf{s}_1)},\dots,\frac{W(\mathbf{s}_m)-W(\mathbf{s}_1)}{W(\mathbf{s}_2)-W(\mathbf{s}_1)}\right)^\top.
\end{split}
\end{equation*}
Note that, to compute the density function of $\bZ$, we can combine the formulas of the two previous cases, since this vector can be seen as a vector of ratios of differences of the elements of $\bW$.
From the location mixture case above, we know that $\left(W(\mathbf{s}_2)-W(\mathbf{s}_1),\dots,W(\mathbf{s}_m)-W(\mathbf{s}_1)\right)\sim \mathcal{N}_{m-1}(\bzero,\bA\bSigma_{\theta_W} \bA^\top)$, where $\bA$ is the $(m-1)\times m$ matrix in \eqref{eq:matrix_A}.
%$$ \bA = \left[ \begin{matrix} -1&1&0&\cdots&0\\ -1&0&1&\cdots&0 \\ \vdots&\vdots&&\ddots&\vdots\\ -1&0&0&\cdots&1 \ \end{matrix} \right].$$
Then, we can use the results on the ratio of components of a multivariate Gaussian, shown in the scale mixture case, to obtain 
\begin{equation*}
    f_{\bZ}(\bz;\,\theta_W) = \pi^{-(m-1)/{2}} |\bA\bSigma_{\theta_W} \bA^\top|^{-{1}/{2}} \Gamma\left(\frac{m-1}{2}\right)\left[\dot{\bz}^\top (\bA\bSigma_{\theta_W} \bA^\top)^{-1}\dot{\bz} \right]^{-(m-1)/{2}},
\end{equation*}
where $\bz\in\R^{m-2}$ and $\dot{\bz}=(1,\bz^\top)^\top\in\R^{m-1}$.


%%       \cite{<label>} ==> [1]

%%


\bibliography{references}
\bibliographystyle{apalike-citesort}



\renewcommand{\theequation}{S\arabic{equation}}
\renewcommand{\thefigure}{S\arabic{figure}}
\renewcommand{\thetable}{S\arabic{table}}
\renewcommand{\thesection}{S\arabic{section}}

%\title{Supplementary material for:\\Flexible space-time models for extreme data}

%\author{Lorenzo Dell'Oro \\
%Dipartimento di Scienze Statistiche, \\
%Universit\`a di Padova, Padova, Italy
%\\
%\,
%\\
%Carlo Gaetan\\
%Dipartimento di Scienze Ambientali, Informatica e Statistica,\\
%Universit\`a Ca' Foscari di Venezia, Venezia, Italy}

%\date{28 March 2025}
%\date{26 May 2025}

%\MakeTitle{Supplementary material}{}{}
\clearpage
\setcounter{equation}{0}
\setcounter{figure}{0}
\setcounter{table}{0}
\setcounter{section}{0}

\begin{center}
{\LARGE\bfseries Supplementary Material}
\end{center}
\vspace{2em}

\section{Invariance of the estimators to the selection of the reference locations}\label{suppl:section:example_deference_locations}
\subsection{Proof}
To prove the invariance of the estimators to the selection of the reference locations, it is sufficient to show that, given two data transformations $\bZ_a$ and $\bZ_b$ (see the three cases below for details; in the third case, they are denoted as $\bZ_{a,b}$ and $\bZ_{c,d}$), the function  $\bh_{a\to b}:\bZ_a\mapsto \bZ_b$ is a diffeomorphism, so that the corresponding likelihood functions are proportional to each other, with proportionality factor given by the Jacobian term (see \citealp{davison2003statistical}, 4.1.2). 
In the sequel %$\be^d_i$,
$\bone_d$ and $\bzero_d$ will denote the %unit vector, the 
ones vector and the zero vector in $\R^d$, respectively. The matrix $\bI_d$ is the $d\times d$ identity matrix.
Let $\bW=(W_i,\,{i\in M})$, $M=\{1,\ldots,m\}$, be the $m$-dimensional centred Gaussian  vector with positive definite covariance matrix $\bSigma$. 

%Note that the Gaussian location mixtures case cane be seen as a variation of the REstricted Maximum Likelihood (REML) method \citep{patterson1971recovery}, for which \cite{harville1974bayesian} shows that the likelihood function is proportional when different sets of contrasts $\bA$ are chosen, such that $E(\bA\bX)=\bzero$.

\subsubsection*{Gaussian location mixtures}
For $k\in M=\{1,\dots,m\}$, let $\bZ_k=\bigl(Z^{(k)}_i\bigr)_{i\in M\backslash\{k\}}$, with $Z^{(k)}_i={W_i}-{W_k}$. We study the transformation $\bh_{a\to b}:\bZ_a\mapsto \bZ_b$, for $a,b\in M$.
The component-wise map $\bh_{a\to b}$ is given by
\begin{eqnarray*}
Z^{(b)}_a&=&{W_a}-{W_b}=-(W_b-W_a)=-Z^{(a)}_b, \\
Z^{(b)}_i&=&W_i-W_b=(W_i-W_a)-(W_b-W_a)=Z^{(a)}_i-Z^{(a)}_b,
\quad i \in M \setminus \{a,b\}.
\end{eqnarray*}
The inverse map $\bh_{a\to b}^{-1}$ from $\bZ_b$ to $\bZ_a$ is given by
\begin{eqnarray*}
Z^{(a)}_b&=&W_b-W_a=-(W_a-W_b)=-Z^{(b)}_a,	\\
Z^{(a)}_i&=&W_i-W_a=(W_i-W_b)-(W_a-W_b)=Z^{(b)}_i-Z^{(b)}_a,
\quad i \in M \setminus \{a,b\}.
\end{eqnarray*}
The map $\bh_{a\to b}$ is a diffeomorphism, and we derive the Jacobian matrix and its determinant. 
%We use the notation $\bz_a = (x_i)_{i\ne a}$ and $\bz_b = (y_i)_{i\ne b}$ to denote compacting. 
We order the  coordinates of $\bz_{a}$ as $(z^{(a)}_b,\bz^{(a)}_D)$, where $\bz^{(a)}_D\in\mathbb R^{m-2}$ and $D=M\setminus\{a,b\}$. Similarly we order the coordinates of $\bz_{b}$  as $(z^{(b)}_a,\bz^{(b)}_D)$, $\bz^{(b)}_D\in\mathbb R^{m-2}$.
The inverse map can be written in block form as
$ z^{(a)}_b = -z^{(b)}_a$ and $\bz^{(a)}_D =  \bz^{(b)}_D-z^{(b)}_a\bone_{m-2}$.
The Jacobian matrix is
$$
\bJ=
\frac{\partial\, \bh_{a\to b}^{-1}(\bz_b)}{\partial \bz_b}
=
\frac{\partial(z^{(a)}_b,\bz^{(a)}_D)}{\partial(z^{(b)}_a,\bz^{(b)}_D)}
=
\begin{pmatrix}
	- 1 & \bzero_{m-2}^\top \\[4pt]
	- \bone_{m-2}  &  \bI_{m-2}
\end{pmatrix}.
$$
Since this matrix is block lower triangular, we have that  $|\det(\bJ)|=1$.
Therefore, $f_{\bZ_a}(\bz_a;\bSigma)=f_{\bZ_b}(\bz_b;\bSigma)$, for any $a$ and $b$, i.e.\ the likelihood is the same, namely
$$
\mathcal{L}(\bSigma;\bz_{b})
= f_{\bZ_b}(\bz_b;\bSigma) =
f_{\bZ_a}(\bh^{-1}_{a\to b}(\bz_b);\bSigma)
= f_{\bZ_a}(\bz_a;\bSigma)
=  \mathcal{L}(\bSigma;\bz_{a}).
$$

\subsubsection*{Gaussian scale mixtures}
For $k\in M=\{1,\dots,m\}$, let $\bZ_k=\bigl(Z^{(k)}_i\bigr)_{i\in M\backslash\{k\}}$, with $Z^{(k)}_i={W_i}/{W_k}$. We study the transformation $\bh_{a\to b}:\bZ_a\mapsto \bZ_b$, for $a,b\in M$.
The component-wise map $\bh_{a\to b}$ is given by
\begin{eqnarray*}
Z^{(b)}_a&=&{W_a}/{W_b}=\cfrac{1}{{W_b}/{W_a}}=\frac{1}{Z^{(a)}_b},\\
Z^{(b)}_i&=&W_i/W_b=\frac{W_i/W_a}{W_b/W_a}=\frac{Z^{(a)}_i}{Z^{(a)}_b},
\quad i\in M\setminus\{a,b\}.
\end{eqnarray*}
The inverse map, $\bh_{a\to b}^{-1}$, from $\bZ_b$ to $\bZ_a$
is given by
\begin{eqnarray*}
Z^{(a)}_b&=&{W_b}/{W_a}=\frac{1}{W_a/W_b}=\frac{1}{Z^{(b)}_a},
\\
Z^{(a)}_i&=&W_i/W_a=\frac{W_i/W_b}{W_a/W_b}=\frac{Z^{(b)}_i}{Z^{(b)}_a},
\quad i\in M\setminus\{a,b\}.
\end{eqnarray*}
Therefore the map $\bh_{a\to b}$ is a diffeomorphism on the open set $\{W_aW_b\neq 0$, a.s.\}. We calculate the Jacobian determinant of the inverse transformation.
Once again %we use the notation $\bz_a = (x_i)_{i\ne a}$ and $\bz_b = (y_i)_{i\ne b}$ to denote compacting and  
we order the coordinates of $\bz_{a}$ as $(z^{(a)}_b,\bz^{(a)}_D)$ and the  coordinates of $\bz_{b}$ as  $(z^{(b)}_a,\bz^{(b)}_D)$, with $\bz^{(a)}_D, \bz^{(b)}_D\in\mathbb R^{m-2}$.
The inverse map can be written in block form as
$ z^{(a)}_b = (z^{(b)}_a)^{-1}$ and $\bz^{(a)}_D = (z^{(b)}_a)^{-1} \bz^{(b)}_D$.
The Jacobian matrix is
$$
\bJ
=
\frac{\partial\, \bh^{-1}_{a\to b}(\bz_b)}{\partial \bz_b}
=
\frac{\partial(z^{(a)}_b,\bz^{(a)}_D)}{\partial(z^{(b)}_a,\bz^{(b)}_D)}
=
\begin{pmatrix}
- (z^{(b)}_a)^{-2} & \bzero_{m-2}^\top \\[4pt]
- (z^{(b)}_a)^{-2} \by_D & (z^{(b)}_a)^{-1} \bI_{m-2}
\end{pmatrix}.
$$
Since this matrix is block lower triangular,
$
|\det(\bJ)|
=
|-(z^{(b)}_a)^{-2}\,(z^{(b)}_a)^{-(m-2)}|
=
%|-\,z^{(b)\,-m}_a|=
|z^{(b)}_a|^{-m}.
$
%If we observed $\bw$, then we can transform it into  $\bz_a$ or $\bz_b=\bh_{a\to b}(\bz_a)$. The relation above implies
Therefore
$$
\mathcal{L}(\bSigma;\bz_{b})
= f_{\bZ_b}(\bz_b;\bSigma)
= |z^{(b)}_a|^{-m}
f_{\bZ_a}(\bh^{-1}_{a\to b}(\bz_b);\bSigma)
= |z^{(b)}_a|^{-m}f_{\bZ_a}(\bz_a;\bSigma)
\propto
\mathcal{L}(\bSigma;\bz_{a}),
$$
so the likelihoods for $\bSigma$  are proportional, with proportionality factor $|z^{(b)}_a|^{-m}=|w_a/w_b|^{-m}$, 
that depends only on the observation $\bw=(w_1,\dots,w_m)^T$, not on $\bSigma$.

\subsubsection*{Gaussian location-scale mixtures}
For $j,k\in M=\{1,\dots,m\}$, $j\ne k$, let $\bZ_{j,k} = \bigl( Z_i^{(j,k)}\bigr)_{i\in M\setminus\{j,k\}}$
with $Z^{(j,k)}_i = (W_i-W_j)/(W_k-W_j)$. We study the transformation $\bh_{(a,b)\to(c,d)}:\bZ_{a,b}\mapsto \bZ_{c,d}$, for $a,b,c,d\in M$.
It is convenient to introduce the {extended} components
$$
\widetilde Z^{(j,k)}_j = 0,\qquad \widetilde Z^{(j,k)}_k = 1,\qquad
\widetilde Z^{(j,k)}_i = Z^{(j,k)}_i\ \ (i\neq j,k),
$$
such that, for {all} $i\in M$,
$$
\widetilde Z^{(j,k)}_i=\frac{W_i-W_j}{W_k-W_j}.
$$
Assume $W_b\neq W_a$ and $W_d\neq W_c$ a.s.\,. For any $i\in M\setminus\{c,d\}$, the component-wise map $\bh_{(a,b)\to(c,d)}$ is given by

\begin{align*}
	\widetilde Z^{(c,d)}_i
	&=\cfrac{W_i-W_c}{W_d-W_c}
	=\cfrac{(W_i-W_a)-(W_c-W_a)}{(W_d-W_a)-(W_c-W_a)}\\
	&=\cfrac{\cfrac{W_i-W_a}{W_b-W_a}-\cfrac{W_c-W_a}{W_b-W_a}}
	{\cfrac{W_d-W_a}{W_b-W_a}-\cfrac{W_c-W_a}{W_b-W_a}}=\frac{\widetilde Z^{(a,b)}_i-\widetilde Z^{(a,b)}_c}
	{\widetilde Z^{(a,b)}_d-\widetilde Z^{(a,b)}_c},
\end{align*}
valid whenever $\widetilde Z^{(a,b)}_d\neq \widetilde Z^{(a,b)}_c$ a.s.\ (or equivalently $W_d\neq W_c$ a.s.).
By symmetry, swapping the roles of $(a,b)$ and $(c,d)$, the inverse map
$\bh_{(a,b)\to(c,d)}^{-1}$ is
\begin{equation}\label{eq:proof_z_ab}
    	\widetilde Z^{(a,b)}_i
	=\frac{\widetilde Z^{(c,d)}_i-\widetilde Z^{(c,d)}_a}
	{\widetilde Z^{(c,d)}_b-\widetilde Z^{(c,d)}_a},
	\qquad i\in M\setminus\{a,b\},
\end{equation}
valid whenever $\widetilde Z^{(c,d)}_b\neq \widetilde Z^{(c,d)}_a$ a.s.\ (or equivalently $W_b\neq W_a$ a.s.).
We remark that the formulas above remain correct even if some indices overlap (e.g.\ $c=a$), provided one uses the extended values
$\widetilde Z^{(a,b)}_a=0$, $\widetilde Z^{(a,b)}_b=1$, $\widetilde Z^{(c,d)}_c=0$, $\widetilde Z^{(c,d)}_d=1$,
and restricts to the events where the relevant denominators are non-zero.

Finally, we show that the map $\bh_{(a,b)\to(c,d)}$ is a diffeomorphism, starting from the case in which the four indices are all distinct: $
\{a,b\}\cap\{c,d\}=\varnothing.
$
In this case, $\bZ_{c,d}$ contains the components $Z^{(c,d)}_a$ and $Z^{(c,d)}_b$. %Let $y_i=\widetilde Z^{(c,d)}_i$. % and set
%$$D=y_b-y_a.$$
%In this generic case, 
The inverse map $\bh^{-1}_{(a,b)\to(c,d)}$ can be written for all $i\in M\setminus\{a,b\}$ as in \eqref{eq:proof_z_ab}.
%$$\widetilde Z^{(a,b)}_i=\frac{\widetilde Z^{(c,d)}_i-\widetilde Z^{(c,d)}_a}{\widetilde Z^{(c,d)}_b-\widetilde Z^{(c,d)}_a}.%$$
Here $\widetilde Z^{(c,d)}_d=1$ and $\widetilde Z^{(c,d)}_c=0$ are constants; all other $\widetilde Z^{(c,d)}_i$ with $i\notin \{c,d\}$ are coordinates of $\bZ_{(c,d)}$. We conclude that the map $\bh_{(a,b)\to(c,d)}$ is a diffeomorphism on the open set $\{(W_b-W_a)(W_d-W_c)\neq 0$, a.s.\} and we derive the Jacobian matrix and its determinant.

Consider the Jacobian matrix of $\bh^{-1}_{(a,b)\to(c,d)}$ with respect to the $(m-2)$ variables
$( Z^{(c,d)}_i)_{i\in S\setminus\{c,d\}}$ (which include $Z^{(c,d)}_a$ and $Z^{(c,d)}_b$).
A direct determinant computation yields
$$
\det\left(\frac{\partial (Z^{(a,b)}_i)_{i\neq a,b}}{\partial (Z^{(c,d)}_u)_{u\neq c,d}}\right)
=\frac{1}{(Z^{(c,d)}_b-Z^{(c,d)}_a)^{\,m-1}}.
$$
Therefore the absolute Jacobian determinant of the inverse map is
$|Z^{(c,d)}_b-Z^{(c,d)}_a|^{-(m-1)}$.

If the four indices are not all distinct (e.g.\ $a=c$), the same component-wise formulas for the map and inverse still hold,
but the Jacobian expression must be written using a pair of {variable} coordinates present in the source vector
(one typically chooses two indices whose extended values are known constants in the target coordinate system).
The generic formula above is the cleanest closed form.

%If we observed $\bw$, then we can transform it into $\bz_{(1,2)}$ or  $\bz_{(c,d)}$, $ (c,d)\ne (1,2)$. However $\bz_{(c,d)}=
%\bh_{(1,2)\to(c,d)}(\bz_{(1,2)})$ and the relations above imply
Therefore
\begin{eqnarray*}
\mathcal{L}(\bSigma;\bz_{c,d})
&= &f_{\bZ_{c,d}}(\bz_{c,d};\bSigma)\\
&=& |Z^{(c,d)}_b-Z^{(c,d)}_a|^{-(m-1)}
f_{\bZ_{a,b}}(\bh_{(a,b)\to(c,d)}^{-1}(\bz_{c,d});\bSigma)\\
&=& |Z^{(c,d)}_b-Z^{(c,d)}_a|^{-(m-1)}f_{\bZ_{a,b}}(\bz_{a,b};\bSigma)
\propto  \mathcal{L}(\bSigma;\bz_{a,b}),
\end{eqnarray*}
%We conclude that 
so the likelihoods for $\bSigma$  are proportional, with proportionality factor $|Z^{(c,d)}_b-Z^{(c,d)}_a|^{-(m-1)}=|(w_b-w_a)/(w_d-w_c)|^{-(m-1)}$, 
that depends only on the observation $\bw=(w_1,\dots,w_m)^T$, not on $\bSigma$.

\subsection{Numerical example}
Let $$(X_1,X_2,X_3)^\top=R \; (W_1,W_2,W_3)^\top$$ be a trivariate Gaussian scale mixture, with $R\sim\text{Gamma}(1,1)$ and
$$
\begin{pmatrix}W_1 \\W_2 \\W_3\end{pmatrix} \sim \mathcal{N}_3 \left(\begin{pmatrix}0 \\0 \\0\end{pmatrix}, \Sigma_{\theta_W} = \begin{pmatrix}1&\rho_1&\rho_2 \\\rho_1&1&\rho_3 \\\rho_2&\rho_3&1\end{pmatrix}\right).
$$
This non-spatial toy example can be used to show that the selection of the reference locations (or of the reference variable, in this case) has no effect on the estimated parameter values.
The following R code produces the maximum likelihood estimates for $\theta_W=(\rho_1,\rho_2,\rho_3)$ by selecting all the possible elements $X_i$, $i=1,2,3$, as the reference one when computing the ratios that cancel the random scale $R$.
Note that the value of the reference random variable has to be positive by construction. For example, if $X_1$ is selected as reference element (\texttt{index=1}), the vector of ratios is
$$\mathbf{Z}=\left(\frac{X_2}{X_1},\frac{X_3}{X_1}\right)$$
and its density function is
$$    f_{\mathbf{Z}}(\mathbf{z};\,\theta_W) =\; \pi^{-m/2} |\Sigma_{\theta_W}|^{-{1}/{2}} \Gamma\left(m/2\right)\left(\dot{\mathbf{z}}^\top \Sigma_{\theta_W}^{-1}\dot{\mathbf{z}} \right)^{-m/2},
$$
with $m=3$ and $\dot{\mathbf{z}}=(1,x_2/x_1,x_3/x_1)^\top$ (referred to as \texttt{z} in the code).
Note that $\pi^{-m/2}$ and $\Gamma\left(m/2\right)$ are constants that can be excluded from the likelihood computations.
By choosing \texttt{index=1,2,3} it is possible to see that the resulting estimated parameter values are numerically equivalent.

\begin{lstlisting}[language=R]
set.seed(123)
cor1 = 0.7
cor2 = 0.5
cor3 = 0.3
Sigma = matrix(c(1,cor1,cor2,cor1,1,cor3,cor2,cor3,1), nrow=3)

library(mvtnorm)
n = 1000
W = rmvnorm(n, sigma = Sigma)
R = rgamma(n, rate=1, shape=1)
X = R * W

nll = function(par, x, index=1){
  if(min(par) < 0) return(Inf)
  if(max(par) > 1) return(Inf)
  
  m = ncol(x)
  S = matrix(c(1,par[1],par[2],par[1],1,par[3],par[2],par[3],1), nrow=3)
  
  # vector of ratios z
  z = x/x[,index]
  
  # computation of z^T Sigma^{-1} z
  L = chol(S)
  y = forwardsolve(t(L), t(z))
  zT_Sigma_z = colSums(y^2)
  
  # negative log-likelihood (excluding constants)
  -sum(-0.5*log(det(S)) -m/2*log(zT_Sigma_z))
}

fit1 = optim(c(0.5,0.5,0.5), nll, method="Nelder-Mead", x=X, index=1)
fit2 = optim(c(0.5,0.5,0.5), nll, method="Nelder-Mead", x=X, index=2)
fit3 = optim(c(0.5,0.5,0.5), nll, method="Nelder-Mead", x=X, index=3)
fit1$par
fit2$par
fit3$par
\end{lstlisting}

\section{Identifiability of Gaussian location-scale mixtures}\label{section:identifiability_issues}

In the general case of Gaussian location-scale mixtures, i.e., when in equation \eqref{eq:gaussian_locationscale_mixtures} in the main text both $S$ and $R$ are non-degenerate random variables, identifiability of the different models and of the parameter $\theta_{S,R}$ deserves some discussions.

\subsection{Parameter identifiability}\label{section:identifiability1_parameter}
A first issue may be the identifiability of $\theta_{S,R}$ from the univariate distribution of $\bar{X}$, used in Step 2 of the proposed inferential method, as detailed in Section \ref{section:new_method_step2} in the main text.
To check that the parameter of a specific model is identifiable, it is possible to compute the characteristic function of $\bar{X}$, keeping in mind that $\bar{X}=S+R\;\bar{W}$, $\bar{W}\sim \mathcal{N}\left(0,\sigma_{\bar{W}}^2\right)$ and $\sigma_{\bar{W}}^2$ is a known variance that depends on the value of $\theta_W$ estimated in Step 1. Then, $R\;\bar{W}\mid R=r \sim \mathcal{N}\left(0,r^2\sigma_{\bar{W}}^2\right)$, and the characteristic function of $\bar{X}$ is 
\begin{equation*}
\begin{split}
    \varphi_{\bar{X}}(t)=\;\E\left[\exp(itS+itR\bar{W})\right] = \E\left[\exp(itS)\right] \cdot \E\left[\exp\left(-\frac{1}{2} t^2R^2\sigma^2_{\bar{W}}\right)\right] =\; \varphi_S(t) \cdot M_{R^2}\left(-\frac{t^2\sigma^2_{\bar{W}}}{2}\right),
\end{split}
\end{equation*}
where $M_{R^2}$ is the moment generating function of $R^2$.
While the identifiability of $\theta_{S,R}$ is not guaranteed in general, we can show that it is ensured in the case for the two location-scale mixture models presented in this work, LSM1 and LSM2.
Recall that, if $Y\sim\text{E}(\lambda)$, then $\varphi_Y(t)=\lambda/(\lambda-it)$ and $M_Y(t)=\lambda/(\lambda-t)$. In model LSM1, $S\sim \text{E}(\lambda)$ and $R^2\sim \text{E}(1/2)$, so
\begin{equation*}
    \varphi_{\bar{X}}(t)=\; \varphi_S(t) \cdot M_{R^2}\left(-\frac{t^2\sigma^2_{\bar{W}}}{2}\right)=\frac{\lambda}{\lambda-it}\cdot \frac{1}{1+t^2\sigma^2_{\bar{W}}}.
\end{equation*}
In model LSM2, $S=S_1-S_2$, with $S_1\sim \text{E}(\lambda_1)$ and $S_2\sim \text{E}(\lambda_2)$, and $R^2\sim \text{E}(1/2)$, so
\begin{equation*}
\begin{split}
    \varphi_{\bar{X}}(t)=&\; \varphi_S(t) \cdot M_{R^2}\left(-\frac{t^2\sigma^2_{\bar{W}}}{2}\right)=\E\left[\exp(itS_1-itS_2)\right] \cdot M_{R^2}\left(-\frac{t^2\sigma^2_{\bar{W}}}{2}\right)\\=&\; \varphi_{S_1}(t) \cdot \varphi_{S_2}(-t) \cdot M_{R^2}\left(-\frac{t^2\sigma^2_{\bar{W}}}{2}\right)=\frac{\lambda_1}{\lambda_1-it}\cdot\frac{\lambda_2}{\lambda_2+it}\cdot \frac{1}{1+t^2\sigma^2_{\bar{W}}}.
\end{split}
\end{equation*}
The identifiability of $\theta_{S,R}$ may not be guaranteed for more complex models involving a higher-dimensional parameter. The study of the characteristic function of $\bar{X}$ is a tool that can help in these cases.

\subsection{Model identifiability}\label{section:identifiability2_model}
\begin{figure}[t]
    \centering
    \includegraphics[width=\linewidth]{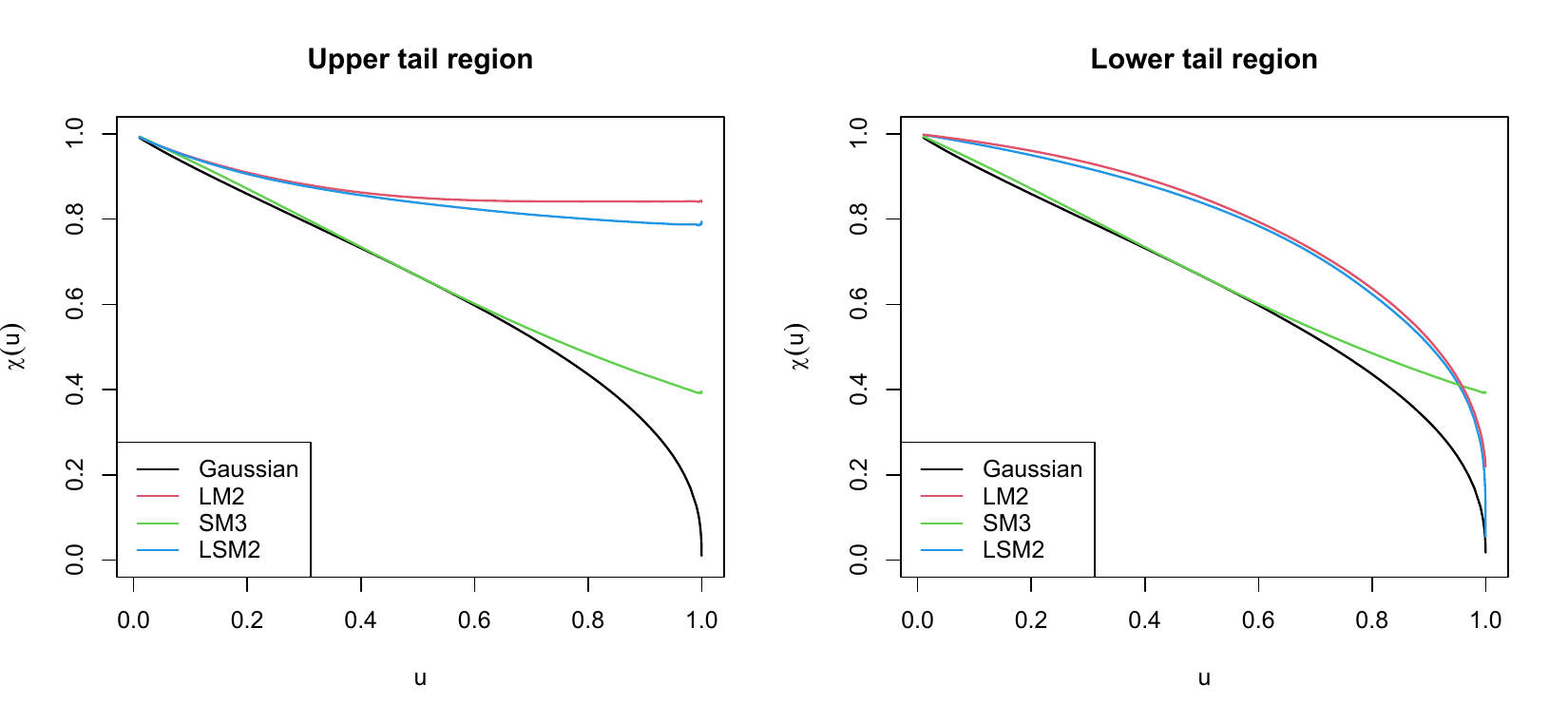}
    \caption{Values of $\chi(u)$, for $u\in(0,1)$, for the upper and lower tail regions of the four models described in Section \ref{section:identifiability2_model}.}
    \label{fig:chi_biv_figure1}
\end{figure}

A separate issue is the identifiability of the model itself, as different pairs $(S,R)$ may potentially lead to similar finite-dimensional distributions. For simplicity, we limit our discussion here to the bivariate case shown in Figure \ref{fig:biv_plots} in the main text and described in Section \ref{section:general_model_definition}.
Recall that $(W_1,W_2)^T$ is a bivariate Gaussian and, for $i=1,2$,
\addtolength{\leftmargini}{1cm}
\begin{itemize}
	\item[(LM2)] $X_i=S+W_i$, with $S=E_1-E_2$, $E_1\sim\text{E}(0.4)$, $E_2\sim\text{E}(2.5)$;
	\item[(SM3)] $X_i= R \; W_i$  with $R=1/\sqrt{E_3}$, $E_3\sim$ E$(1)$; 
	\item[(LSM2)] $X_i=S+R \; W_i$ with $S=E_1-E_2$ and  $R=\sqrt{E_4}$, $E_4\sim$ E$(1/2)$.
\end{itemize} 
\addtolength{\leftmargini}{-1cm}
Figure \ref{fig:chi_biv_figure1} shows the values of $\chi(u)$, for $u\in(0,1)$, for the upper and lower tail regions of the four models, fixing $\text{cor}(W_1,W_2)=0.5$.
A first difference we can observe is that asymmetry between the upper and lower tail regions is possible only when $S$ is not degenerate, while Gaussian models and Gaussian scale-mixture models are symmetric by construction.
The dependence structure looks very similar for models LM2 and LSM2, although the former has a slight stronger dependence and is AD in both tail regions, while the latter is AD in the upper tail region and AI in the lower tail region.
This difference, however, may not be evident when looking to the sub-asymptotic dependence structures of the two models. For this reason, we study here the bivariate characteristic functions of LM2 and LSM2, to inspect the difference between their distributions.

Let $\mathbf{t}=(t_1,t_2)^\top$ and $\mathbf{W}=(W_1,W_2)^\top\sim(\bzero,\Sigma_2)$. Note that the characteristic function of $\mathbf{W}$ is $\varphi_{\mathbf{W}}(\mathbf{t})=\exp(-\mathbf{t}^\top\Sigma_2\mathbf{t}/2)$. Then, the characteristic function of a general bivariate Gaussian location-scale mixture $\mathbf{X}=(X_1,X_2)^\top$ is 
\begin{equation*}
\begin{split}
    \varphi_{\mathbf{X}}(\mathbf{t})=&\; \E\left[\exp\{-i(t_1X_1+t_2X_2)\}\right] = \E\left[\exp\{-i(t_1S+t_1RW_1+t_2S+t_2RW_2)\}\right]\\
    =&\; \varphi_S(t_1+t_2) \cdot \E_R\left[\varphi_{\mathbf{W}}(R\,\mathbf{t}) \right] = \varphi_S(t_1+t_2) \cdot \E_R\left[\exp\left(-\frac{1}{2} R^2\,\mathbf{t}^\top\Sigma_2\mathbf{t}\right) \right].
\end{split}
\end{equation*}
In the case of model LM2, $R=1$, so %and, recalling the expansion $\exp(x)=\sum_{k=0}^\infty x^k/k$, we can write
\begin{equation*}
\begin{split}
    \varphi_{\mathbf{X}}(\mathbf{t})=&\; \varphi_S(t_1+t_2) \cdot \exp\left(-\frac{1}{2} \mathbf{t}^\top\Sigma_2\mathbf{t}\right). %= \varphi_S(t_1+t_2) \cdot \frac{1}{1+\mathbf{t}^\top\Sigma_2\mathbf{t}/2+(\mathbf{t}^\top\Sigma_2\mathbf{t}/2)^2/2+\cdots}.
\end{split}
\end{equation*}
In the case of model LSM2, since $R^2\sim\text{E}(1/2)$ (see also Section \ref{section:identifiability1_parameter}),
\begin{equation*}
\begin{split}
    \varphi_{\mathbf{X}}(\mathbf{t})=&\; \varphi_S(t_1+t_2) \cdot M_{R^2}\left(-\frac{1}{2} \mathbf{t}^\top\Sigma_2\mathbf{t}\right) = \varphi_S(t_1+t_2) \cdot \frac{1}{1+\mathbf{t}^\top\Sigma_2\mathbf{t}}.
\end{split}
\end{equation*}
The bivariate distributions of the two models are different because the presence of the random variable $R$ modifies the effect of the dependence of the bivariate Gaussian $\mathbf{W}$, with respect to the random variable $S$, common to the two models.

In general, the study of bivariate or multivariate characteristic functions can help to check the differences between the distributions of models characterized by different pairs $(S,R)$, in case these differences are not clear when visualizing their dependence structure through some graphics.

\section{Consistency of simulated minimum distance estimator }\label{section:smde_supplementary}

Let $X_1,\ldots,X_n$ be i.i.d.\ observations from a continuous $F_0=F_{\theta_0}$, and define $
Y_b(\theta)=F_\theta^{-1}(U_b)$, $b=1,\ldots,B$, where $U_1,\ldots,U_B$ are i.i.d.\ uniform variables independent of the observations. We remark that the same uniforms are used for every $\theta$. Consider the following estimator 
\begin{equation*} \hat\theta_{n,B}=\argmin_\theta  \frac{1}{n}\sum_{i=1}^n \left[\frac{i-1/2}{n}-\widehat{F}_{B}({X}_{(i)})\right]^2. \end{equation*} 
Here $\widehat{F}_B$ is the empirical cdf of $Y_b(\theta)$ and ${X}_{(i)}$, $i=1,\dots,n$, are the ordered values of ${X}_{i}$, $i=1,\dots,n$. 
\begin{prop} We assume that
\begin{enumerate}
\item $\Theta\subset\mathbb R^p$ is compact and $\theta_0\in\operatorname{int}(\Theta)$.
\item $F_0$ is continuous and strictly increasing. The functions $
G_\theta(t)=F_\theta(F_0^{-1}(t))$, $0<t<1,$
extend continuously to $[0,1]$, and $\theta\mapsto G_\theta$ is continuous in the uniform norm.
\item The model is identified by $
\int_0^1[G_\theta(t)-t]^2\,dt>0$
whenever $\theta\ne\theta_0$.
\end{enumerate}
Then
$$
\hat\theta_{n,B}\xrightarrow{p}\theta_0\quad
\text{for } n\to\infty, B=B_n\to\infty.$$
\end{prop}
\textbf{Proof.}
The $Z_i=F_0(X_i)$, $i=1,\dots,n$, are i.i.d.\ uniform random variables on $[0,1]$. Write
$
E_n(t)=n^{-1}\sum_{i=1}^n\mathbf1\{Z_i\le t\}$ and
$S_B(t)=B^{-1}\sum_{b=1}^B\mathbf1\{U_b\le t\}$.
Inverse-transform simulation gives $
F_{B,\theta}(X_i)=S_B(G_\theta(Z_i))$.
Because the observations have no ties almost surely, $
E_n(Z_{(i)})= i/n.
$
Thus the contrast function can be written as
$$
Q_{n,B}(\theta)
=
\int_0^1
\left[
E_n(t)-S_B(G_\theta(t))-\frac1{2n}
\right]^2\,dE_n(t).
$$
Let $\widehat\theta_{n,B}$ be any measurable sequence satisfying
$$
Q_{n,B}(\widehat\theta_{n,B})
\le
\inf_{\theta\in\Theta} Q_{n,B}(\theta)+\frac1n.
$$
Set $
e_n(t)=E_n(t)-t$, $s_B(t)=S_B(t)-t$. The Glivenko--Cantelli theorem gives
$
\|e_n\|_\infty+\|s_B\|_\infty\xrightarrow{p}0.
$
Moreover, $\sup_{\theta\in\Theta}
\|S_B\circ G_\theta-G_\theta\|_\infty
\le \|s_B\|_\infty.
$
Hence
$$
\sup_{\theta\in\Theta}\sup_{t\in[0,1]}
\left|
E_n(t)-S_B(G_\theta(t))-\frac1{2n}
-
\{t-G_\theta(t)\}
\right|
\le
\|e_n\|_\infty+\|s_B\|_\infty+\frac1{2n}
\xrightarrow{p}0.
$$
Since the functions involved are uniformly bounded, this implies
\begin{equation*}
\sup_{\theta\in\Theta}
\left|
Q_{n,B}(\theta)
-\int_0^1[t-G_\theta(t)]^2\,dE_n(t)
\right|\xrightarrow{p}0.
\end{equation*}
The class $\mathcal F=\{t\mapsto[t-G_\theta(t)]^2:\theta\in\Theta\}$
is compact in the uniform norm. Indeed, it is a continuous image of compact $\Theta$. Covering $\mathcal F$ by a finite uniform $\varepsilon$-net and applying the law of large numbers to its elements yields
\begin{equation*}
\sup_{\theta\in\Theta}
\left|
\int[t-G_\theta(t)]^2\,dE_n(t)
-\int[t-G_\theta(t)]^2\,dt
\right|\xrightarrow{p}0.
\end{equation*}
Therefore,
\begin{equation*}
\sup_{\theta\in\Theta}|Q_{n,B}(\theta)-Q(\theta)|
\xrightarrow{p}0,
\qquad
Q(\theta)=\int_0^1[t-G_\theta(t)]^2\,dt.
\end{equation*}
For every $\varepsilon>0$ for which the set below is nonempty, compactness, continuity, and identification imply $
c_\varepsilon
=\inf_{\|\theta-\theta_0\|\ge\varepsilon}Q(\theta)>0.
$
On the event
$
\sup_\theta|Q_{n,B}(\theta)-Q(\theta)|<c_\varepsilon/3,
$
every $\theta$ outside the $\varepsilon$-ball has a larger empirical criterion than $\theta_0$. Hence
\begin{equation*}
\Pr(\|\hat\theta_{n,B}-\theta_0\|\ge\varepsilon)\to0.
\end{equation*}
\hfill$\square$

\section{Additional simulation studies}\label{section:simul_supplementary}

%This section is complementary to Section \ref{section:simulation_studies} in the main text and covers

\subsection{Estimation of different models}\label{section:simul_consistency}

\begin{figure}[ht!]
    \centering
    \includegraphics[width=1\linewidth]{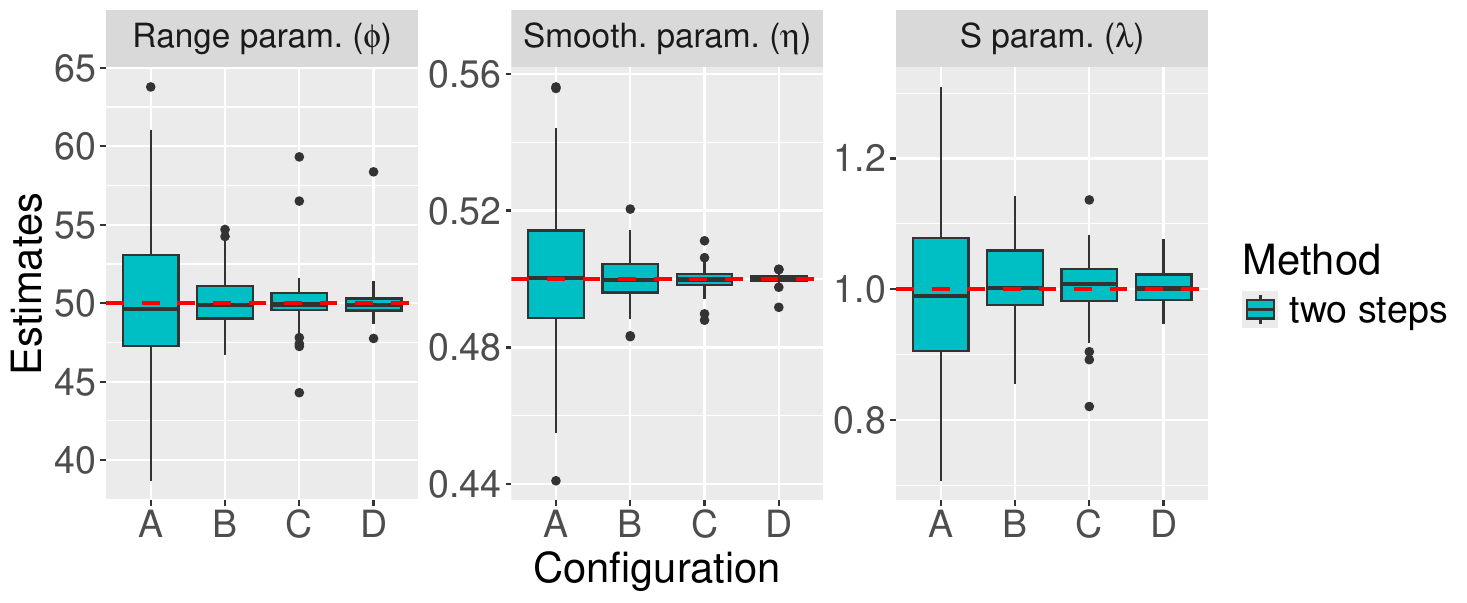}
    \caption{Estimation of Gaussian location-scale mixture 1, with $S\sim\text{Exp}(\lambda)$ and $R=\sqrt{E}$, $E\sim\text{Exp}(1/2)$, on 100 simulated datasets for each configuration, with true values $\varphi=50$, $\eta=0.5$ and $\lambda=1$ (red lines). The boxplots refer to the two-step solution for inference explained in Section \ref{section:estimation_methods}.}
    \label{fig:boxplots_sim_exp_lapl}
\end{figure}

\begin{figure}[ht!]
    \centering
    \includegraphics[width=1\linewidth]{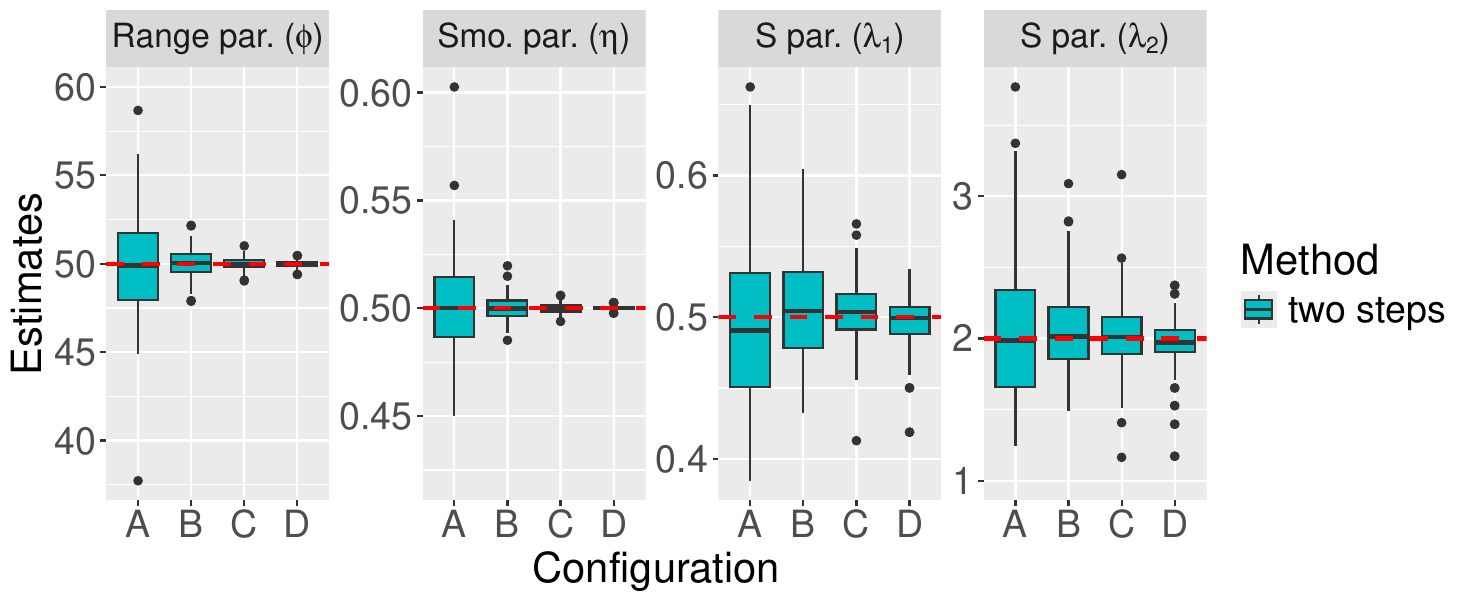}
    \caption{Estimation of Gaussian location mixture 2, with $S=S_1-S_2$, $S_1\sim\text{Exp}(\lambda_1)$ and $S_2\sim\text{Exp}(\lambda_2)$, on 100 simulated datasets for each configuration, with true values $\varphi=50$, $\eta=0.5$, $\lambda_1=0.5$ and $\lambda_2=2$ (red lines). The boxplots refer to the two-step solution for inference explained in Section \ref{section:estimation_methods}.}
    \label{fig:boxplots_sim_aslapl}
\end{figure}

This simulation study aims to show the consistency of the two-step method for inference explained in Section \ref{section:estimation_methods} (in the main text) as we increase $m$ and $n$, i.e.\ the number of spatial locations and data replications.
In all the following cases, inference is performed on 100 simulated datasets for each of the configurations A, B, C and D (see Section \ref{section:simulation_studies} in the main text).
This simulation study covers the three classes of Gaussian location, scale and location-scale mixtures described in the main text; for reference about the names of the models, see Section \ref{section:example_models_gaussiansclocmixt} and Table \ref{table:examples_gaussian_locscmix_models}.
\begin{figure}[ht!]
    \centering
    \includegraphics[width=1\linewidth]{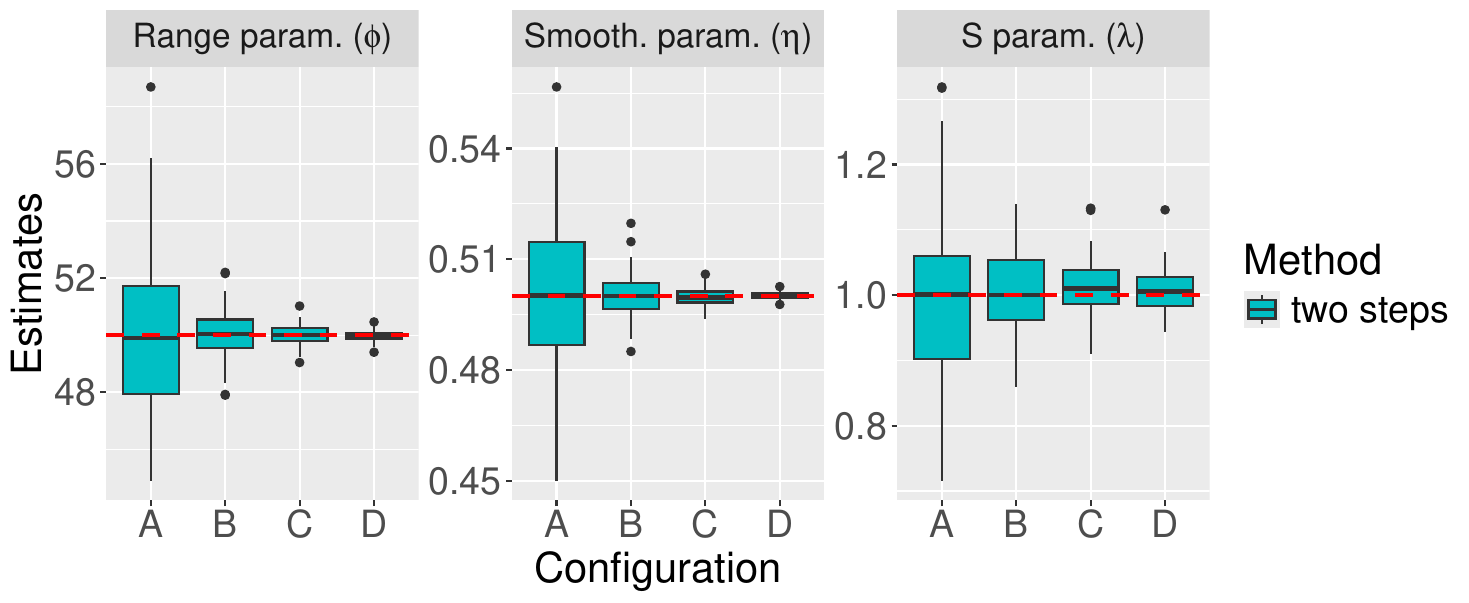}
    \caption{Estimation of Gaussian location mixture 1, with $S\sim\text{Exp}(\lambda)$, on 100 simulated datasets for each configuration, with true values $\varphi=50$, $\eta=0.5$ and $\lambda=1$ (red lines). The boxplots refer to the two-step solution for inference explained in Section \ref{section:estimation_methods}.}
    \label{fig:boxplots_sim_exp}
\end{figure}
\begin{figure}[ht!]
    \centering
    \includegraphics[width=1\linewidth]{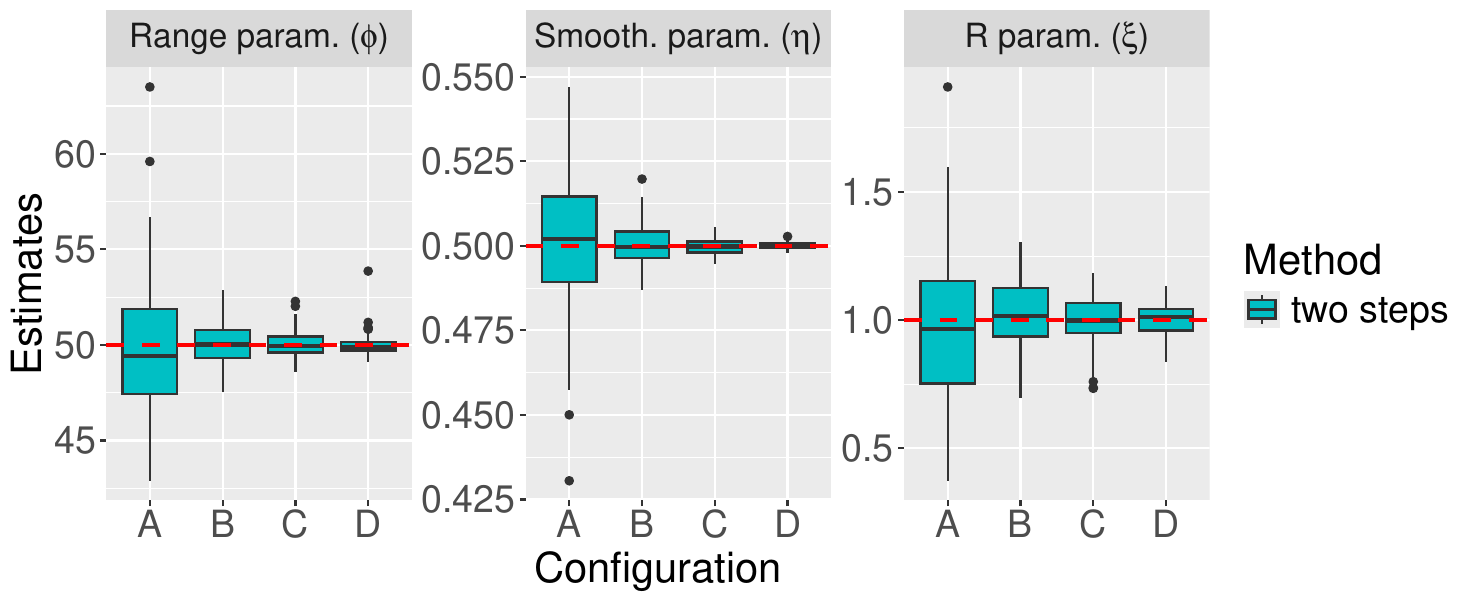}
    \caption{Estimation of Gaussian scale mixture 4, with $R=\sqrt{E}/G$, $E\sim\text{Exp}(1/2)$ and $G\sim\text{Gamma}(1/\xi,1/\xi)$, on 100 simulated datasets for each configuration, with true values $\varphi=50$, $\eta=0.5$ and $\lambda=1$ (red lines). The boxplots refer to the two-step solution for inference explained in Section \ref{section:estimation_methods}.}
    \label{fig:boxplots_sim_sgpd}
\end{figure}
In particular, Figure \ref{fig:boxplots_sim_exp} shows the estimation results for the parameters of Gaussian location mixture 1, with $S\sim\text{Exp}(\lambda)$; Figure \ref{fig:boxplots_sim_sgpd} is about the estimation of the parameters of Gaussian scale mixture 4, with $R=\sqrt{E}/G$, $E\sim\text{Exp}(1/2)$ and $G\sim\text{Gamma}(1/\xi,1/\xi)$; Figure \ref{fig:boxplots_sim_exp_lapl} refers to Gaussian location-scale mixture 1, with $S\sim\text{Exp}(\lambda)$ and $R=\sqrt{E}$, $E\sim\text{Exp}(1/2)$; finally, Figure \ref{fig:boxplots_sim_aslapl} shows the estimation results for Gaussian location mixture 2, with $S=S_1-S_2$, $S_1\sim\text{Exp}(\lambda_1)$ and $S_2\sim\text{Exp}(\lambda_2)$, i.e.\ a case in which $S$ has two parameters.
In all these cases, the estimators seem to be unbiased and consistent for all parameters. Nevertheless, the statistical efficiency is higher for the correlation parameter $\theta_W=(\varphi,\eta)$ of the Gaussian process $\mathbf{W}$, while the estimators for the parameter $\theta_{S,R}$ of the random variables $S$ and $R$ display higher variability, although they still appear to be empirically unbiased and consistent.

\subsection{Comparison with numerical integration}\label{section:simul_integration}

As stated in Section \ref{section:estimation_methods} in the main text, the standard solution for inference on Gaussian location and scale mixtures involves the numerical integration of density functions \eqref{eq:chapter2_integral_density}.
Note that in the cases of Gaussian location mixtures and Gaussian scale mixtures, the formula \eqref{eq:chapter2_integral_density} reduces to a single integral with respect to $S$ or $R$, while a double integral must be evaluated in the case of Gaussian location-scale mixtures.
In this section, we estimate two models for which a single integral is sufficient and compare the results with the newly proposed method and the closed-form maximum likelihood estimates, when
available. All estimations are performed on 50 datasets and only for Configurations A and B, to reduce the computational cost of the integration-based method.
Integration points are chosen adaptively by the R function \texttt{integrate()} within the interval $(0,30)$.

\begin{figure}[ht!]
    \centering
    \includegraphics[width=1\linewidth]{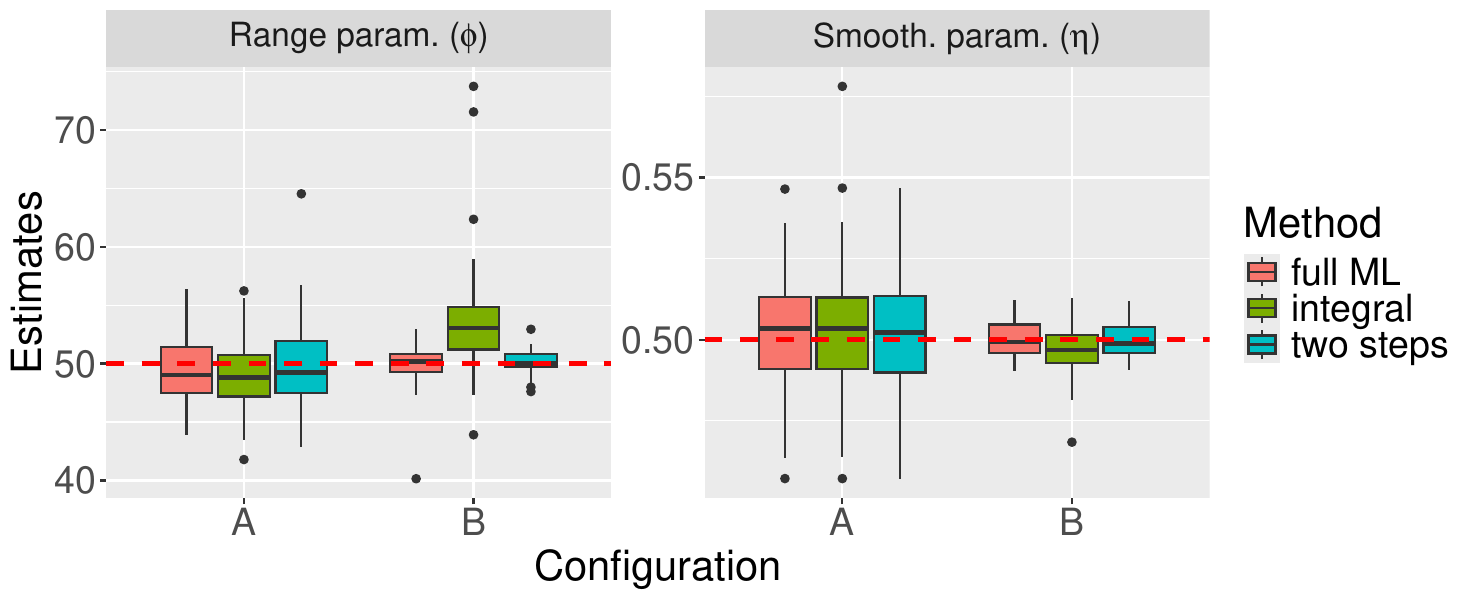}
    \caption{Estimation of Laplace process on 50 simulated datasets for each configuration, with true values $\varphi=50$ and $\eta=0.5$ (red lines). Orange boxplots refer to full maximum likelihood estimates, while green boxplots refer to estimates obtained via numerical integration and blue boxplots refer to the two-step solution for inference explained in Section \ref{section:estimation_methods}.}
    \label{fig:boxplots_sim_laplace_int}
\end{figure}

\begin{figure}[ht!]
    \centering
    \includegraphics[width=1\linewidth]{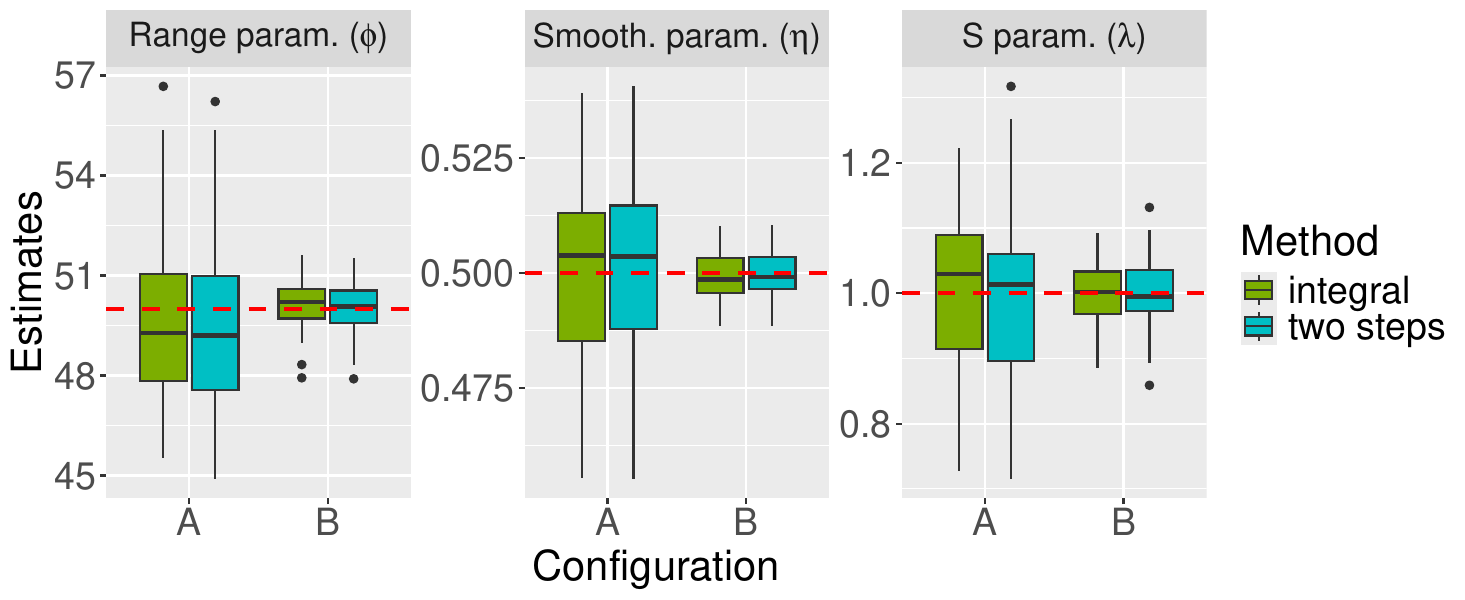}
    \caption{Estimation of Gaussian location mixture 1, with $S\sim\text{Exp}(\lambda)$, on 50 simulated datasets for each configuration, with true values $\varphi=50$, $\eta=0.5$ and $\lambda=1$ (red lines). Green boxplots refer to estimates obtained via numerical integration, while blue boxplots refer to the two-step solution for inference explained in Section \ref{section:estimation_methods}.}
    \label{fig:boxplots_sim_exp_int}
\end{figure}

Figure \ref{fig:boxplots_sim_laplace_int} shows estimates for a Laplace process (model SM1 in Table \ref{table:examples_gaussian_locscmix_models}), comparing the new method (in blue) with the maximum likelihood in closed form (in orange) and the maximum likelihood obtained by numerical integration (in green), similarly to Figure \ref{fig:boxplots_sim_laplace} in the main text.
Figure \ref{fig:boxplots_sim_exp_int} refers to Gaussian location mixture 1, similarly to Figure \ref{fig:boxplots_sim_exp}, and in this case the maximum likelihood is not computed in closed form.
In both cases, the new estimation method seems to perform as well as the classical ones. However, the estimation based on numerical integration of densities \eqref{eq:chapter2_integral_density} has a much higher computational cost. A comparison of computation times is provided in Section \ref{section:simul_computation} in the main text.


\newpage

\subsection{Estimation with copula}\label{section:simul_copula}
\noindent
The simulation studies in Section \ref{section:simul_consistency} and in Section \ref{section:simul_comparison} in the main text focus on the estimation of the parameters of Gaussian location-scale mixtures when data are simulated directly from these models.
However, as explained in Section \ref{section:new_method_copula}, data are often assumed to be observed in another scale, $Y(\mathbf{s})$, $\mathbf{s}\in\mathcal{S}$; after marginal transformations, a vector of uniforms $\mathbf{U}$ can be reconstructed as in equation \eqref{eq:tranform_copula_UY} in the main text. Typically, a copula approach involves estimating the parameters of $\mathbf{X}$ on the vector $\mathbf{U}$; this approach is followed in the simulation studies presented in this section. To construct the vector $\mathbf{U}$, a non-parametric marginal transformation is applied to the data vector $\mathbf{X}$, simulated from the Gaussian location-scale mixture models. In particular, a location-wise rank transformation is employed.
Then, the new estimation method is applied to $\mathbf{U}$ as described in Section \ref{section:new_method_copula} in the main text.

\begin{figure}[ht!]
    \centering
    \includegraphics[width=1\linewidth]{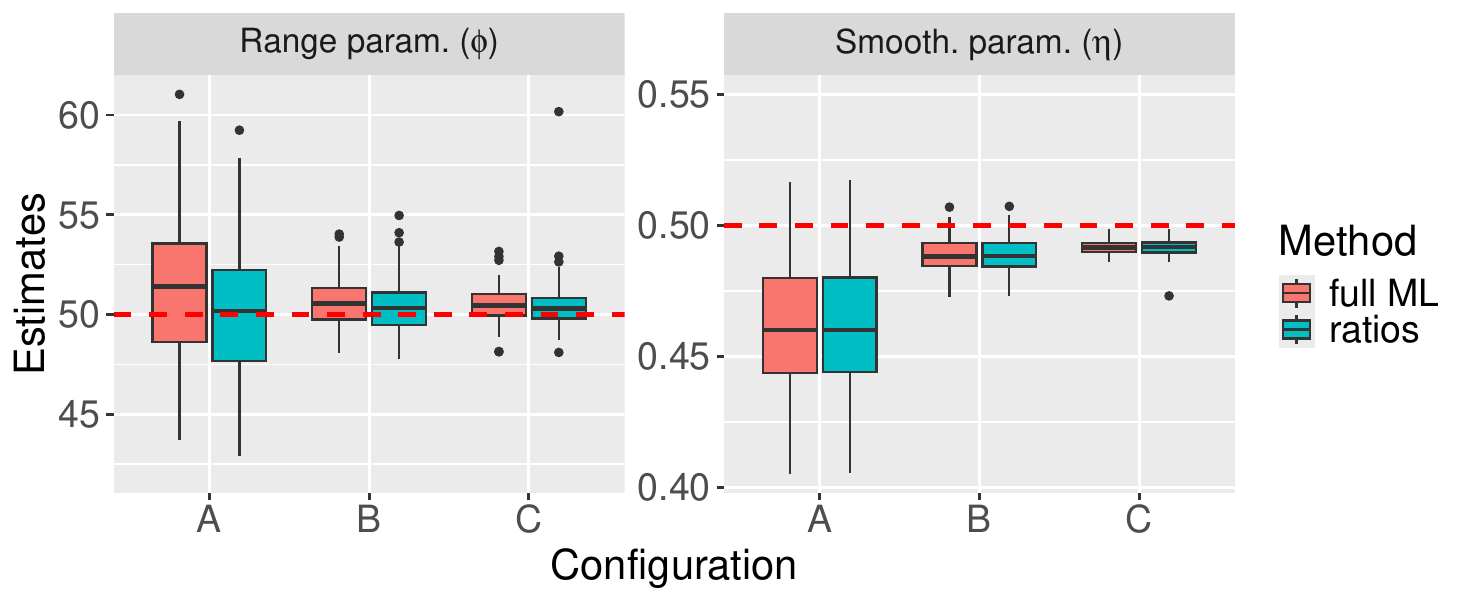}
    \caption{Estimation of Laplace process with copula on 100 simulated datasets for each configuration, with true values $\varphi=50$ and $\eta=0.5$ (red lines). Orange boxplots refer to full maximum likelihood estimates, while blue boxplots refer to the estimates based on the likelihood of ratios explained in Section \ref{section:new_method_step1}.}
    \label{fig:boxplots_sim_laplace_copula}
\end{figure}

\begin{figure}[ht!]
    \centering
    \includegraphics[width=1\linewidth]{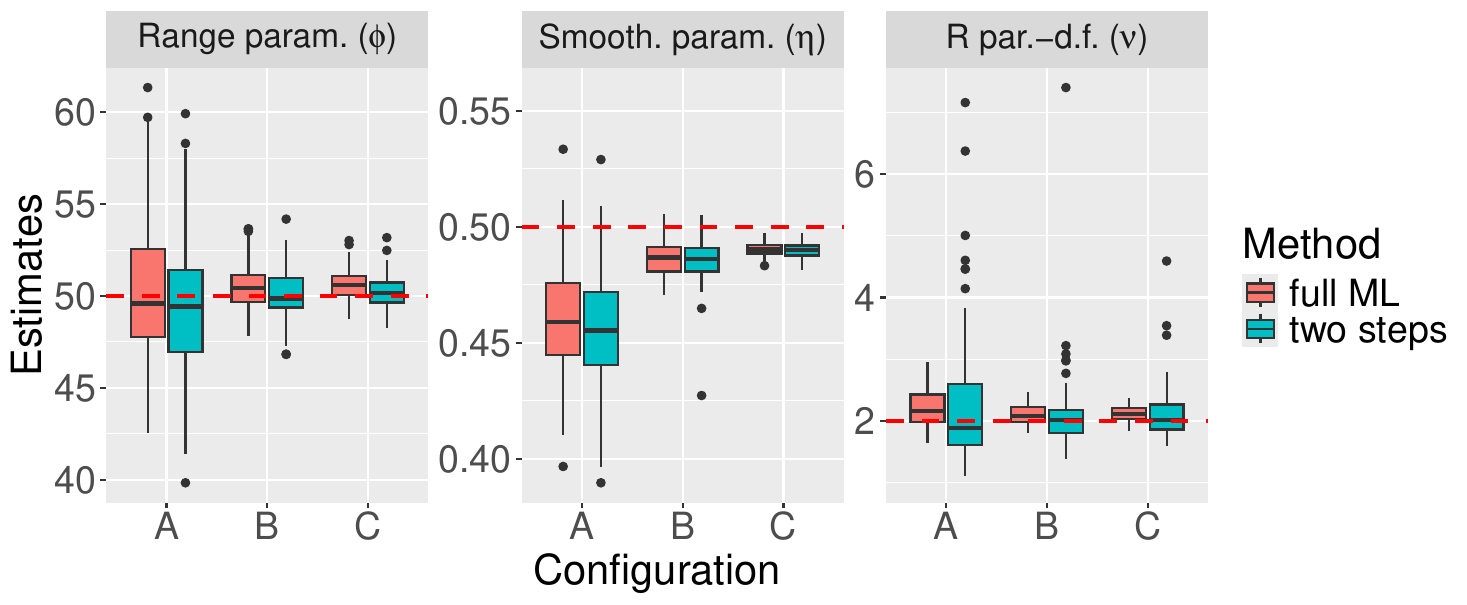}
    \caption{Estimation of Student's \textit{t} process with copula on 100 simulated datasets for each configuration, with true values $\varphi=50$, $\eta=0.5$ and $\nu=2$ (red lines). Orange boxplots refer to full maximum likelihood estimates, while blue boxplots refer to the two-step solution for inference explained in Section \ref{section:estimation_methods}.}
    \label{fig:boxplots_sim_t_2_copula}
\end{figure}

Figures \ref{fig:boxplots_sim_laplace_copula} and \ref{fig:boxplots_sim_t_2_copula} %and \ref{fig:boxplots_sim_t_8_copula}
show analogue simulations to those in Section \ref{section:simul_comparison}, for which a comparison with the full-likelihood case is possible. The estimators of $\theta_W=(\varphi,\eta)$ show some bias in the lower-dimensional configurations, in particular for the smoothness parameter $\eta$. However, the bias reduces as $m$ and $n$ are increased, and it appears to be the same for the two estimation methods (see Figure \ref{fig:boxplots_sim_laplace_copula}).
This bias seems to be caused by the non-parametric rank transformation mentioned above, and it is not present when parametric marginal transformations are employed, as is the case in real data applications.
In the three-parameter case of Figures \ref{fig:boxplots_sim_t_2_copula}% and \ref{fig:boxplots_sim_t_8_copula}
, the new estimators of $\theta_W$ are again comparable to those based on the full likelihood, but the estimator of the third parameter $\theta_d=\nu$ displays higher variability than that based on the full likelihood, as in the non-copula case. Again, the consistency of the new method has to be shown on each of the three classes of models.
\begin{comment}
    \begin{figure}[ht!]
    \centering
    \includegraphics[width=1\linewidth]
    \caption{Estimation of Student's \textit{t} process with copula on 100 simulated datasets for each configuration, with true values $\varphi=50$, $\eta=0.5$ and $\nu=8$ (red lines). Orange boxplots refer to full maximum likelihood estimates, while blue boxplots refer to the two-step solution for inference explained in Section \ref{section:estimation_methods}.}
    \label{fig:boxplots_sim_t_8_copula}
\end{figure}
\end{comment}

\begin{figure}[ht!]
    \centering
    \includegraphics[width=1\linewidth]{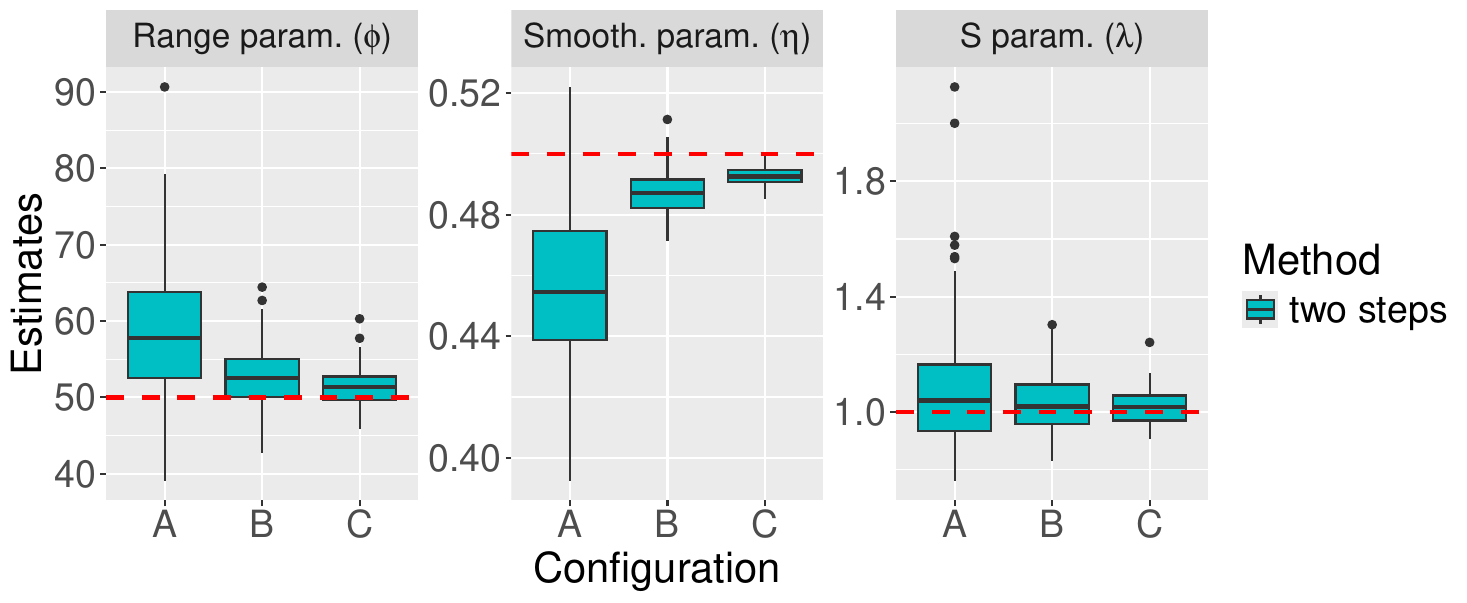}
    \caption{Estimation of Gaussian location mixture 1 with copula, with $S\sim\text{Exp}(\lambda)$, on 100 simulated datasets for each configuration, with true values $\varphi=50$, $\eta=0.5$ and $\lambda=1$ (red lines). The boxplots refer to the two-step solution for inference explained in Section \ref{section:estimation_methods}.}
    \label{fig:boxplots_sim_exp_copula}
\end{figure}

\begin{figure}[ht!]
    \centering
    \includegraphics[width=1\linewidth]{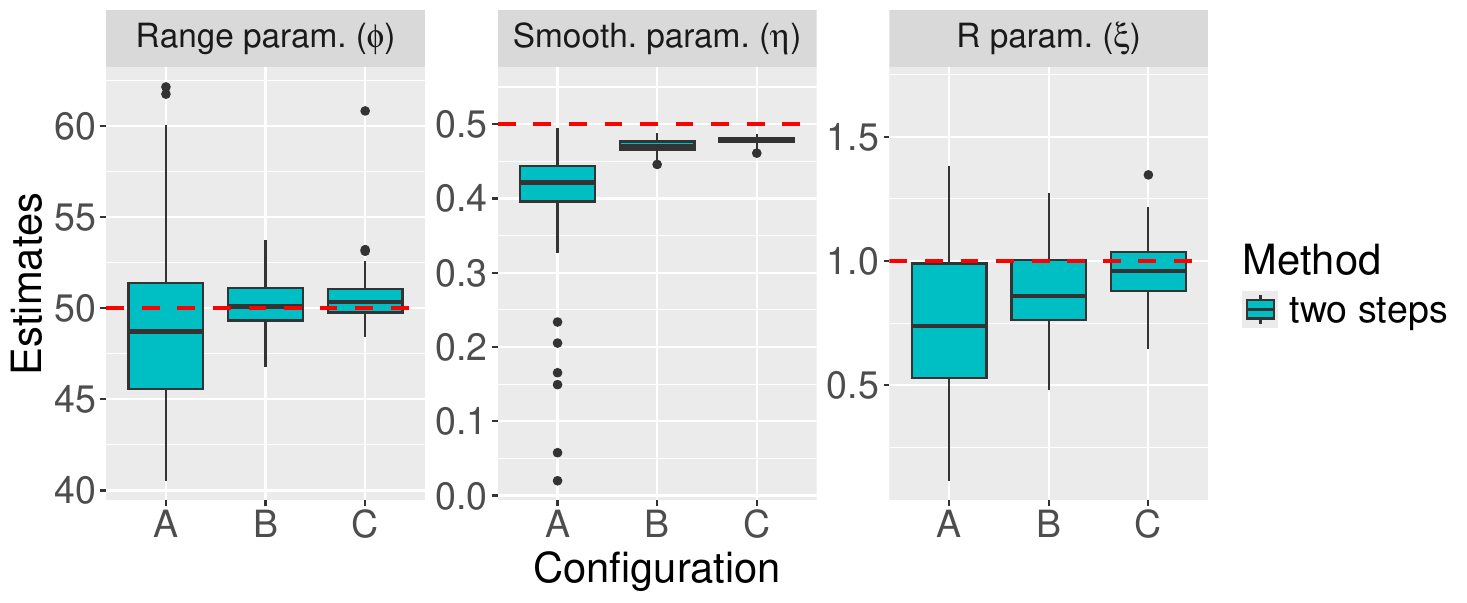}
    \caption{Estimation of Gaussian scale mixture 4 with copula, with $R=\sqrt{E}/G$, $E\sim\text{Exp}(1/2)$ and $G\sim\text{Gamma}(1/\xi,1/\xi)$, on 100 simulated datasets for each configuration, with true values $\varphi=50$, $\eta=0.5$ and $\lambda=1$ (red lines). The boxplots refer to the two-step solution for inference explained in Section \ref{section:estimation_methods}.}
    \label{fig:boxplots_sim_sgpd_copula}
\end{figure}

\begin{figure}[ht!]
    \centering
    \includegraphics[width=1\linewidth]{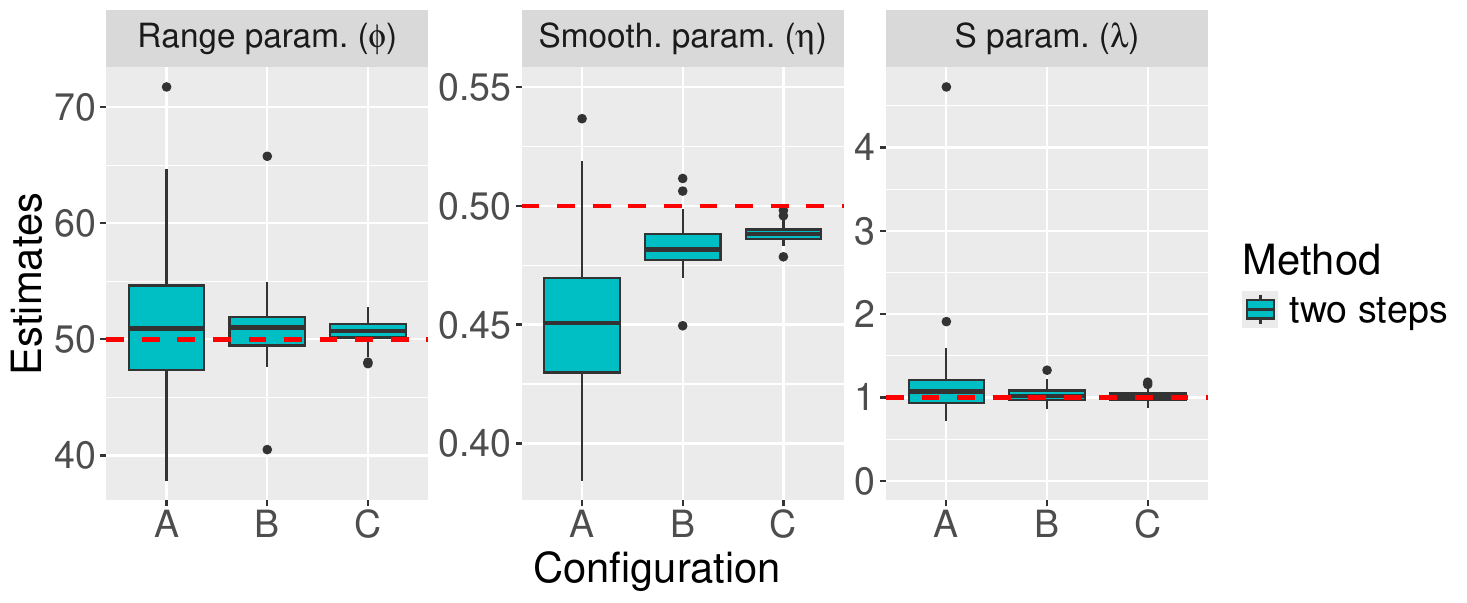}
    \caption{Estimation of Gaussian location-scale mixture 1 with copula, with $S\sim\text{Exp}(\lambda)$ and $R=\sqrt{E}$, $E\sim\text{Exp}(1/2)$, on 100 simulated datasets for each configuration, with true values $\varphi=50$, $\eta=0.5$ and $\lambda=1$ (red lines). The boxplots refer to the two-step solution for inference explained in Section \ref{section:estimation_methods}.}
    \label{fig:boxplots_sim_exp_lapl_copula}
\end{figure}

Figures \ref{fig:boxplots_sim_exp_copula}, \ref{fig:boxplots_sim_sgpd_copula} and \ref{fig:boxplots_sim_exp_lapl_copula} display analogue simulations to those of Section \ref{section:simul_consistency}, to show the consistency of the new estimation method on models belonging to the classes of Gaussian location, scale and location-scale mixtures, respectively.
As above, a systematic bias is present due to the employed rank transformation, in particular for the correlation smoothness parameter $\eta$, although this bias reduces as $m$ and $n$ are increased, e.g.\ in Configuration C.
The higher-dimensional Configuration D is not studied in these copula-based simulations, due to the increasing computational costs. An overview of the computation times for each data configuration and parameter settings is provided in Section \ref{section:simul_computation} in the main text.

\clearpage
\section{Data application}

\begin{figure}[h!]
    \centering
    \includegraphics[width=0.8\linewidth]{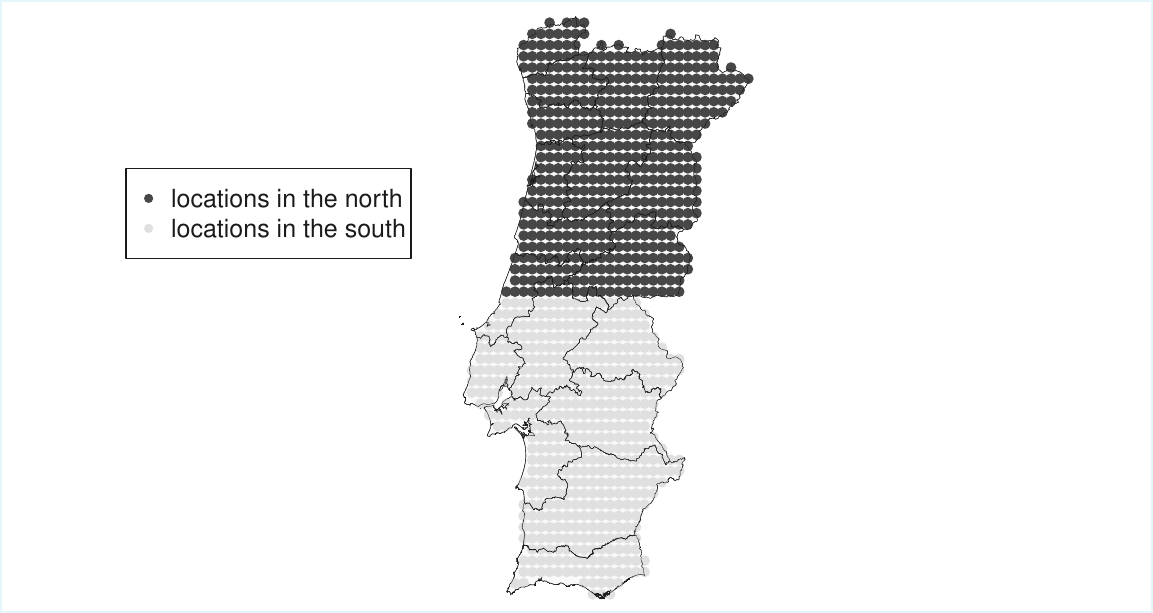}
    \caption{Division of the spatial locations into north and south of Portugal.}
    \label{fig:portugal_northsouth}
\end{figure}

Figure \ref{fig:portugal_northsouth} shows the division of the spatial locations into north and south of Portugal. The rest of the analysis focuses on the north for the reasons explained in Section \ref{section:application_datadescription} in the main text.

%Figure \ref{fig:pt_fitted_egpd_hist} shows the fitted EGPD, with all parameters constant over the spatial domain, compared to the histogram of all observed FWI values in the north of Portugal.

%\begin{figure}[h!]
%    \centering
%    \includegraphics[width=0.7\linewidth]{Figures/pt_fitted_egpd_hist.pdf}
%    \caption{Histogram of FWI over the north of Portugal. The blue line is the fitted EGPD density from an independence likelihood approach, with estimated parameter values $\hat{\sigma}=16.44$, $\hat{\xi}_{EGPD}=-0.22$, $\hat{p}=0.22$, $\hat{\kappa}_1=0.74$ and $\hat{\kappa}_2=5.81$.}
%    \label{fig:pt_fitted_egpd_hist}
%\end{figure}

\begin{figure}[h!]
\centering
\begin{subfigure}{0.195\linewidth}
\includegraphics[width=\linewidth]{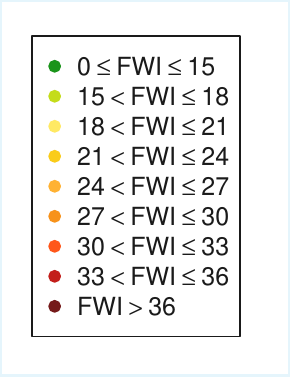}
\end{subfigure}
\begin{subfigure}{0.325\linewidth}
\includegraphics[width=\linewidth]{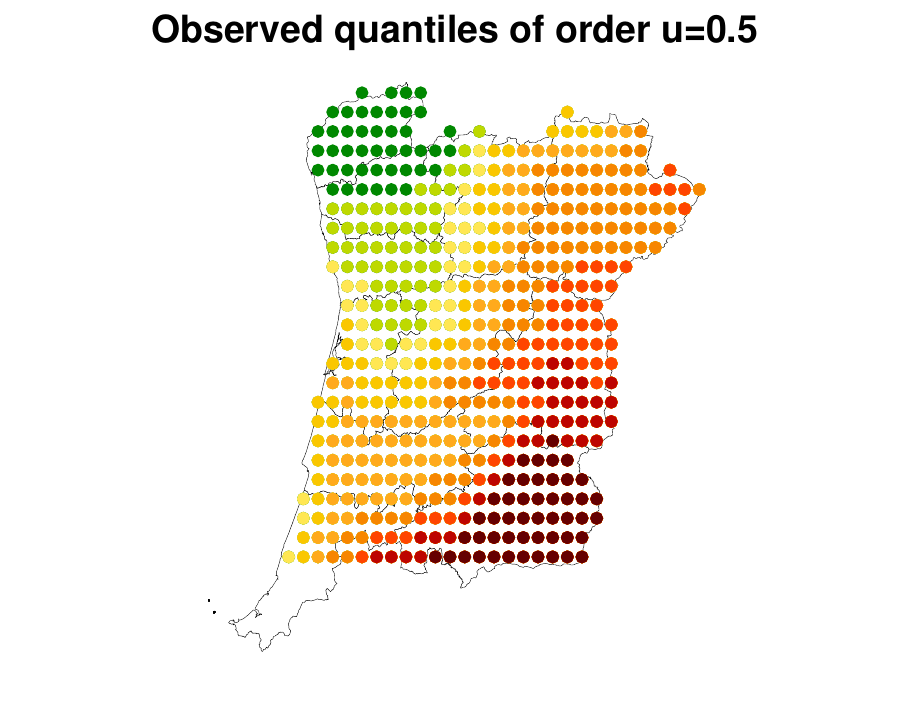}
\end{subfigure}
\begin{subfigure}{0.325\linewidth}
\includegraphics[width=\linewidth]{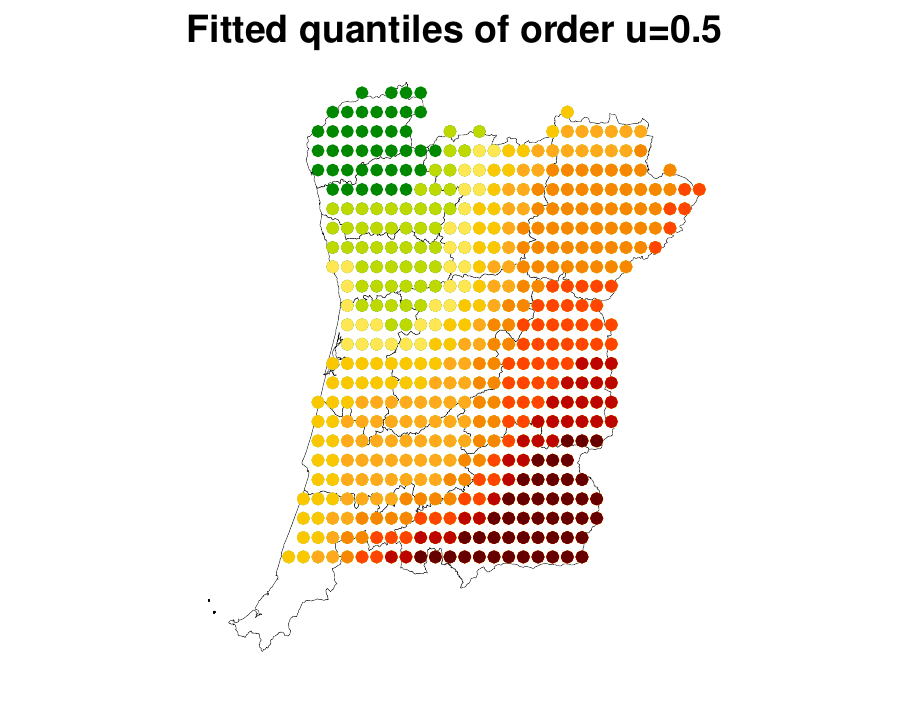}
\end{subfigure}
\begin{subfigure}{0.195\linewidth}
\includegraphics[width=\linewidth]{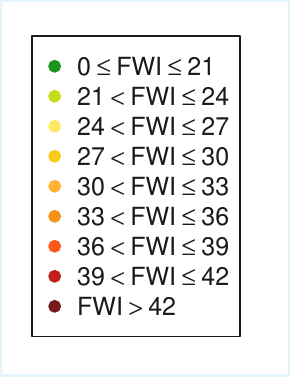}
\end{subfigure}
\begin{subfigure}{0.325\linewidth}
\includegraphics[width=\linewidth]{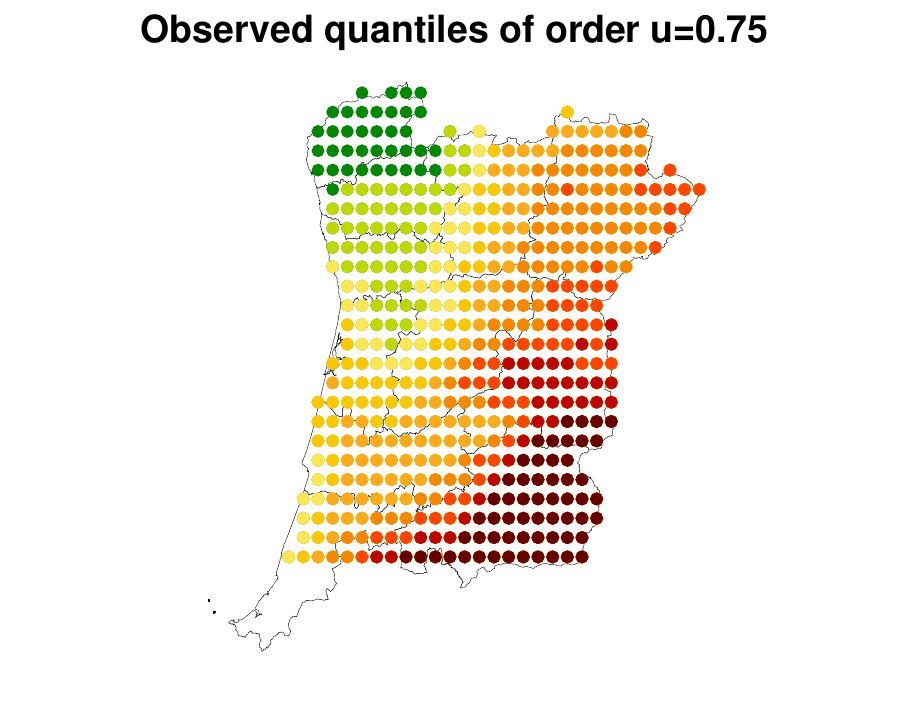}
\end{subfigure}
\begin{subfigure}{0.325\linewidth}
\includegraphics[width=\linewidth]{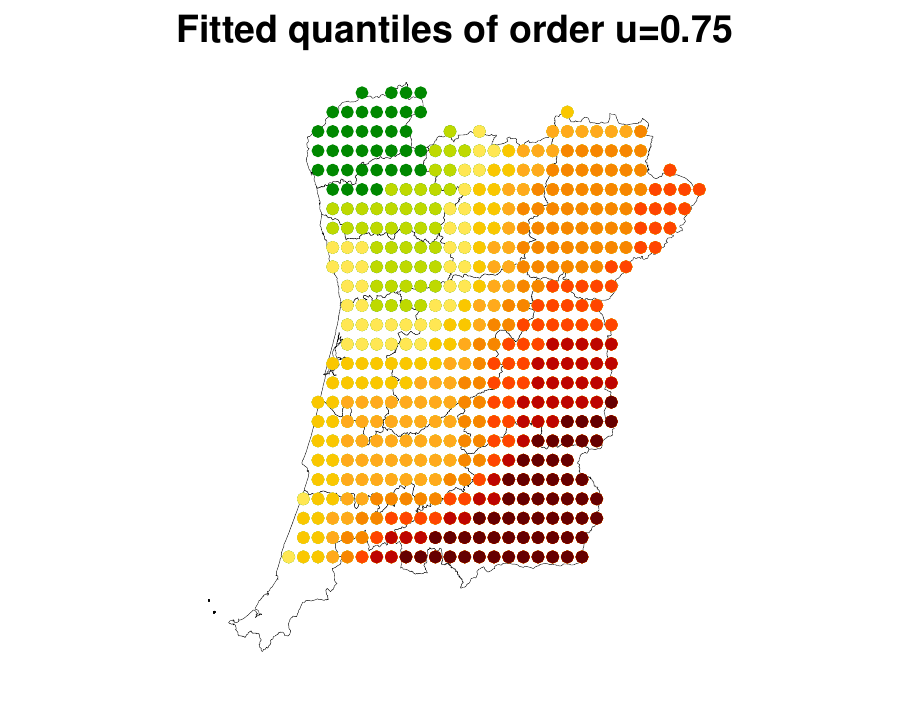}
\end{subfigure}
\begin{subfigure}{0.195\linewidth}
\includegraphics[width=\linewidth]{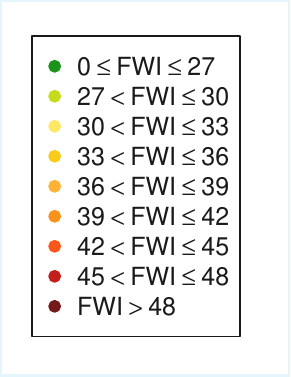}
\end{subfigure}
\begin{subfigure}{0.325\linewidth}
\includegraphics[width=\linewidth]{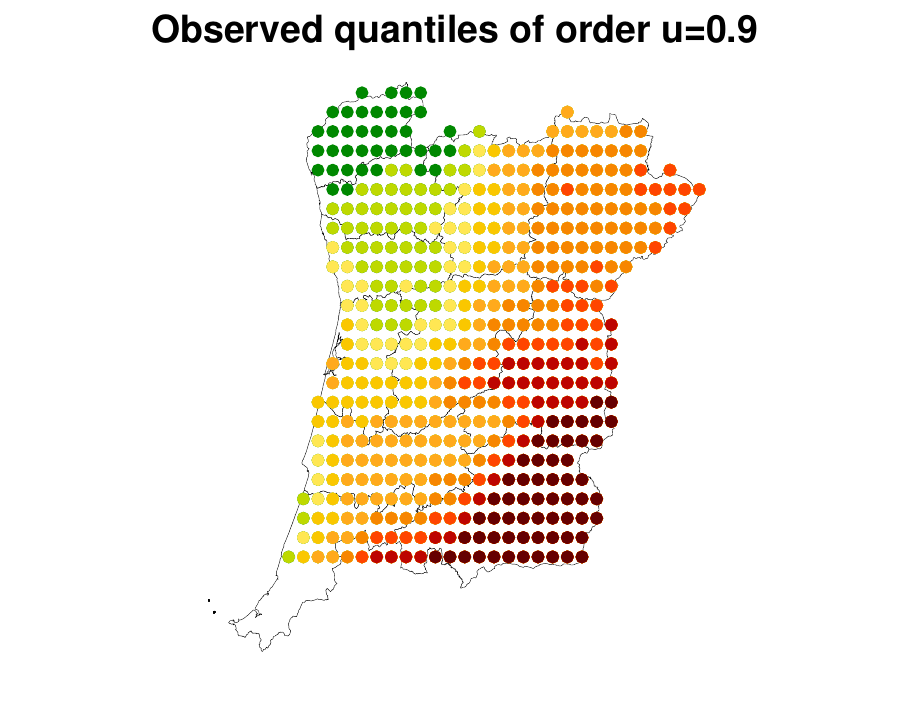}
\end{subfigure}
\begin{subfigure}{0.325\linewidth}
\includegraphics[width=\linewidth]{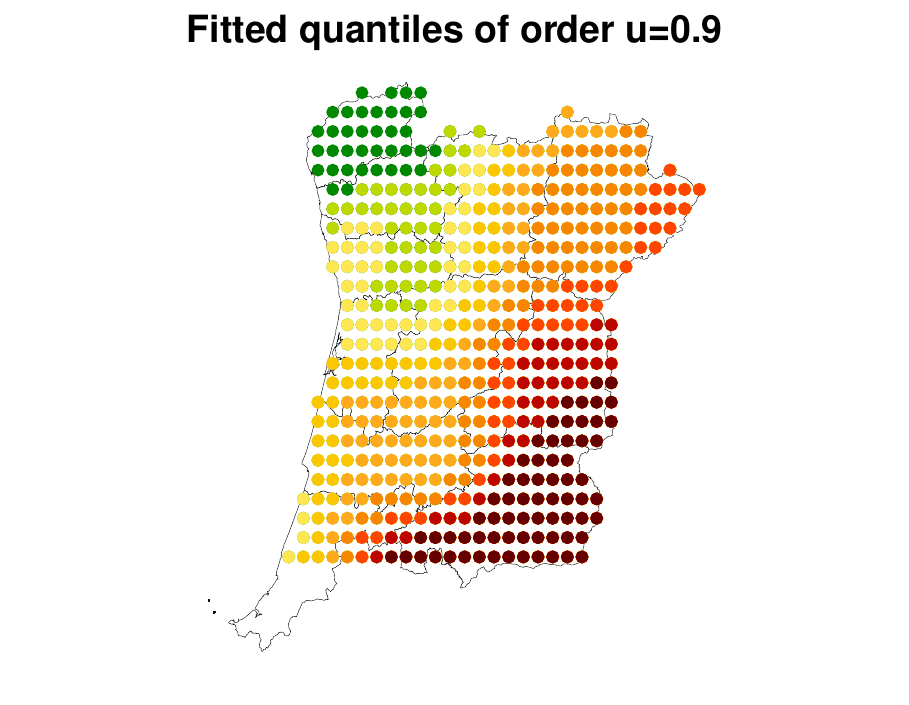}
\end{subfigure}
\begin{subfigure}{0.195\linewidth}
\includegraphics[width=\linewidth]{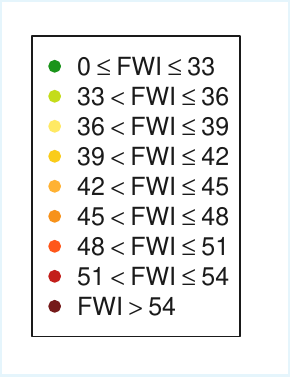}
\end{subfigure}
\begin{subfigure}{0.325\linewidth}
\includegraphics[width=\linewidth]{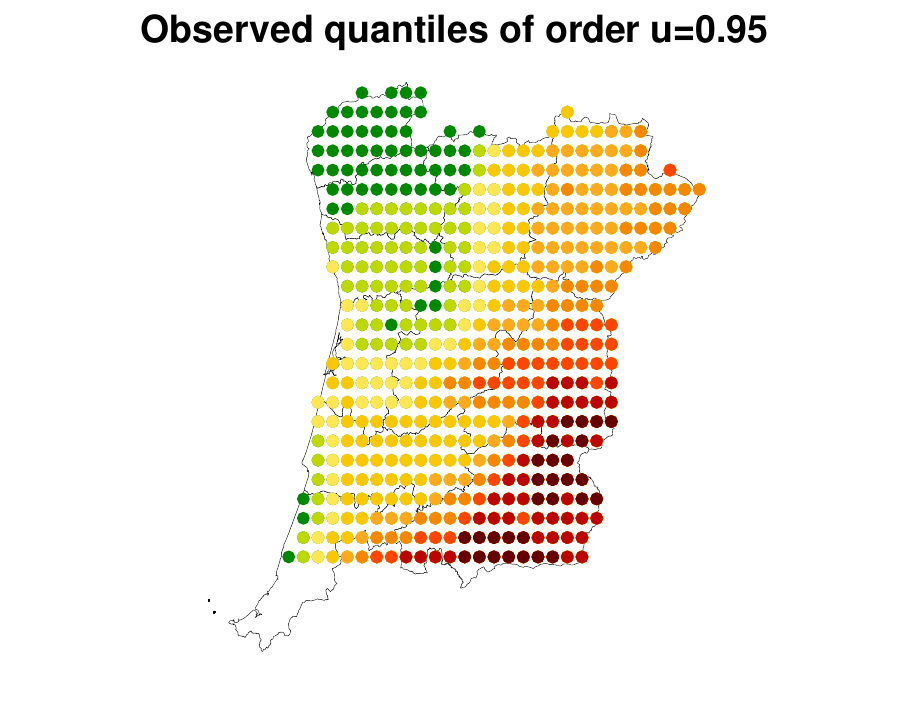}
\end{subfigure}
\begin{subfigure}{0.325\linewidth}
\includegraphics[width=\linewidth]{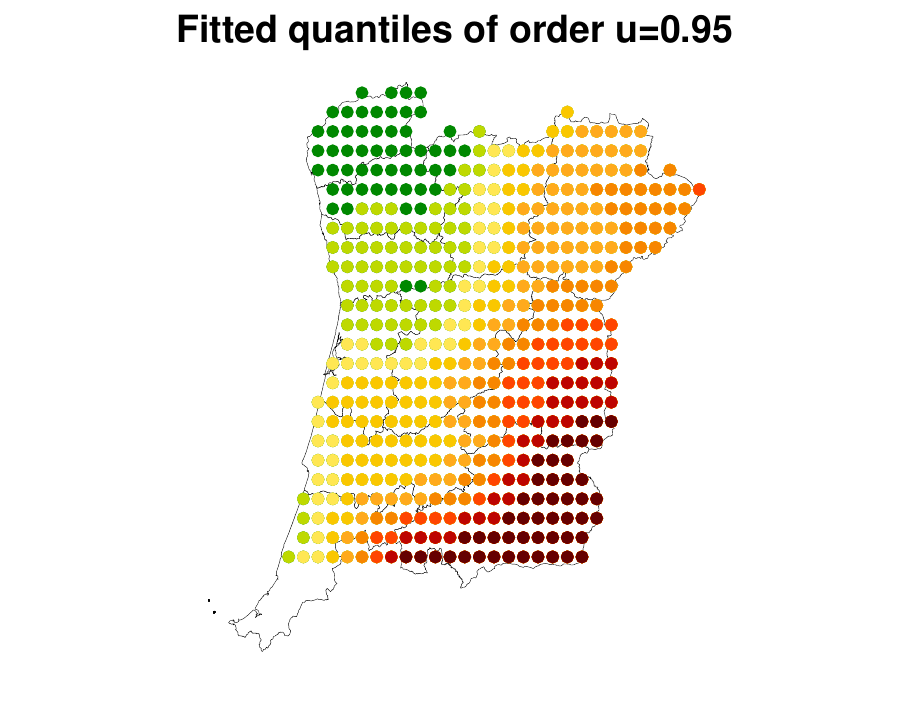}
\end{subfigure}
\caption{Comparison between empirical and fitted quantiles in the north of Portugal. Note that the colour legends are different in each row.}
\label{fig:pt_quantile_maps}
\end{figure}

Figure \ref{fig:pt_quantile_maps} compares empirical and fitted quantiles of order $u\in\{0.5,0.75,0.9,0.95\}$ for all the locations in the north of Portugal.
\clearpage

\clearpage
\subsection{Alternative dependence models}\label{section_alternative_dep_models}
For comparison, two additional copula models are fitted on the data  (and displayed in Figure \ref{fig:portugal_chi_dist} in the main text):
\begin{itemize}
    \item A Huser-Wadsworth random-scale model \citep{Huser:Wadsworth:2019}:
    $$X(\mathbf{s})=R^\delta \times W(\mathbf{s})^{1-\delta}, \quad \mathbf{s}\in\mathcal{S},$$
    where $\delta\in[0,1]$, $R$ is a standard Pareto random variable, and $W(\mathbf{s})$ Gaussian process with Matérn correlation function, transformed marginally into the standard Pareto scale. Due to the unfeasibility of likelihood evaluations for this model on $m=491$ spatial locations, inference is performed through a neural estimator \citep{zammit2024neural}, adapting the methodology of \cite{DellOro:Gaetan:2025} by replacing the grids of empirical $\chi$ statistics with five vectors of $\hat\chi(u)$, computed for $u\in\{0.5,0.75,0.9,0.95,0.99\}$ on 24 regular bins covering distances from $0$ to $240$ km. The estimated parameter values are $\hat\delta=0.492$ and, in the Matérn correlation function, $\hat\varphi=137.7$ and $\hat\eta=1.057$. The value of $\delta$, close to $0.5$, confirms the ambiguity commented above between asymptotic dependence and independence, but the model seems to overestimate the $\chi$ values for the highest quantiles;
    \item A max-stable Brown-Resnick process \citep{Brown:Resnick:1977}, estimated with pairwise likelihood using the R package \texttt{SpatialExtremes} \citep{Ribatet2022:SpatialExtremes}, with covariance model \texttt{"brown"}. The estimated range parameter is $163.0$ and the estimated smoothness is $1.130$. This model overestimates the sub-asymptotic extremal dependence for high quantiles and seems too rigid to capture the decay of $\chi$ values for increasing spatial distances.
\end{itemize}

\end{document}